\documentclass[12pt,english]{article}
\pdfoutput=1

\usepackage[T1]{fontenc}
\usepackage[utf8]{inputenc}
\usepackage{subfigure,lmodern, amsmath,amssymb, graphicx, pifont, adjustbox, bm, xcolor}
\usepackage{amsfonts}
\usepackage{geometry}
\usepackage{comment}
\usepackage{mathtools}
\usepackage{float}
\usepackage{slashed}
\usepackage{cite}
\usepackage{ragged2e}
\usepackage{etoolbox}
\usepackage{array}
\usepackage[normalem]{ulem}

\usepackage[compat=1.1.0]{tikz-feynman}
\usepackage{tikz}
\usetikzlibrary{shapes.geometric}

\tikzset{
bgsource/.style={decorate, thick, empty dot, scale=2.00},
bgraviton/.style={decorate, decoration={complete sines, amplitude=1mm, segment length=2mm, pre length=0mm, post length=0mm}, double, thick},
recoil/.style={decorate, thick, empty dot, scale=1.25, inner sep = 1pt},
sourceone/.style={decorate, thick, empty dot, scale=1.25, inner sep = 1pt},
sourcetwo/.style={decorate, thick, empty dot, scale=1.15},
binsertion/.style={decorate, thick, crossed dot, scale=1.25},
bscalar/.style={decorate, double, thick},
flatscalar/.style={decorate, thick},
myscalar/.style={decorate, dashed, thick},
gvert/.style={
},
vert/.style={decorate, dot, scale=0.75},
source/.style={decorate, thick, blob, scale = 0.37}
}

\newcommand{\duffone}{\begin{tikzpicture}
        \begin{feynman}
          \vertex (f) at (0, 0);
          \vertex[source] (c) at (0,1) {};
          \diagram*{
            (c) -- [photon, thick] (f) 
          };
        \end{feynman}
    \end{tikzpicture}
}

\newcommand{\dufftwo}{
\begin{tikzpicture}
        \begin{feynman}
          \vertex (f) at (0, 0);
          \vertex (i) at (0,0.5);
          \vertex[source] (c) at (-0.5,1) {};
          \vertex[source] (d) at (0.5,1) {};
          \diagram*{
            (f) -- [photon, thick] (i) -- [photon, thick] (c),
            (i) -- [photon, thick] (d)
          };
        \end{feynman}
    \end{tikzpicture}
}

\newcommand{\duffthreeone}{
\begin{tikzpicture}
        \begin{feynman}
          \vertex (f) at (0, 0);
          \vertex (i) at (0,0.5);
          \vertex[source] (b) at (-0.5,1) {};
          \vertex[source] (c) at (0,1) {};
          \vertex[source] (d) at (0.5,1) {};
          \diagram*{
            (f) -- [photon, thick] (i) -- [photon, thick] (b),
            (i) -- [photon, thick] (c),
            (i) -- [photon, thick] (d)
          };
        \end{feynman}
    \end{tikzpicture}
}

\newcommand{\duffthreetwo}{
\begin{tikzpicture}
        \begin{feynman}
          \vertex (f) at (0, 0);
          \vertex (i) at (0,0.3);
          \vertex (j) at (0.25,0.6);
          \vertex[source] (b) at (-0.5,1) {};
          \vertex[source] (c) at (0,1) {};
          \vertex[source] (d) at (0.5,1) {};
          \diagram*{
            (f) -- [photon, thick] (i) -- [photon, thick] (b),
            (i) -- [photon, thick] (j) -- [photon, thick] (c),
            (j) -- [photon, thick] (d)
          };
        \end{feynman}
    \end{tikzpicture}
}

\newcommand{\selfenergy}{
\begin{tikzpicture}
        \begin{feynman}
        
          \vertex (m) at (-1,0);
          \vertex (n) at (1,0);
          \vertex (c) at (-0.5,0);
          \vertex (d) at (0.5, 0);
          \diagram*{
            (m) -- [flatscalar] (n),
            (c) -- [photon, thick, half left] (d)           
          };
        \end{feynman}
    \end{tikzpicture}
}
    
\newcommand{\flatproparrow}{
\begin{tikzpicture}
        \begin{feynman}
          \vertex (i) at (-1, 0);
          \vertex (f) at (1, 0);
          \diagram*{
            (i) -- [photon, momentum={[
                  arrow shorten=0.25, 
                  xshift=0mm,
                  yshift=0mm
                ]$p$}, thick]  (f) 
          };
        \end{feynman}
    \end{tikzpicture}
}

\newcommand{\flatrecoil}{
\begin{tikzpicture}
        \begin{feynman}
          \vertex (i) at (-1, 0);
          \vertex (f) at (1, 0);
          \vertex[recoil] (c) at (0,0) {\tiny{\(H\)}};
          \diagram*{
            (i) -- [photon, thick, momentum={[
                  arrow shorten=0.25, 
                  xshift=0mm,
                  yshift=0mm
                ]$p_1$}] (c),
            (f) -- [photon, thick, momentum={[
                  arrow shorten=0.25, 
                  xshift=0mm,
                  yshift=0mm
                ]$p_2$}] (c) 
          };
        \end{feynman}
    \end{tikzpicture}
}

\newcommand{\source}{
\begin{tikzpicture}
        \begin{feynman}
          \vertex (f) at (0, 1);
          \vertex[sourceone] (c) at (0,0) {\tiny{\(L\)}};
          \diagram*{
            (c) -- [photon, thick, momentum={[
                  arrow shorten=0.25, 
                  xshift=0mm,
                  yshift=0mm
                ]$p$}] (f) 
          };
        \end{feynman}
    \end{tikzpicture}
}

\newcommand{\sourcezero}{
\begin{tikzpicture}
        \begin{feynman}
          \vertex (f) at (0, 1);
          \vertex[sourcetwo] (c) at (0,0) {\tiny{\(L_0\)}};
          \diagram*{
            (c) -- [photon, thick, momentum={[
                  arrow shorten=0.25, 
                  xshift=0mm,
                  yshift=0mm
                ]$p$}] (f) 
          };
        \end{feynman}
    \end{tikzpicture}
}

\newcommand{\sourceone}{
\begin{tikzpicture}
        \begin{feynman}
          \vertex (f) at (0, 1);
          \vertex[sourcetwo] (c) at (0,0) {\tiny{\(L_1\)}};
          \diagram*{
            (c) -- [photon, thick, momentum={[
                  arrow shorten=0.25, 
                  xshift=0mm,
                  yshift=0mm
                ]$p$}] (f) 
          };
        \end{feynman}
    \end{tikzpicture}
}

\newcommand{\brecoilEM}{
\begin{tikzpicture}
        \begin{feynman}
          \vertex[recoil] (c) at (0,1.0) {\tiny{\(H\)}};
          \vertex[sourceone] (i) at (-1, 0 ) {\tiny{\(L\)}};
          \vertex[sourceone] (f) at (1, 0) {\tiny{\(L\)}};
          \diagram*{
            (c) -- [photon, thick, quarter right] (i), (c) --  [photon, thick, quarter left] (f),
          };
        \end{feynman}
    \end{tikzpicture}
    }

\newcommand{\scalarprop}{
\begin{tikzpicture}
        \begin{feynman}
          \vertex (top) at (0, 1);
          \vertex (i) at (-1, 0);
          \vertex (f) at (1, 0);
          \diagram*{
            (i) -- [flatscalar, momentum={[
                  arrow shorten=0.25, 
                  xshift=0mm,
                  yshift=0mm
                ]$k$}]  (f) 
          };
        \end{feynman}
    \end{tikzpicture}
}

\newcommand{\backgroundone}{
\begin{tikzpicture}
        \begin{feynman}
          \vertex (top) at (0, 1);
          \vertex (i) at (-1, 0);
          \vertex (c) at (0, 0);
          \vertex (f) at (1, 0);
          \vertex[binsertion] (c) at (0,0) {}; 
          \diagram*{
            (i) -- [flatscalar] (c) -- [flatscalar] (f) 
          };
        \end{feynman}
    \end{tikzpicture}
}

\newcommand{\backgroundonearrow}{
\begin{tikzpicture}
        \begin{feynman}
          \vertex (top) at (0, 1);
          \vertex (i) at (-1, 0);
          \vertex (c) at (0, 0);
          \vertex (f) at (1, 0);
          \vertex[binsertion] (c) at (0,0) {};
          \vertex (a) at (0,1.0);
          \diagram*{
            (c) -- [flatscalar,rmomentum={[
                  arrow shorten=0.25, 
                  xshift=0mm,
                  yshift=0mm
                ]$k_1$}] (i), (f) -- [flatscalar,momentum={[
                  arrow shorten=0.25, 
                  xshift=0mm,
                  yshift=0mm
                ]$k_2$}] (c),
                (a) -- [draw=none, momentum={[
                  arrow shorten=0.25, 
                  xshift=-3.0mm,
                  yshift=0mm]$q$}](c)
          };
        \end{feynman}
    \end{tikzpicture}
}

\newcommand{\backgroundoneone}{
\begin{tikzpicture}
        \begin{feynman}
          \vertex (top) at (0, 1);
          \vertex (i) at (-1, 0);
          \vertex (c) at (0, 0);
          \vertex (f) at (1, 0);
          \vertex[binsertion] (c) at (0,0) {};
          \vertex [below right=0.75em of c] {\tiny\(1\)};
          \diagram*{
            (i) -- [flatscalar] (c) -- [flatscalar] (f) 
          };
        \end{feynman}
    \end{tikzpicture}
}

\newcommand{\backgroundonetwo}{
\begin{tikzpicture}
        \begin{feynman}
          \vertex (top) at (0, 1);
          \vertex (i) at (-1, 0);
          \vertex (c) at (0, 0);
          \vertex (f) at (1, 0);
          \vertex[binsertion] (c) at (0,0) {};
          \vertex [below right=0.75em of c] {\tiny\(2\)};
          \diagram*{
            (i) -- [flatscalar] (c) -- [flatscalar] (f) 
          };
        \end{feynman}
    \end{tikzpicture}
}
\newcommand{\backgroundonethree}{
\begin{tikzpicture}
        \begin{feynman}
          \vertex (top) at (0, 1);
          \vertex (i) at (-1, 0);
          \vertex (c) at (0, 0);
          \vertex (f) at (1, 0);
          \vertex[binsertion] (c) at (0,0) {};
          \vertex [below right=0.75em of c] {\tiny\(3\)};
          \diagram*{
            (i) -- [flatscalar] (c) -- [flatscalar] (f) 
          };
        \end{feynman}
    \end{tikzpicture}
}

\newcommand{\tree}{
\begin{tikzpicture}
        \begin{feynman}
          \vertex[source] (top) at (0, 1) {};
          \vertex (top1) at (-0.5, 1);
          \vertex (top2) at (0.5, 1);
          \vertex (i) at (-1, 0);
          \vertex[gvert] (c) at (0, 0) {};
          \vertex (f) at (1, 0);
          \vertex[thick] (c) at (0,0); 
          \diagram*{
            (i) -- [flatscalar] (c) -- [flatscalar] (f), 
            (top) -- [photon, thick] (c)
          };
        \end{feynman}
    \end{tikzpicture}
    }

\newcommand{\contacttriangle}{
    \begin{tikzpicture}
        \begin{feynman}
          \vertex[source] (top1) at (-0.5, 1) {};
          \vertex[source] (top2) at (0.5, 1) {};
          \vertex (i) at (-1, 0);
          \vertex[gvert] (c) at (0, 0);
          \vertex (f) at (1, 0);
          \diagram*{
            (i) -- [flatscalar] (c) -- [flatscalar] (f), 
            (top1) -- [photon, thick] (c),
            (top2) -- [photon, thick] (c) 
          };
        \end{feynman}
    \end{tikzpicture}
    }

\newcommand{\triangl}{
    \begin{tikzpicture}
        \begin{feynman}
          \vertex[source] (top1) at (-0.5, 1) {};
          \vertex[source] (top2) at (0.5, 1) {};
          \vertex (i) at (-1, 0);
          \vertex[gvert] (c) at (0, 0);
          \vertex[gvert] (c2) at (0, 0.5);
          \vertex (f) at (1, 0);
          \vertex[thick] (c) at (0,0); 
          \diagram*{
            (i) -- [flatscalar] (c) -- [flatscalar] (f), 
            (top1) -- [photon, thick] (c2),
            (top2) -- [photon, thick] (c2),
            (c2) -- [photon, thick] (c)
          };
        \end{feynman}
    \end{tikzpicture}
    }

\newcommand{\wdiagram}{
    \begin{tikzpicture}
        \begin{feynman}
          \vertex[source] (top) at (0, 1) {};
          \vertex[source] (top1) at (-0.75, 1) {};
          \vertex[source] (top2) at (0.75, 1) {};
          \vertex (i) at (-1, 0);
          \vertex[gvert] (c) at (0, 0);
          \vertex (f) at (1, 0);
          \vertex[thick] (c) at (0,0); 
          \diagram*{
            (i) -- [flatscalar] (c) -- [flatscalar] (f), 
            (top) -- [photon, thick] (c),
            (top1) -- [photon, thick] (c),
            (top2) -- [photon, thick] (c)
          };
        \end{feynman}
    \end{tikzpicture}
}

\newcommand{\triY}{
    \begin{tikzpicture}
        \begin{feynman}
          \vertex[source] (top) at (0, 1) {};
          \vertex[source] (top1) at (-0.75, 1) {};
          \vertex[source] (top2) at (0.75, 1) {};
          \vertex (i) at (-1, 0);
          \vertex[gvert] (c) at (0, 0);
          \vertex[gvert] (c2) at (0, 0.5);
          \vertex (f) at (1, 0);
          \vertex[thick] (c) at (0,0); 
          \diagram*{
            (i) -- [flatscalar] (c) -- [flatscalar] (f), 
            (top) -- [photon, thick] (c2),
            (top1) -- [photon, thick] (c2),
            (top2) -- [photon, thick] (c2),
            (c2) -- [photon, thick] (c)
          };
        \end{feynman}
    \end{tikzpicture}
}

\newcommand{\othertriY}{
    \begin{tikzpicture}
        \begin{feynman}
          \vertex[source] (top) at (0, 1) {};
          \vertex[source] (top1) at (-0.75, 1) {};
          \vertex[source] (top2) at (0.75, 1) {};
          \vertex (i) at (-1, 0);
          \vertex[gvert] (c) at (0, 0);
          \vertex[gvert] (c2) at (0, 0.5);
          \vertex[gvert] (c3) at (-0.375, 0.75);
          \vertex (f) at (1, 0);
          \vertex[thick] (c) at (0,0); 
          \diagram*{
            (i) -- [flatscalar] (c) -- [flatscalar] (f), 
            (top) -- [photon, thick] (c3),
            (top1) -- [photon, thick] (c3) -- [photon, thick] (c2),
            (top2) -- [photon, thick] (c2),
            (c2) -- [photon, thick] (c)
          };
        \end{feynman}
    \end{tikzpicture}
    }
    
\newcommand{\scalarfan}{
     \begin{tikzpicture}
        \begin{feynman}
          \vertex[source] (top) at (-0.25, 1) {};
          \vertex[source] (top1) at (-1, 1) {};
          \vertex[source] (top2) at (1, 1) {};
          \vertex (dots) at (0.3,0.75) {$\cdots$};
          \vertex (i) at (-1, 0);
          \vertex[gvert] (c) at (0, 0) {};
          \vertex (f) at (1, 0);
          \vertex[thick] (c) at (0,0); 
          \diagram*{
            (i) -- [flatscalar] (c) -- [flatscalar] (f), 
            (top) -- [flatscalar] (c),
            (top1) -- [flatscalar] (c),
            (top2) -- [flatscalar] (c)
          };
        \end{feynman}
    \end{tikzpicture}
    }

\newcommand{\trees}{
\begin{tikzpicture}
        \begin{feynman}
          \vertex[source] (top) at (0, 1) {};
          \vertex (top1) at (-0.5, 1);
          \vertex (top2) at (0.5, 1);
          \vertex (i) at (-1, 0);
          \vertex[gvert] (c) at (0, 0) {};
          \vertex (f) at (1, 0);
          \vertex[thick] (c) at (0,0); 
          \diagram*{
            (i) -- [flatscalar] (c) -- [flatscalar] (f), 
            (top) -- [flatscalar] (c)
          };
        \end{feynman}
    \end{tikzpicture}
    }

    \newcommand{\contacttriangles}{
    \begin{tikzpicture}
        \begin{feynman}
          \vertex[source] (top1) at (-0.5, 1) {};
          \vertex[source] (top2) at (0.5, 1) {};
          \vertex (i) at (-1, 0);
          \vertex[gvert] (c) at (0, 0);
          \vertex (f) at (1, 0);
          \diagram*{
            (i) -- [flatscalar] (c) -- [flatscalar] (f), 
            (top1) -- [flatscalar] (c),
            (top2) -- [flatscalar] (c) 
          };
        \end{feynman}
    \end{tikzpicture}
    }

    \newcommand{\wdiagrams}{
    \begin{tikzpicture}
        \begin{feynman}
          \vertex[source] (top) at (0, 1) {};
          \vertex[source] (top1) at (-0.75, 1) {};
          \vertex[source] (top2) at (0.75, 1) {};
          \vertex (i) at (-1, 0);
          \vertex[gvert] (c) at (0, 0);
          \vertex (f) at (1, 0);
          \vertex[thick] (c) at (0,0); 
          \diagram*{
            (i) -- [flatscalar] (c) -- [flatscalar] (f), 
            (top) -- [flatscalar] (c),
            (top1) -- [flatscalar] (c),
            (top2) -- [flatscalar] (c)
          };
        \end{feynman}
    \end{tikzpicture}
}

\newcommand{\bprop}{
\begin{tikzpicture}
        \begin{feynman}
          \vertex[sourceone] (i) at (-1, 0 ) {\tiny{\(L\)}};
          \vertex[sourceone] (f) at (1, 0) {\tiny{\(L\)}}; 
          \diagram*{
            (i) -- [bgraviton, half left]  (f)
          };
        \end{feynman}
    \end{tikzpicture}
    }

\newcommand{\propzerozero}{
\begin{tikzpicture}
        \begin{feynman}
          \vertex[sourcetwo] (i) at (-1, 0 ) {\tiny{\(L_0\)}};
          \vertex[sourcetwo] (f) at (1, 0) {\tiny{\(L_0\)}};
          \diagram*{
            (i) -- [photon, thick, half left]  (f)
          };
        \end{feynman}
    \end{tikzpicture}
    }

\newcommand{\proponeone}{
\begin{tikzpicture}
        \begin{feynman}
          \vertex[sourcetwo] (i) at (-1, 0 ) {\tiny{\(L_0\)}};
          \vertex[sourcetwo] (f) at (1, 0) {\tiny{\(L_1\)}};
          \diagram*{
            (i) -- [photon, thick, half left]  (f)
          };
        \end{feynman}
    \end{tikzpicture}
    }

\newcommand{\proponetwo}{
\begin{tikzpicture}
        \begin{feynman}
          \vertex[sourcetwo] (i) at (-1, 0 ) {\tiny{\(L_1\)}};
          \vertex[sourcetwo] (f) at (1, 0) {\tiny{\(L_0\)}};
          \diagram*{
            (i) -- [photon, thick, half left]  (f)
          };
        \end{feynman}
    \end{tikzpicture}
    }

\newcommand{\proponethree}{
\begin{tikzpicture}
        \begin{feynman}
          \vertex[binsertion] (c) at (0,1.0) {};
          \vertex [below right=0.75em of c] {\tiny\(1\)};
          \vertex[sourcetwo] (i) at (-1, 0 ) {\tiny{\(L_0\)}};
          \vertex[sourcetwo] (f) at (1, 0) {\tiny{\(L_0\)}};
          \diagram*{
            (c) -- [photon, thick, quarter right] (i), (c) --  [photon, thick, quarter left] (f),
          };
        \end{feynman}
    \end{tikzpicture}
    }    

\newcommand{\proptwotwo}{
\begin{tikzpicture}
        \begin{feynman}
          \vertex[binsertion] (c) at (0,1.0) {};
          \vertex [below right=0.75em of c] {\tiny\(1\)};
          \vertex[sourcetwo] (i) at (-1, 0 ) {\tiny{\(L_0\)}};
          \vertex[sourcetwo] (f) at (1, 0) {\tiny{\(L_1\)}};
          \diagram*{
            (c) -- [photon, thick, quarter right] (i), (c) --  [photon, thick, quarter left] (f),
          };
        \end{feynman}
    \end{tikzpicture}
    }

\newcommand{\proptwothree}{
\begin{tikzpicture}
        \begin{feynman}
          \vertex[binsertion] (c) at (0,1.0) {};
          \vertex [below right=0.75em of c] {\tiny\(1\)};
          \vertex[sourcetwo] (i) at (-1, 0 ) {\tiny{\(L_1\)}};
          \vertex[sourcetwo] (f) at (1, 0) {\tiny{\(L_0\)}};
          \diagram*{
            (c) -- [photon, thick, quarter right] (i), (c) --  [photon, thick, quarter left] (f),
          };
        \end{feynman}
    \end{tikzpicture}
    } 

\newcommand{\proptwofour}{
\begin{tikzpicture}
        \begin{feynman}
          \vertex[binsertion] (d) at (0.60,0.80) {};
          \vertex [below right=0.75em of d] {\tiny\(1\)};
          \vertex[binsertion] (c) at (-0.60,0.80) {};
          \vertex [below right=0.75em of c] {\tiny\(1\)};
          \vertex[sourcetwo] (i) at (-1, 0 ) {\tiny{\(L_0\)}};
          \vertex[sourcetwo] (f) at (1, 0) {\tiny{\(L_0\)}};
          \diagram*{
            (i) -- [photon, thick, /tikz/out=90 ,/tikz/in=200] (c) -- [photon, thick, /tikz/out=40 ,/tikz/in=140] (d) --  [photon, thick,/tikz/out=340 ,/tikz/in=90] (f),
          };
        \end{feynman}
    \end{tikzpicture}
    }

\newcommand{\proptwofive}{
\begin{tikzpicture}
        \begin{feynman}
          \vertex[binsertion] (c) at (0,1.0) {};
          \vertex [below right=0.75em of c] {\tiny\(2\)};
          \vertex[sourcetwo] (i) at (-1, 0 ) {\tiny{\(L_0\)}};
          \vertex[sourcetwo] (f) at (1, 0) {\tiny{\(L_0\)}};
          \diagram*{
            (c) -- [photon, thick, quarter right] (i), (c) --  [photon, thick, quarter left] (f),
          };
        \end{feynman}
    \end{tikzpicture}
    }

\newcommand{\brecoil}{
\begin{tikzpicture}
        \begin{feynman}
          \vertex[recoil] (c) at (0,1.0) {\tiny{\(H\)}};
          \vertex[sourceone] (i) at (-1, 0 ) {\tiny{\(L\)}};
          \vertex[sourceone] (f) at (1, 0) {\tiny{\(L\)}};
          \diagram*{
            (c) -- [bgraviton, quarter right] (i), (c) --  [bgraviton, quarter left] (f),
          };
        \end{feynman}
    \end{tikzpicture}
    }

\newcommand{\recoilone}{
\begin{tikzpicture}
        \begin{feynman}
          \vertex[recoil] (c) at (0,1.0) {\tiny{\(H\)}};
          \vertex[sourcetwo] (i) at (-1, 0 ) {\tiny{\(L_0\)}};
          \vertex[sourcetwo] (f) at (1, 0) {\tiny{\(L_0\)}};
          \diagram*{
            (c) -- [photon, thick, quarter right] (i), (c) --  [photon, thick, quarter left] (f),
          };
        \end{feynman}
    \end{tikzpicture}
    }

\newcommand{\recoiltwoone}{
\begin{tikzpicture}
        \begin{feynman}
          \vertex[recoil] (c) at (0,1.0) {\tiny{\(H\)}};
          \vertex[sourcetwo] (i) at (-1, 0 ) {\tiny{\(L_0\)}};
          \vertex[sourcetwo] (f) at (1, 0) {\tiny{\(L_1\)}};
          \diagram*{
            (c) -- [photon, thick, quarter right] (i), (c) --  [photon, thick, quarter left] (f),
          };
        \end{feynman}
    \end{tikzpicture}
    }    

\newcommand{\recoiltwotwo}{
\begin{tikzpicture}
        \begin{feynman}
          \vertex[recoil] (c) at (0,1.0) {\tiny{\(H\)}};
          \vertex[sourcetwo] (i) at (-1, 0 ) {\tiny{\(L_1\)}};
          \vertex[sourcetwo] (f) at (1, 0) {\tiny{\(L_0\)}};
          \diagram*{
            (c) -- [photon, thick, quarter right] (i), (c) --  [photon, thick, quarter left] (f),
          };
        \end{feynman}
    \end{tikzpicture}
    }

\newcommand{\recoiltwothree}{
\begin{tikzpicture}
        \begin{feynman}
          \vertex[recoil] (c) at (-0.60,0.80) {\tiny{\(H\)}};
          \vertex[binsertion] (d) at (0.60,0.80) {};
          \vertex [below right=0.75em of d] {\tiny\(1\)};
          \vertex[sourcetwo] (i) at (-1, 0 ) {\tiny{\(L_0\)}};
          \vertex[sourcetwo] (f) at (1, 0) {\tiny{\(L_0\)}};
          \diagram*{
            (i) -- [photon, thick, /tikz/out=90 ,/tikz/in=200] (c) -- [photon, thick, /tikz/out=40 ,/tikz/in=140] (d) --  [photon, thick,/tikz/out=340 ,/tikz/in=90] (f),
          };
        \end{feynman}
    \end{tikzpicture}
    }

\newcommand{\recoiltwofour}{
\begin{tikzpicture}
        \begin{feynman}
          \vertex[recoil] (d) at (0.60,0.80) {\tiny{\(H\)}};
          \vertex[binsertion] (c) at (-0.60,0.80) {};
          \vertex [below right=0.75em of c] {\tiny\(1\)};
          \vertex[sourcetwo] (i) at (-1, 0 ) {\tiny{\(L_0\)}};
          \vertex[sourcetwo] (f) at (1, 0) {\tiny{\(L_0\)}};
          \diagram*{
            (i) -- [photon, thick, /tikz/out=90 ,/tikz/in=200] (c) -- [photon, thick, /tikz/out=40 ,/tikz/in=140] (d) --  [photon, thick,/tikz/out=340 ,/tikz/in=90] (f),
          };
        \end{feynman}
    \end{tikzpicture}
    }

\newcommand{\fpropone}{
\begin{tikzpicture}
    \begin{feynman}
        \vertex[] (i) at (-1.5, 0 );
        \vertex[] (a) at (-0.5, 0); 
        \vertex[] (b) at (0.5, 0);
        \vertex[] (f) at (1.5, 0);
        \vertex[] (j) at (-1.5,1.5);
        \vertex[] (c) at (0, 0.75);
        \vertex[] (d) at (0,1.5);
        \vertex[] (k) at (1.5, 1.5);

        \diagram*{
            (i) -- [scalar] (a) -- [scalar] (b) -- [scalar] (f),
            (a) -- [photon, thick] (c) -- [photon, thick] (b),
            (d) -- [photon, thick] (c),
            (j) -- [scalar] (k)
          };
    \end{feynman}
\end{tikzpicture}
}

\newcommand{\fproptwo}{
\begin{tikzpicture}
    \begin{feynman}
        \vertex[] (i) at (-1.5, 0 );
        \vertex[] (a) at (-0.5, 0); 
        \vertex[] (b) at (0.5, 0);
        \vertex[] (f) at (1.5, 0);
        \vertex[] (j) at (-1.5,1.5);
        \vertex[] (c) at (0, 0.75);
        \vertex[] (d) at (0,1.5);
        \vertex[] (k) at (1.5, 1.5);

        \diagram*{
            (i) -- [scalar] (a) -- [scalar] (b) -- [scalar] (f),
            (a) -- [photon, thick] (d) -- [photon, thick] (b),
            (j) -- [scalar] (k)
          };
    \end{feynman}
\end{tikzpicture}
}

\newcommand{\fpropthree}{
\begin{tikzpicture}
    \begin{feynman}
        \vertex[] (i) at (-1.5, 0 );
        \vertex[] (a) at (-0.5, 0); 
        \vertex[] (b) at (0.5, 0);
        \vertex[] (f) at (1.5, 0);
        \vertex[] (j) at (-1.5,1.5);
        \vertex[] (c) at (-0.5, 1.5);
        \vertex[] (d) at (0.5,1.5);
        \vertex[] (k) at (1.5, 1.5);
        \vertex[] (g) at (-0.5,0.75);
        \vertex[] (h) at (0.5, 0.75);

        \diagram*{
            (i) -- [scalar] (a) -- [scalar] (b) -- [scalar] (f),
            (a) -- [photon, thick] (g)-- [photon, thick] (c),
            (b) -- [photon, thick] (h)-- [photon, thick] (d),
            (g) -- [photon, thick] (h),
            (j) -- [scalar] (k)
          };
    \end{feynman}
\end{tikzpicture}
}

\newcommand{\fpropfour}{
\begin{tikzpicture}
    \begin{feynman}
        \vertex[] (i) at (-1.5, 0 );
        \vertex[] (a) at (-0.5, 0); 
        \vertex[] (b) at (0.5, 0);
        \vertex[] (f) at (1.5, 0);
        \vertex[] (j) at (-1.5,1.5);
        \vertex[] (c) at (-0.5, 1.5);
        \vertex[] (d) at (0.5,1.5);
        \vertex[] (k) at (1.5, 1.5);
        \vertex[] (g) at (0,0.5);
        \vertex[] (h) at (0, 1.0);

        \diagram*{
            (i) -- [scalar] (a) -- [scalar] (b) -- [scalar] (f),
            (a) -- [photon, thick] (g)-- [photon, thick] (b),
            (c) -- [photon, thick] (h)-- [photon, thick] (d),
            (g) -- [photon, thick] (h),
            (j) -- [scalar] (k)
          };
    \end{feynman}
\end{tikzpicture}
}

\newcommand{\fpropfive}{
\begin{tikzpicture}
    \begin{feynman}
        \vertex[] (i) at (-1.5, 0 );
        \vertex[] (a) at (-0.5, 0); 
        \vertex[] (b) at (0.5, 0);
        \vertex[] (f) at (1.5, 0);
        \vertex[] (j) at (-1.5,1.5);
        \vertex[] (c) at (-0.5, 1.5);
        \vertex[] (d) at (0.5,1.5);
        \vertex[] (k) at (1.5, 1.5);
        \vertex[] (g) at (0,0.75);

        \diagram*{
            (i) -- [scalar] (a) -- [scalar] (b) -- [scalar] (f),
            (a) -- [photon, thick] (g)-- [photon, thick] (c),
            (b) -- [photon, thick] (g)-- [photon, thick] (d),
            (j) -- [scalar] (k)
          };
    \end{feynman}
\end{tikzpicture}
}

\newcommand{\fpropsix}{
\begin{tikzpicture}
    \begin{feynman}
        \vertex[] (i) at (-1.5, 0 );
        \vertex[] (a) at (-0.75, 0); 
        \vertex[] (b) at (0.75, 0);
        \vertex[] (f) at (1.5, 0);
        \vertex[] (j) at (-1.5,1.5);
        \vertex[] (c) at (-0.375, 1.5);
        \vertex[] (d) at (0.75,1.5);
        \vertex[] (k) at (1.5, 1.5);
        \vertex[] (g) at (-0.375,0.75);
        \vertex[] (h) at (0, 0);

        \diagram*{
            (i) -- [scalar] (h) -- [flatscalar] (b) -- [scalar] (f),
            (a) -- [photon, thick] (g)-- [photon, thick] (h),
            (b) -- [photon, thick] (d),
            (g) -- [photon, thick] (c),
            (j) -- [scalar] (k)
          };
    \end{feynman}
\end{tikzpicture}
}

\newcommand{\fpropseven}{
\begin{tikzpicture}
    \begin{feynman}
        \vertex[] (i) at (-1.5, 0 );
        \vertex[] (a) at (-0.75, 0); 
        \vertex[] (b) at (0.75, 0);
        \vertex[] (f) at (1.5, 0);
        \vertex[] (j) at (-1.5,1.5);
        \vertex[] (c) at (-0.375, 1.5);
        \vertex[] (d) at (0.75,1.5);
        \vertex[] (k) at (1.5, 1.5);
        \vertex[] (h) at (0, 0);

        \diagram*{
            (i) -- [scalar] (h) -- [flatscalar] (b) -- [scalar] (f),
            (a) -- [photon, thick] (c)-- [photon, thick] (h),
            (b) -- [photon, thick] (d),
            (j) -- [scalar] (k)
          };
    \end{feynman}
\end{tikzpicture}
}

\newcommand{\frecone}{
\begin{tikzpicture}
    \begin{feynman}
        \vertex[] (i) at (-1.5, 0 );
        \vertex[] (a) at (-0.5, 0); 
        \vertex[] (b) at (0.5, 0);
        \vertex[] (f) at (1.5, 0);
        \vertex[] (j) at (-1.5,1.5);
        \vertex[] (c) at (-0.5,1.5);
        \vertex[] (d) at (0.5,1.5);
        \vertex[] (k) at (1.5, 1.5);

        \diagram*{
            (i) -- [scalar] (a) -- [scalar] (b) -- [scalar] (f),
            (a) -- [photon, thick] (c),
            (b) -- [photon, thick] (d),
            (j) -- [scalar] (c) -- [flatscalar] (d) -- [scalar] (k)
          };
    \end{feynman}
\end{tikzpicture}
}

\newcommand{\frectwo}{
\begin{tikzpicture}
    \begin{feynman}
        \vertex[] (i) at (-1.5, 0 );
        \vertex[] (a) at (-0.75, 0); 
        \vertex[] (b) at (0.75, 0);
        \vertex[] (f) at (1.5, 0);
        \vertex[] (j) at (-1.5,1.5);
        \vertex[] (c) at (-0.75,1.5);
        \vertex[] (d) at (0.75,1.5);
        \vertex[] (k) at (1.5, 1.5);
        \vertex[] (g) at (0,0);
        \vertex[] (h) at (0, 1.5);

        \diagram*{
            (i) -- [scalar] (g) -- [flatscalar] (b) -- [scalar] (f),
            (a) -- [photon, thick] (c),
            (b) -- [photon, thick] (d),
            (g) -- [photon, thick] (h),
            (j) -- [scalar] (c) -- [flatscalar] (h) -- [scalar] (k)
          };
    \end{feynman}
\end{tikzpicture}
}

\newcommand{\frecthree}{
\begin{tikzpicture}
    \begin{feynman}
        \vertex[] (i) at (-1.5, 0 );
        \vertex[] (a) at (-0.75, 0); 
        \vertex[] (b) at (0.375, 0);
        \vertex[] (f) at (1.5, 0);
        \vertex[] (j) at (-1.5,1.5);
        \vertex[] (c) at (-0.75,1.5);
        \vertex[] (d) at (0.75,1.5);
        \vertex[] (k) at (1.5, 1.5);
        \vertex[] (g) at (0.375,0.75);
        \vertex[] (h) at (0, 1.5);

        \diagram*{
            (i) -- [scalar] (a) -- [scalar] (b) -- [scalar] (f),
            (a) -- [photon, thick] (c),
            (b) -- [photon, thick] (g)-- [photon, thick] (h),
            (g) -- [photon, thick] (d),
            (j) -- [scalar] (c) -- [flatscalar] (h) -- [scalar] (k)
          };
    \end{feynman}
\end{tikzpicture}
}
    
\newcommand{\bpropscalar}{
\begin{tikzpicture}
        \begin{feynman}
          \vertex[sourceone] (i) at (-1, 0 ) {\tiny{\(L\)}};
          \vertex[sourceone] (f) at (1, 0) {\tiny{\(L\)}}; 
          \diagram*{
            (i) -- [bscalar, half left]  (f)
          };
        \end{feynman}
    \end{tikzpicture}
    }

\newcommand{\bspropone}{
\begin{tikzpicture}
        \begin{feynman}
          \vertex[binsertion] (c) at (0,1.0) {};
          \vertex [below right=0.75em of c] {\tiny\(1\)};
          \vertex[sourcetwo] (i) at (-1, 0 ) {\tiny{\(L_0\)}};
          \vertex[sourcetwo] (f) at (1, 0) {\tiny{\(L_0\)}};
          \diagram*{
            (c) -- [flatscalar, quarter right] (i), (c) --  [flatscalar, quarter left] (f),
          };
        \end{feynman}
    \end{tikzpicture}
    }

\newcommand{\bsproptwoone}{
\begin{tikzpicture}
        \begin{feynman}
          \vertex[binsertion] (d) at (0.5,0.95) {};
          \vertex [below right=0.75em of d] {\tiny\(1\)};
          \vertex[binsertion] (c) at (-0.5,0.95) {};
          \vertex [below right=0.75em of c] {\tiny\(1\)};
          \vertex[sourcetwo] (i) at (-1, 0 ) {\tiny{\(L_0\)}};
          \vertex[sourcetwo] (f) at (1, 0) {\tiny{\(L_0\)}};
          \vertex[] (a) at (0,1); 
          \diagram*{
           (i) -- [flatscalar, /tikz/out=90 ,/tikz/in=210] (c) -- [flatscalar, /tikz/out=15 ,/tikz/in=165]  (d) --  [flatscalar,/tikz/out=330 ,/tikz/in=90] (f),
          };
        \end{feynman}
    \end{tikzpicture}
    }
    
\newcommand{\bsproptwotwo}{
\begin{tikzpicture}
        \begin{feynman}
          \vertex[binsertion] (c) at (0,1.0) {};
          \vertex [below right=0.75em of c] {\tiny\(2\)};
          \vertex[sourcetwo] (i) at (-1, 0 ) {\tiny{\(L_0\)}};
          \vertex[sourcetwo] (f) at (1, 0) {\tiny{\(L_0\)}};
          \diagram*{
            (c) -- [flatscalar, quarter right] (i), (c) --  [flatscalar, quarter left] (f),
          };
        \end{feynman}
    \end{tikzpicture}
    }    

\newcommand{\bsproptwothree}{
\begin{tikzpicture}
        \begin{feynman}
          \vertex[binsertion] (c) at (0,1.0) {};
          \vertex [below right=0.75em of c] {\tiny\(1\)};
          \vertex[sourcetwo] (i) at (-1, 0 ) {\tiny{\(L_0\)}};
          \vertex[sourcetwo] (f) at (1, 0) {\tiny{\(L_1\)}};
          \diagram*{
            (c) -- [flatscalar, quarter right] (i), (c) --  [flatscalar, quarter left] (f),
          };
        \end{feynman}
    \end{tikzpicture}
    }

\newcommand{\bint}{
\begin{tikzpicture}
        \begin{feynman}
          \vertex[sourceone] (i) at (-1, 0 ) {\tiny{\(L\)}};
          \vertex[sourceone] (f) at (1, 0) {\tiny{\(L\)}};
          \vertex[sourceone] (g) at (0, 0) {\tiny{\(L\)}};
          \vertex[] (c) at (0,1.0);
          \diagram*{
            (i) -- [bgraviton, thick, half left]  (f),
            (c) -- [bgraviton, thick]  (g)
          };
        \end{feynman}
    \end{tikzpicture}
    }

\newcommand{\brecoiltwoone}{
\begin{tikzpicture}
        \begin{feynman}
          \vertex[recoil] (c) at (0,1.0) {\tiny{\(H\)}};
          \vertex[sourceone] (i) at (-1, 0 ) {\tiny{\(L\)}};
          \vertex[sourceone] (f) at (1, 0) {\tiny{\(L\)}};
          \vertex[sourceone] (g) at (0, 0 ) {\tiny{\(L\)}};
          \diagram*{
            (c) -- [bgraviton, quarter right] (i), (c) --  [bgraviton, quarter left] (f),
            (c) --  [bgraviton] (g),
          };
        \end{feynman}
    \end{tikzpicture}
    }

\newcommand{\brecoiltwotwo}{
\begin{tikzpicture}
        \begin{feynman}
          \vertex[recoil] (c) at (-0.33, 1.0) {\tiny{\(H\)}};
          \vertex[sourceone] (i) at (-1, 0) {\tiny{\(L\)}};
          \vertex[sourceone] (f) at (1, 0) {\tiny{\(L\)}};
          \vertex[sourceone] (g) at (0, 0 ) {\tiny{\(L\)}};
          \vertex[] (h) at (0.5, 0.66);
          \diagram*{
            (c) -- [bgraviton, quarter right] (i), (h) --  [bgraviton, bend right] (c),
            (g) -- [bgraviton, half left] (f),
          };
        \end{feynman}
    \end{tikzpicture}
    }

\newcommand{\traj}{
    \begin{tikzpicture}
        \begin{feynman}
          \vertex[source] (top1) at (-0.5, 1) {};
          \vertex[source] (top2) at (0.5, 1) {};
          \vertex (i) at (-1, 0);
          \vertex[gvert] (c) at (0, 0);
          \vertex (f) at (1, 0);
          \diagram*{
            (i) -- [myscalar] (c) -- [flatscalar] (f), 
            (top1) -- [photon, thick] (c),
            (top2) -- [photon, thick] (c) 
          };
        \end{feynman}
    \end{tikzpicture}
    }

\newcommand{\bgprop}{
\begin{tikzpicture}
        \begin{feynman}
          \vertex (i) at (-1, 0);
          \vertex (f) at (1, 0);
          \diagram*{
            (i) -- [bgraviton]  (f) 
          };
        \end{feynman}
    \end{tikzpicture}
}

\newcommand{\flatprop}{
\begin{tikzpicture}
        \begin{feynman}
          \vertex (i) at (-1, 0);
          \vertex (f) at (1, 0);
          \diagram*{
            (i) -- [photon, thick]  (f) 
          };
        \end{feynman}
    \end{tikzpicture}
}

\newcommand{\propone}{
\begin{tikzpicture}
        \begin{feynman}
          \vertex (i) at (-1, 0);
          \vertex (f) at (1, 0);
          \vertex[binsertion] (c) at (0,0) {};
          \diagram*{
            (i) -- [photon, thick](c) -- [photon, thick]  (f) 
          };
        \end{feynman}
    \end{tikzpicture}
}

\newcommand{\proptwo}{
\begin{tikzpicture}
        \begin{feynman}
          \vertex (i) at (-1, 0);
          \vertex (f) at (1, 0);
          \vertex[binsertion] (c) at (-0.4,0) {};
          \vertex[binsertion] (d) at (0.4,0) {};
          \diagram*{
           (i) -- [photon, thick] (c) -- [photon, thick] (d) -- [photon, thick]  (f) 
          };
        \end{feynman}
    \end{tikzpicture}
}

\newcommand{\sbgprop}{
\begin{tikzpicture}
        \begin{feynman}
          \vertex (i) at (-1, 0);
          \vertex (f) at (1, 0);
          \diagram*{
            (i) -- [bscalar]  (f) 
          };
        \end{feynman}
    \end{tikzpicture}
}

\newcommand{\sflatprop}{
\begin{tikzpicture}
        \begin{feynman}
          \vertex (top) at (0, 1);
          \vertex (i) at (-1, 0);
          \vertex (f) at (1, 0);
          \diagram*{
            (i) -- [flatscalar]  (f) 
          };
        \end{feynman}
    \end{tikzpicture}
}

\newcommand{\spropone}{
\begin{tikzpicture}
        \begin{feynman}
          \vertex (i) at (-1, 0);
          \vertex (f) at (1, 0);
          \vertex[binsertion] (c) at (0,0) {};
          \diagram*{
            (i) -- [flatscalar](c) -- [flatscalar]  (f) 
          };
        \end{feynman}
    \end{tikzpicture}
}

\newcommand{\sproptwo}{
\begin{tikzpicture}
        \begin{feynman}
          \vertex (i) at (-1, 0);
          \vertex (f) at (1, 0);
          \vertex[binsertion] (c) at (-0.4,0) {};
          \vertex[binsertion] (d) at (0.4,0) {};
          \diagram*{
           (i) -- [flatscalar] (c) -- [flatscalar] (d) -- [flatscalar]  (f) 
          };
        \end{feynman}
    \end{tikzpicture}
}

\newcommand{\gbackgroundoneone}{
\begin{tikzpicture}
        \begin{feynman}
          \vertex (top) at (0, 1);
          \vertex (i) at (-1, 0);
          \vertex (c) at (0, 0);
          \vertex (f) at (1, 0);
          \vertex[binsertion] (c) at (0,0) {};
          \vertex [below right=0.75em of c] {\tiny\(1\)};
          \diagram*{
            (i) -- [photon, thick] (c) -- [photon, thick] (f) 
          };
        \end{feynman}
    \end{tikzpicture}
}

\newcommand{\gbackgroundonetwo}{
\begin{tikzpicture}
        \begin{feynman}
          \vertex (top) at (0, 1);
          \vertex (i) at (-1, 0);
          \vertex (c) at (0, 0);
          \vertex (f) at (1, 0);
          \vertex[binsertion] (c) at (0,0) {};
          \vertex [below right=0.75em of c] {\tiny\(2\)};
          \diagram*{
            (i) -- [photon, thick] (c) -- [photon, thick] (f) 
          };
        \end{feynman}
    \end{tikzpicture}
}
\newcommand{\gbackgroundonethree}{
\begin{tikzpicture}
        \begin{feynman}
          \vertex (top) at (0, 1);
          \vertex (i) at (-1, 0);
          \vertex (c) at (0, 0);
          \vertex (f) at (1, 0);
          \vertex[binsertion] (c) at (0,0) {};
          \vertex [below right=0.75em of c] {\tiny\(3\)};
          \diagram*{
            (i) -- [photon, thick] (c) -- [photon, thick] (f) 
          };
        \end{feynman}
    \end{tikzpicture}
}

\apptocmd{\thebibliography}{\justifying\setlength{\leftskip}{7.4mm}}{}{}

\makeatletter\g@addto@macro\bfseries{\boldmath}\makeatother

\usepackage{stackengine}
\usepackage{esint}
\usepackage[unicode=true,pdfusetitle,
 bookmarks=true,bookmarksnumbered=false,bookmarksopen=false,
 breaklinks=false,pdfborder={0 0 1},backref=false,colorlinks=true]
 {hyperref}
\hypersetup{pdfauthor={Clifford Cheung and Grant N. Remmen},
 citecolor=black,linkcolor=black,urlcolor=black}

\newcommand{\appendixref}[1]{\hyperref[#1]{appendix~\ref{#1}}}
\def\equationautorefname~#1\null{eq.\,(#1)\null}
\usepackage[hang,flushmargin]{footmisc} 
\allowdisplaybreaks

\usepackage{changepage}

\newcommand{\be}{\begin{equation}}
\newcommand{\ee}{\end{equation}}
\newcommand{\bea}{\begin{eqnarray}}
\newcommand{\eea}{\end{eqnarray}}

\newcommand{\eq}[2]{\be\begin{aligned}#1 \label{#2}\end{aligned}\ee}

\newcommand{\Fig}[1]{Fig.~\ref{#1}}

\newcommand{\Eq}[1]{Eq.~(\ref{#1})}
\newcommand{\Eqs}[2]{Eqs.~(\ref{#1}) and (\ref{#2})}
\newcommand{\Sec}[1]{Sec.~\ref{#1}}

\newcommand{\App}[1]{App.~\ref{#1}}

\newcommand{\OO}{\mathcal{O}}

\usepackage{xcolor}

\definecolor{author1color}{RGB}{255,0,0}    
\definecolor{author2color}{RGB}{0,0,255}    
\definecolor{author3color}{RGB}{0,128,0}    

\newcolumntype{P}[1]{>{\centering\arraybackslash}p{#1}}

\newcommand{\mL}{m_L}
\newcommand{\mH}{m_H}

\renewcommand{\v}{u}
\newcommand{\vL}{u_L}
\newcommand{\vH}{u_H}
\newcommand{\zL}{z_L}
\newcommand{\zH}{z_H}
\newcommand{\qL}{q_L}
\newcommand{\qH}{q_H}
\newcommand{\Rc}{r_{c}}
\newcommand{\RS}{r_S}
\newcommand{\Rsc}{r_{\Phi}}
\newcommand{\Rvc}{r_{A}}

\newcommand{\yL}{y_L}

\newcommand{\xL}{x_L}
\newcommand{\xH}{x_H}
\newcommand{\dotxL}{\dot{x}_L}
\newcommand{\dotxH}{\dot{x}_H}

\newcommand{\ddotxH}{\ddot{x}_H}

\newcommand{\dotdxH}{\delta\dot{x}_H}

\newcommand{\dotbarxL}{\dot{\bar{x}}_L}
\newcommand{\dotbarxH}{\dot{\bar{x}}_H}

\newcommand{\ddotbarxL}{\ddot{\bar{x}}_L}
\newcommand{\ddotbarxH}{\ddot{\bar{x}}_H}

\newcommand{\barx}{\bar{x}}
\newcommand{\barxL}{\bar{x}_L}
\newcommand{\barxH}{\bar{x}_H}

\newcommand{\barJL}{\bar{J}_L}

\newcommand{\barTL}{\bar{T}_L}

\newcommand{\dx}{\delta x}
\newcommand{\dA}{\delta A}
\newcommand{\dF}{\delta F}

\newcommand{\dg}{\delta g}
\newcommand{\dG}{\delta \Gamma}

\newcommand{\dxL}{\delta x_L}
\newcommand{\dxH}{\delta x_H}

\begin{document}

\interfootnotelinepenalty=10000
\baselineskip=18pt
\hfill

\thispagestyle{empty}


\begin{flushright}
    \footnotesize
    CALT-TH 2024-023
\end{flushright}

\vspace*{0.4in}

\begin{center}
{\LARGE \bf
Gravitational Scattering and Beyond from \\ \medskip\smallskip
 Extreme Mass Ratio Effective Field Theory
}\\
\bigskip\vspace{1cm}

\begin{center}{\large Clifford Cheung${}^{a}$, Julio Parra-Martinez${}^{b}$, Ira Z. Rothstein${}^{c}$, \\ \smallskip
Nabha Shah${}^{a}$, Jordan Wilson-Gerow${}^{a}$
}\end{center}
\vspace{7mm}

{
\it ${}^a$Walter Burke Institute for Theoretical Physics,\\[-1mm]
California Institute of Technology, Pasadena, California 91125, USA\\[1.5mm]
${}^b$Institut des Hautes \'{E}tudes Scientifiques, 91440 Bures-sur-Yvette, France\\[1.5mm]
${}^c$Department of Physics,\\[-1mm] Carnegie Mellon University, Pittsburgh, PA 15213, USA\\[1.5mm]
}
\end{center}

\bigskip

\centerline{\large\bf Abstract}
\begin{quote} \small


We explore a recently proposed effective field theory describing electromagnetically or gravitationally interacting massive particles in an expansion about their mass ratio, also known as the self-force (SF) expansion.  By integrating out the deviation of the heavy particle about its inertial trajectory, we obtain an effective action whose only degrees of freedom are the lighter particle together with the photon or graviton, all propagating in a Coulomb or Schwarzschild background.  The 0SF dynamics are described by the usual background field method, which at 1SF is supplemented by a ``recoil operator'' that encodes the wobble of the heavy particle, and similarly computable corrections appearing at 2SF and higher. Our formalism exploits the fact that the analytic expressions for classical backgrounds and particle trajectories encode dynamical information to all orders in the couplings, and from them we extract multiloop integrands for perturbative scattering.  As a check, we study the two-loop classical scattering of scalar particles in electromagnetism and gravity, verifying known results.  We then present new calculations for the two-loop classical scattering of dyons, and of particles interacting with an additional scalar or vector field coupling directly to the lighter particle but only gravitationally to the heavier particle.

\end{quote}
	
\setcounter{footnote}{0}

\setcounter{tocdepth}{3}

\newpage
\tableofcontents

\newpage

\section{Introduction}

The natural domain of perturbative quantum field theory is a dilute ensemble of particles evolving atop a quiescent vacuum state.   However, as those degrees of freedom pile up, this naive picture breaks down. Their collective behavior is instead more aptly described by an ambient coherent background, for instance, as characterized by the Coulomb potential or the Schwarzschild metric.  In this condensed regime, one usually adopts a background field method in which each degree of freedom is decomposed into a fixed  background and the perturbation about it.  However, it is obvious that this description will also  fail as soon as the perturbation carries charge or energy-momentum of order of the background field itself.

At what point does quantum field theory on the vacuum ``end’’ and quantum field theory on a background ``begin’’?  Here, an important clue can be gleaned from the seminal work of \cite{Duff} which, following \cite{Boulware:1968zz}, showed how the Schwarzschild metric is constructed order by order in the gravitational constant, $G$, by iteratively solving the equation of motions in a {\it flat space background}.  Specifically, the metric is computed from the one-point function of the graviton sourced by a massive scalar point particle.   The relevant Feynman diagrams for this calculation are shown in \Fig{fig:Duff}, and were recently resummed to all orders in \cite{Damgaard:2024fqj, Mougiakakos:2024nku}. 
Note that this procedure is only possible because the Schwarzschild metric has a regular series expansion in $G$.

The broader takeaway of \cite{Duff} is that certain classical solutions can be computed from perturbation theory in flat space.  Conversely, it is natural to ask whether the reverse procedure---extracting perturbative data from classical solutions---can be used to reorganize or simplify perturbative calculations in a trivial background.  After all, the Schwarzschild metric is known in closed form and compactly encodes perturbative information at all orders in $G$.  The same is true of the geodesic trajectories of particles in a nontrivial background, which in some cases are known at finite coupling $G$.

We thus arrive upon our earlier question, here framed more sharply: which flat space Feynman diagrams are accounted for by a given classical background and its geodesics?   Conversely, which  contributions are not encoded by these elements, and can they be computed systematically?

In this paper, we explore an effective field theory, first described in \cite{sfshort}, that rigorously delineates between these contributions while exploiting classical backgrounds and trajectories to extract all-orders perturbative data.  Our system of interest is composed of a light and heavy point particle of mass $\mL$ and $\mH$, respectively, interacting via a long-range force carrier.  We construct an effective field theory that can be applied to electromagnetism (EM), general relativity (GR), and beyond.  However, for clarity, we describe our results here in terms of GR.  A treatment similar in spirit to ours, though distinct in methodology, was also proposed in \cite{2308.15304}.

The expansion parameter of our effective field theory is the ratio of masses of two interacting bodies,
\eq{
\lambda = \frac{\mL}{\mH}.
}{}
In the parlance of existing approaches to the black hole binary inspiral problem for extreme mass ratios, we refer to the expansion in $\lambda$ as the self-force (SF) expansion.  
While the SF expansion is reminiscent of heavy quark mass expansion \cite{Isgur:1989vq,Georgi:1990um,Eichten:1989zv,Shifman:1987rj}, our setup differs in its stated aim of extracting certain contributions to all orders in the coupling. Naively, such a feat might seem impossible but, as we have just described, classical GR does precisely this.

\begin{figure*}
    \centering
    $\bar g_{\mu \nu} \ = \ \eta_{\mu \nu} \ + \vcenter{\hbox{\duffone}} + \vcenter{\hbox{\dufftwo}} + \vcenter{\hbox{\duffthreeone}} + \vcenter{\hbox{\duffthreetwo}} + \ \ \cdots$
    \caption{The expansion of the Schwarzschild metric, $\bar g_{\mu \nu}$, in powers of the gravitational constant, $G$, corresponds to a perturbative computation, in flat space, of the graviton one-point function in the presence of a massive scalar point particle source \cite{Duff}.}
    \label{fig:Duff}
\end{figure*}

Let us briefly describe the structure of our effective field theory for extreme mass ratios.  At 0SF order, we use the well-known result that the full gravitational dynamics of the heavy and light particle are described by a probe evolving in a Schwarzschild background.  Taking our cues from \cite{Duff}, we interpret the Schwarzschild background as the 0SF graviton one-point function sourced perturbatively by a heavy particle traversing an inertial 0SF trajectory.  Meanwhile, the 0SF trajectory of the light particle is dictated by the geodesic equation.


At 1SF order, the deviation of the heavy particle from its 0SF trajectory becomes important.  However, this perturbation enters quadratically into the action so we can integrate it out exactly.  The resulting effective action is  the action for a probe particle interacting with graviton perturbations in a Schwarzschild background---plus an additional operator describing the {\it recoil} of the heavy particle.
At a technical level, the recoil operator is simply a correction to the two-point function of the graviton.  It is nonlocal in time, precisely because the heavy particle is  a propagating degree of freedom.  Physically, the recoil operator describes how graviton propagation is modified by wobbles of the heavy particle  sourcing the Schwarzschild background.  Any background field calculation must be supplemented with the recoil operator in order  to correctly reproduce the results of perturbation theory in a flat background.   By expanding the action order by order in $\lambda$, one can systematically derive the effective action to 2SF order and higher.

As a demonstration of the power of our formalism we present several old and new calculations describing the elastic scattering of massive particles in EM and GR. In particular, we show how the conservative dynamics---as encoded by the on-shell radial action---can be computed with relative ease in explicit examples at 0SF, 1SF, and 2SF. The radial action \cite{Landau:1975pou} encodes the  same dynamical information as more familiar on-shell scattering amplitudes \cite{PM4,Kol:2021jjc,Bern:2021yeh,Damgaard:2021ipf}. For EM and GR, we work in the post-Lorentzian (PL) and post-Minkowskian (PM) expansions, which  correspond to perturbation theory in the fine structure constant, $\alpha$, and the gravitational constant, $G$, respectively. In the language of quantum field theory, all of our calculations include contributions up to and including two-loop order, i.e., to 3PL and 3PM order in EM and GR respectively. 
 Our checks match known results \cite{Saketh:2021sri,Bern:2021xze,Cachazo:2017jef,BBDamFestPlaVan,PM1,PM2}.  Here, we emphasize that this agreement relies critically on the inclusion of the recoil operator---the background field method alone does not yield correct expressions.
Afterwards, we present new calculations of the radial action for dyonic scattering at 3PL, and for massive particle scattering in GR where the light particle
couples to a massless scalar or vector field that itself interacts gravitationally.  The latter theories are of particular relevance to previous SF studies \cite{poissontoy,Barack:2022pde,scalartoy}.

Of course, there is a long history of studying perturbative classical gravitational dynamics in the relativistic weak-field regime, also known as the PM expansion \cite{Bertotti:1956pxu,Kerr:1959zlt,BERTOTTI1960169,Westpfahl:1979gu,MPortilla_1980,PhysRevD.94.104015,PhysRevD.97.044038}. In recent years, the PM program has received an infusion of new ideas from the scattering amplitudes program and effective field theory, retooled to explore the binary inspiral problem and push the state of the art in PM calculations \cite{NeillRoth,PM1,PM2,PM3,PM4,Bern:2021yeh,Bern:2022jvn,Luna:2017dtq,Kosower:2018adc,christofoligonzo,Cheung:2020gyp,Kalin:2020mvi,PhysRevLett.125.261103,PhysRevD.102.124025,Mogull:2020sak,PhysRevLett.126.201103,Dlapa:2021npj,PhysRevLett.128.161104,BBDamFestPlaVan,Bjerrum-Bohr:2021wwt,Bjerrum-Bohr:2021din,Brandhuber:2021eyq,DIVECCHIA2020135924,DiVecchia:2021bdo,DIVECCHIA2021136379,DIVECCHIA2023138049,heissenberg,Bern:2020uwk,Herrmann:2021lqe,Herrmann:2021tct, HBET1,HBET2,Damgaard:2021ipf,Damgaard:2023vnx,huber,Brandhuber:2023hhy,Herderschee:2023fxh,Caron-Huot:2023vxl,PhysRevD.104.024041,PhysRevLett.130.101401,Dlapa:2023hsl,Riva:2021vnj,PhysRevLett.129.121101,Jakobsen:2021zvh,Jakobsen:2022psy,Bhattacharyya:2024aeq,PhysRevD.109.L041504,PhysRevD.108.024050,Driesse:2024xad,Bern:2024adl}. The effective field theory described here is essentially a reorganization of this perturbative PM approach into the language of classical solutions.  As we will see, this change of perspective offers some advantages for practical calculations.  Firstly, it accommodates a procedure for extracting  multiloop integrands {\it directly} from the time-domain geodesic trajectories of probe particles and the Schwarzschild solution.   Second, it allows for the trivial elimination of well-known self-energy divergences that appear in classical EM and GR.  As we will show explicitly, since all classical dynamics are secretly resummations of perturbative diagrams in a trivial background, we can regulate these divergences using standard dimensional regularization.  While this is standard in treatments of point particle effective theory \cite{GoldRoth, GilmoreRoss,Foffa:2011ub,Foffa2,PortoRoth,Foffa:2019hrb,Blumlein:2019zku,FoffaPortoRoth,Blumlein:2020pog,Blumlein:2020znm}, our framework allows us to apply these ideas to dynamics in a nontrivial background. Note that a central theme of this work---that there is an important distinction between Schwarzschild as a vacuum solution of nonperturbative GR as opposed to the field generated by point particle in perturbation theory---was studied closely previous work~\cite{Pfenning:2000zf, Gralla:2021qaf}.


Finally, we note that while the present work focuses solely on classical dynamics, this is  not required for the validity of our effective field theory.  In particular, our general approach can also be applied to quantum or thermal corrections that arise from  loops of the photon, graviton or other light particles.

The remainder of this paper will be structured as follows.  We begin with an extensive derivation of our effective field theory in the context of EM in \Sec{sec:EM}.  This discussion includes the derivation of the effective action at 0SF, 1SF, and 2SF orders, followed by a presentation of the perturbative Feynman rules.  We describe the physics of classical resummation, whereby known expressions for EM trajectories can be used to extract perturbative multi-loop integrands.  We then present a calculation of the radial action for scattering charged particles and dyons.  Afterwards, in \Sec{sec:GR} we generalize all of these results to the case of GR and the perturbative multi-loop integrands are extracted from the Schwarzschild metric and its geodesic trajectories.  Furthermore, we present explicit calculations of the radial action for massive particles scattering in GR, with and without additional scalar and vector fields.

\medskip

\noindent {\bf Notation and Conventions:} We work in mostly minus metric signature and natural units where $c=\hbar=1$. We also make use of the notation $\hat{\delta}(x)=2\pi\delta(x)$. Where we are not explicitly working in $D$ dimensions, divergent integrals are defined via dimensional regularization. We use the integral notation $\int_{\ell_{1},...,\ell_{n}}=\int\frac{d^{D}\ell_{1}}{(2\pi)^{D}}\cdots\int\frac{d^{D}\ell_{n}}{(2\pi)^{D}}$, and will often take the $D=4$ limit implicitly after integration when the context is unambiguous.

\medskip

\section{Electromagnetism}  
\label{sec:EM}

In this section, we construct an effective field theory for electromagnetically interacting particles in which the expansion parameter is the ratio of their masses.  Let us briefly outline our plan of attack.
Our starting point is the worldline action for a pair of charged particles interacting via the photon.
The 0SF dynamics, corresponding to the limit in which the mass of the
lighter particle is negligible compared to that of the heavier one, are exactly solvable.  In this case, the latter moves in a straight line trajectory, forming a background Coulomb field that governs the probe motion of the former.  

At 1SF, the deviation of the heavy particle away from its inertial motion must be taken into account.  However, as we will see, the deviation of the light particle from its probe motion can be ignored.
By integrating out the perturbations of the heavy particle about its inertial 0SF trajectory, we derive a 1SF recoil operator encoding the back-reaction of the heavy particle.  This recoil operator is a nonlocal-in-time correction to the propagation of the photon.  It  encodes the leading correction to Rutherford scattering that scales as the ratio between the masses of the light and heavy particles.

This approach generalizes systematically to higher orders. Power counting in the mass ratio, we integrate out perturbations in the heavy trajectory at higher orders, and explicitly derive the 2SF recoil operator.

As a check of our formalism, we compute the radial action for scattering at 0SF, 1SF, and 2SF---up to 3PL order. The 0SF radial action is a known quantity \cite{Bern:2021xze} to all PL orders, which we present in generality in \App{app:probe_action}. Meanwhile, we compute the 1SF radial action at 2PL and 3PL order and find perfect agreement with known results \cite{Saketh:2021sri,Bern:2021xze}. The first appearance of 2SF contributions are at 3PL order and they match, as expected, the 0SF-3PL terms upon exchanging the light and heavy particles.
Finally, we present new results for the 1SF radial action for the scattering of dyonically charged particles at 2PL and 3PL.

\subsection{Effective Theory}

To begin, consider the action describing a pair of massive charged particles interacting via an EM field.  As described in \App{app:probe_traj}, we can gauge fix the worldline einbein so that the action takes the simple form,
\eq{
S &= \ - \sum_{i=H,L} m_i \int d\tau \left[\tfrac{1}{2}+ \tfrac{1}{2} \dot x_i^2 + z_i\dot x_i^\mu A_\mu(x_i) \right] -   \int d^4x\left[ \tfrac14 F_{\mu\nu}(x)F^{\mu\nu}(x)\right]\, ,
}{S_EM}
where $x_i^\mu(\tau)$ are worldline trajectories and $A_{\mu}(x)$ is the photon field.   Our worldline gauge fixing enforces the on-shell condition, $\dot x_i^2(\tau)=1$, for physical solutions. Here, \Eq{S_EM} is written in terms of the charge-to-mass ratios, $z_i = q_i/m_i$, which we assume throughout to be of similar size. This is, of course, not generically true. For example, the charge-to-mass ratios of the electron and proton are very different in magnitude.  However, for our purposes, we assume $z_i$ of the same magnitude so that the electric forces scale proportionally to mass and the dynamics more closely parallels that of gravity.

The equations of motion for the particles and fields derived from \Eq{S_EM} are 
\eq{
&\ddot x^\mu_i  - z_i  F^{\mu\nu}(x_i) \dot x_{i\nu}  =0 \qquad \textrm{and} \qquad
\partial_\mu F^{\mu\nu}= J^\nu  \, ,
}{EOM_EM}
where the electromagnetic current is 
\eq{
J^\mu(x)  = \sum_{i=H,L} J_i^\mu(x)= \sum_{i=H,L}  z_i m_i \int d\tau  \, \delta^4(x-x_i) \dot x_i^\mu\, .
}{J_EM}
There are two basic approaches to solving this equation of motion, which we now discuss. As we will see, these different methods yield distinct intermediate expressions on the way to computing physical observables.

\subsubsection{Post-Lorentzian Expansion on Trivial Background}

The standard method for solving the equations of motion is to expand perturbatively in the couplings,  which for EM are the charges, $q_{i}$.  Doing so yields
\eq{
        x_H^\mu &= u_H^\mu \tau + b_{H}^\mu +  \delta x_H^\mu \, ,\\
        x_L^\mu &= u_L^\mu \tau + b_{L}^\mu +  \delta x_L^\mu \, ,\\
        A_\mu &= 0 + \delta A_\mu\, .
}{EM_pert}
Here, $\delta x_H, \delta x_L, \delta A_\mu$ encode deviations away from the inertial trajectories of neutral particles together with the photon.  These perturbations are all implicitly suppressed by powers of the coupling constants, $q_{i}$, and are computed by iteratively solving the equations of motion. For example, at leading order one has
\eq{
&\delta x^\mu_i  = z_i  u_{i\nu}\frac{1}{\partial_{\tau}^{2}} F^{\mu\nu}(b_i+u_{i}\tau)  \qquad \textrm{and} \qquad
\partial_\mu \delta F^{\mu\nu} = \sum_{i=H,L}  z_i m_i u_{i}^{\nu} \int d\tau  \, \delta^4(x-b_i-u_i \tau)  \,,
}{}
and higher orders are obtained by expanding the equations of motion yet further about these deviated solutions. 

Of course, there is nothing intrinsically flawed in this standard approach. However, it does not make use of known results describing the motion of a probe charge in a fixed Coulomb field, which secretly encodes physics to all orders in the PL expansion.  Furthermore, the integrability of the dynamics in this limit is not at all obvious from the perturbative expansion described above, let alone leveraged to simplify computations.

In what follows, we instead build an effective field theory that encodes the solutions to these equations of motion organized in powers of the mass ratio, $\lambda$.  A crucial ingredient is that many of our manipulations will be all orders in the PL expansion. As we will show,  this approach does indeed utilize the exactly known test-particle dynamics.

\subsubsection{Self-Force Expansion on Coulomb Background}


The key observation is that we can solve the equations of motion which describe 0SF dynamics exactly. In this limit,
the heavy particle moves in an inertial, straight line trajectory,
\eq{
\barxH^\mu(\tau) = \vH^\mu \tau \, ,
}{xH_EM}
where $\vH$ is the heavy particle velocity, and sources an ambient boosted Coulomb field as per the equation of motion, 
\eq{
        \partial_\mu \bar F^{\mu\nu}= \square \bar A^{\nu} = \bar J^\nu_H = z_H m_H \int d\tau u_H^\nu \delta^4(x^\mu - u_H^\mu \tau)\, ,
}{xHbkgd_EOM_EM}
written here in Lorenz gauge, $\partial_\mu \bar A^\mu (x) = 0$. This has a well-known solution,
\eq{
        \bar A_\mu(x) =  \frac{\zH \mH \vH{}_\mu}{4\pi r}   \, ,
}{bkgd_EM}
where $r=\sqrt{(\vH x)^2-x^2}$ is the boosted radius.
Meanwhile, the light particle equation of motion at 0SF is
\eq{
\ddotbarxL^\mu  - \zL  \bar F^{\mu\nu}(\barxL) \dotbarxL{}_\nu  =0 \, ,
}{xL_EOM_EM}
which is the usual Lorentz force law in a Coulomb field.

At this point, we perform an expansion about the 0SF dynamics.  The corrections include deviations away from the 0SF solutions,
\eq{
x^\mu_i  = \bar x^\mu_i + \dx_i^\mu\qquad \textrm{and} \qquad
A_\mu = \bar A_\mu + \dA_\mu \, ,
}{pert_EM}
where, hereafter in this paper, all perturbations will refer to the expansion in mass ratio, $\lambda$, rather than the coupling constant. 
Since $\bar x^\mu_i$ and $\bar A_\mu$ are exact 0SF solutions, we know that the deviation degrees of freedom are effectively 1SF objects and scale as 
\eq{
\dx_i^\mu \sim \dA_\mu \sim {\cal O}(\lambda^1) \, ,
}{}
when they are set to their on-shell configurations. 
As we will see, the 0SF trajectory for the light particle is critical for computing the 1SF action.  

\subsubsection{Regularizing Self-Force Divergences}\label{reg}

At this point, we encounter the subtlety of self-energy contributions to the heavy particle.  Consider the heavy particle action at 0SF,
\eq{
\bar{S}_{H}=-m_H \int d\tau\,\left[\tfrac{1}{2}+\tfrac{1}{2} \dotbarxH^2+z_H\dotbarxH^\mu \bar{A}_\mu(\barxH) \right],
}{SEM_singular}
with the corresponding equation of motion,
\eq{
\ddotbarxH^\mu  - \zH  \bar F^{\mu\nu}(\barxH) \dotbarxH{}_\nu  =0 \, .
}{}
Evaluating the heavy effective action on the solution of the 0SF equations of motion yields singular terms. These involve $\bar F^{\mu\nu}(\barxH)$  evaluated at $r=0$, corresponding to the EM force acting on the heavy particle coming from its own Coulomb field.

In the usual approach to classical dynamics, one must devise a regularization scheme to subtract this self-energy or self-force contribution. As we have emphasized, our setup repackages probe trajectories and background field configurations as resummations of flat space perturbative dynamics.  In this picture, $\bar F^{\mu\nu}(\barxH)$ corresponds to a potential photon mode that is emitted and then reabsorbed by the heavy worldline.  As usual, in point-particle effective field theory, such terms yield self-energy contributions which are absorbed through the mass counterterm. Effectively, we can then discard $\bar F^{\mu\nu}(\barxH)$ wherever it appears.  More generally, we are permitted to drop any contributions which arise in the flat space theory from potential photons emitted and reabsorbed by the heavy particle.  The upshot is that the heavy particle equation of motion is effectively $\ddotbarxH^\mu =0$, whose solution is \Eq{xH_EM}.

To explain our choice of regularization scheme it is instructive to pedantically recall how the solution in \Eq{bkgd_EM} arises using the language of Feynman diagrams. In Lorenz gauge, the solution to the equation of motion, \Eq{xHbkgd_EOM_EM}, is given by the single Feynman diagram which equals the propagator integrated against the source,
\eq{
        \bar A_{\nu}(x) = \int d^4 y \, G_{\nu\mu}(x-y) \bar J_H^{\mu}(y) \, .
}{}
As is customary, we can evaluate this in momentum space, where the heavy source is simply 
\eq{
        \bar J_H^\mu(q) =   z_H m_H u_H^\mu\, \hat{\delta}(u_H q) \, ,
}{}
and the Lorenz gauge propagator is\footnote{The choice of $i\epsilon$-prescription is immaterial for this computation.} 
\eq{
        G_{\mu\nu}(q) =   - \frac{\eta_{\mu\nu}}{q^2}\,.
}{}
The solution to the equations of motion is then simply
\eq{
\bar A_{\nu}(q) =  - z_H m_H u_{H\,\nu}  \frac{\hat{\delta}(u_H q)}{q^2} \,,
}{}
whose Fourier transform yields \Eq{bkgd_EM}. In the language of Feynman diagrams, the singular term in \Eq{SEM_singular} is a contribution to the classical self-energy,

\eq{
        \vcenter{\hbox{\selfenergy}} = i \bar S_{H} = \frac{-i}{2}\int d^4 x d^4 y \,  \bar J_H^{\mu}(x) G_{\mu\nu}(x-y) \bar J_H^{\nu}(y) = -i T \frac{z_H^2 m_H^2}{2} \int \frac{d^4q}{(2\pi)^4} \frac{\hat{\delta}(u_H q)}{-q^2},
}{}
where $T=2\pi\delta(0)$ is the total time integral. The coefficient of $-iT$, i.e., the energy, is ultraviolet divergent due to the integration over large values of $q$. For instance, a cutoff regularization gives
\eq{
        \int_{|q|<\Lambda} \frac{d^4q}{(2\pi)^4} \frac{\hat{\delta}(u_H   q)}{q^2} \sim \Lambda \, .
}{}
This linear divergence corresponds to the $r=0$ singularity in \Eq{SEM_singular}. This can be explicitly reabsorbed by a mass counterterm. However, since this divergence is power law, it is most convenient to  use dimensional regularization, whereby the integral is analytically continued to general dimension $D$. The advantage of this choice is that, \emph{by definition}, dimensional regularization sets all power divergences to zero,
\eq{
        \int \frac{d^Dq}{(2\pi)^D} \frac{\hat{\delta}(u_H q)}{q^2} =0 \, .
}{}
Hence, in dimensional regularization we find that
\eq{
        \bar A_{\nu}(\barxH)  = 0\,.
}{}
This is an important simplification, which will become crucial when we study the gravitational case. To reiterate, the renormalized heavy particle action at 0SF is simply that of a free particle,
\eq{\bar{S}_{H}=-m_{H}\int d\tau \tfrac{1}{2}(1+\dotbarxH^{2})\,,
}{}
and we will drop this contribution, since it contains no dynamical information.


\subsubsection{0SF Dynamics}
\label{EM_0SF}

In this work we will focus on conservative scattering dynamics, for which the gauge invariant quantity of interest is the scattering angle. Practically though, we will compute the on-shell action, from which the scattering angle follows by simple differentiation.

Inserting the SF expanded trajectories and fields shown in \Eq{pert_EM} into the action in \Eq{S_EM}, we now compute 
\eq{
S = \bar S +\sum_{n=1}^{\infty} S^{(n)} \, ,
}{}
where $n$ denotes the SF order and $\bar S$ is the 0SF on-shell action obtained by evaluating the action in the probe limit, i.e., on the support of the solution where the heavy particle is on an inertial path and the light particle traverses a probe trajectory in the background sourced by the heavy particle.  After using dimensional regularization to eliminate self-energy divergences, we obtain the renormalized 0SF action,
\eq{
\bar S &= - \mH \int d\tau \left[\tfrac{1}{2}+\tfrac{1}{2} \dotbarxH^2 + \lambda\left(\tfrac{1}{2}+ \tfrac{1}{2} \dotbarxL^2+ \zL\dotbarxL^\mu \bar A_\mu(\barxL)  \right) \right]\, .
}{}
Here, the heavy particle contributions to the 0SF action enter at ${\cal O}(\lambda^0)$  while those of the light particle enter at ${\cal O}(\lambda^1)$ on account of the factor of the light particle mass in the action.  Note that $\bar S$ does not contain any dependence on the dynamical perturbations away from the probe trajectories.  Thus, the 1SF contribution to the action, $S^{(1)}$, starts at ${\cal O}(\lambda^2)$. 

Next, let us study the 0SF dynamics, working in $D=4$ dimensions for concreteness. In the probe limit, the heavy particle simply moves in a straight line, while the light particle dynamics are governed by the single-particle action,
\eq{
S &= - \mL \int d\tau \left[\tfrac{1}{2}+\tfrac{1}{2} \dotxL^2+ \zL\dotxL^\mu \bar A_\mu(\xL)  \right] \, ,
}{}
in a background Coulomb electric field, \Eq{bkgd_EM}. The equation of motion is just the Lorentz force equation. Using conservation of energy and angular momentum, we find that
\eq{
E= \mL\left(\dot t + \zL A_0\right) \qquad \textrm{and} \qquad J= \mL r^2 \dot \phi \, .
}{eq:EMfirstinteq1}
The remaining radial equation of motion is also a first-order ordinary differential equation coming simply from the on-shell condition,
\eq{
\dot{r}^{2}=\left(\frac{E-\qL\bar{A}_{0}(r)}{\mL}\right)^{2}-\frac{J^{2}}{r^{2}}-1 \, .
}{eq:EMfirstinteq2}
In $D=4$, one can write exact solutions for the orbital trajectories, $r(\phi)$, since this system describes Keplerian motion~\cite{sommerfeld, Landau:1975pou}. For general $D$, we outline how to solve these equations perturbatively in \App{app:probe_traj}. In practice, depending on the physical observable in question, the explicit solutions need not be computed.

 To compute the on-shell action to higher SF orders, we will need the explicit probe trajectory solutions. As described in \App{app:probe_traj}, the 0SF dynamics can be derived from textbook classical mechanics by evaluating the on-shell action as a radial action integral,
 \eq{
 \bar{S}=2\int_{r_{\rm min}}^{\infty} dr |p_{r}(r,E,J)|\,.
 }{}
 The radial conjugate momentum is readily solved for from the on-shell condition, giving
 \eq{
 |p_{r}|=\left[\left(E-\qL A_{0}(r)\right)^{2}-\mL-\frac{J^{2}}{r^{2}\mL^{2}}\right]^{1/2}\,.
 }{}
Converting to orbital parameters more convenient for scattering,
\eq{
\sigma=\frac{E}{\mL} \qquad \textrm{and} \qquad b = \frac{J}{\mL(\sigma^{2}-1)^{1/2}}\,,
}{}
we evaluate the radial action integral to 3PL order (see \App{app:probe_action}),
\eq{
\bar{S}=&-\pi b\mL(\sigma^{2}-1)^{1/2}-\frac{2 \mL \Rc \sigma  \log \left(\tilde{\mu}be^{\frac{-1}{D-4}}\right)}{\sqrt{\sigma ^2-1}}+\frac{\pi  \mL \Rc^2}{2 b \sqrt{\sigma
   ^2-1}}+\frac{\mL \Rc^3 \sigma  \left(2
   \sigma ^2-3\right)}{3 b^2 \left(\sigma
   ^2-1\right)^{5/2}}\,,
}{eq:EMproberadialactiontext}
where $\tilde{\mu}$ is a reference mass scale, and the infrared divergence in $D=4$ is the familiar Coulomb logarithm---it is a physical divergence in the time delay but does not affect the scattering angle. With a physical IR cutoff, as is the case for finite time processes, this is no longer divergent but set by the time scale. Our interest in the radial action is that it generates the scattering angle, for which these IR divergent terms do not contribute anyways since they are independent of $b$.

The PL expansion parameter in the above expression is the dimensionless ratio $\Rc/b$, where $b$ is the impact parameter and we've defined the length scale,
 \eq{
 \Rc = -\frac{\zH\zL \mH}{4\pi} \, ,
 }{def_rc}
 which is positive if the particles are oppositely charged, i.e., attracting. This is simply the classical charge radius of a particle with charge, $\sqrt{|\qH\qL|}$, and mass, $\mL$.

As we generalize to higher SF orders, we no longer know the exact expression for the radial momentum. However, we will still be able to compute the on-shell action with perturbative diagrams and the symmetries of the problem ensure that the radial action continues to serve as a generating function for the scattering angle. Since we wish to fix the energy and angular momentum of the solutions but do not, {\it a priori}, know the interaction potential, we must be sure to use causal propagators so that the integrals of motion can be specified by their values in the asymptotic past when the particles are decoupled.

\subsubsection{1SF Dynamics}

As noted earlier, the 0SF action contains contributions at $\mathcal{O}(\lambda^0)$  and $\mathcal{O}(\lambda^1)$, where the scaling comes from the masses of the heavy and light particles respectively.  Consequently,  the 1SF action starts at $\mathcal{O}(\lambda^2)$ and depends on the perturbations away from the probe trajectories and backgrounds.  

First, the action for the electromagnetic field becomes
\eq{
 - \int d^4x\left[\tfrac14 F_{\mu\nu}F^{\mu\nu}\right]=  - \int d^4x\left[\tfrac14 \bar F_{\mu\nu}\bar F^{\mu\nu}   {\color{blue}  +\tfrac12 \dF_{\mu\nu}\bar F^{\mu\nu} }   +\tfrac14 \dF_{\mu\nu}\dF^{\mu\nu}\right]\, ,
}{EMaction_FF}
where $\dF_{\mu\nu} = \partial_\mu \dA_\nu - \partial_\nu \dA_\mu$ and we will highlight in blue terms that will cancel with other terms in the action because $\bar F_{\mu\nu}$ satisfies its equation of motion.

Next, consider the contributions to the action from the worldlines,
\eq{
 - \int d\tau \left[\tfrac{1}{2} + \tfrac{1}{2} \dot x_i^2 + z_i \dot x_i^\mu A_\mu(x_i) \right]
 = 
 - & \int d\tau  \left[\tfrac{1}{2} + \tfrac{1}{2} \dot\barx_i^2 + z_i \dot \barx_i^\mu \bar A_\mu(\barx_i) \right. \\
& \; \left.+  \delta \dot x_i  \dot\barx_i + z_i   \delta x_i^\mu \dot \barx_i^\nu \bar F_{\mu\nu}(\barx_i) + z_i \dot \barx_i^\mu \dA_\mu(\barx_i) \right. \\
& \; \left.+ \tfrac{1}{2} \delta\dot x_i^2 + z_i \delta x_i^\mu \dot \barx_i^\nu    \dF_{\mu\nu}(\barx_i) \right. \\
& \; \left. + \tfrac{1}{2} z_i \delta x_i^\mu \delta \dot x_i^\nu \bar F_{\mu\nu}(\barx_i) + \tfrac{1}{2} z_i \delta x_i^\rho \delta x_i^\mu \dot \barx_i^\nu \partial_\rho \bar F_{\mu \nu} (\barx_i)  +  \cdots\right]
  \, ,
}{}
where the ellipses are higher than quadratic order in the perturbations. It will be useful to separately consider the contributions from the heavy and light worldlines, which will be treated slightly differently.  In particular, for the heavy particle and light particle we have
\eq{
 - \int d\tau \left[\tfrac{1}{2} + \tfrac{1}{2} \dot x_H^2+ z_H \dot x_H^\mu A_\mu(x_H) \right]
 = 
 -& \int d\tau  \left[\tfrac{1}{2} + \tfrac{1}{2} \dot\barx_H^2{\color{red} + z_H \dot \barx_H^\mu \bar A_\mu(\barx_H) }\right. \\
& \; \left.{\color{green} +  \delta \dot x_H  \dot\barx_H} {\color{red} + z_H   \delta x_H^\mu \dot \barx_H^\nu \bar F_{\mu\nu}(\barx_H)}{\color{blue} + z_H \dot \barx_H^\mu \dA_\mu(\barx_H)} \right. \\
& \; \left.+ \tfrac{1}{2} \delta\dot x_H^2 + z_H \delta x_H^\mu \dot \barx_H^\nu    \dF_{\mu\nu}(\barx_H) \right. \\
& \; \left. {\color{red}+ \tfrac{1}{2} z_H \delta x_H^\mu \delta \dot x_H^\nu \bar F_{\mu\nu}(\barx_H)} \right. \\
& \; \left. {\color{red} + \tfrac{1}{2} z_H \delta x_H^\rho \delta x_H^\mu \dot \barx_H^\nu \partial_\rho \bar F_{\mu \nu} (\barx_H)}  +  \cdots \right]
  \, ,
\\
-  \int d\tau \left[\tfrac{1}{2} + \tfrac{1}{2} \dot x_L^2- z_L \dot x_L^\mu A_\mu(x_L) \right]
 = 
 - & \int d\tau  \left[\tfrac{1}{2} + \tfrac{1}{2} \dot\barx_L^2 + z_L \dot \barx_L^\mu \bar A_\mu(\barx_L) \right. \\
& \; \left.{\color{purple}+  \delta \dot x_L  \dot\barx_L + z_L   \delta x_L^\mu \dot \barx_L^\nu \bar F_{\mu\nu}(\barx_L)} + z_L \dot \barx_L^\mu \dA_\mu(\barx_L) \right. \\
& \; \left.{\color{brown} + \tfrac{1}{2} \delta\dot x_L^2 + z_L \delta x_L^\mu \dot \barx_L^\nu    \dF_{\mu\nu}(\barx_L) }\right. \\
& \; \left. {\color{brown} + \tfrac{1}{2} z_L \delta x_L^\mu \delta \dot x_L^\nu \bar F_{\mu\nu}(\barx_L) }\right.\\
& \; \left.{\color{brown} + \tfrac{1}{2} z_L \delta x_L^\rho \delta x_L^\mu \dot \barx_L^\nu \partial_\rho \bar F_{\mu \nu} (\barx_L) } +  \cdots\right]
  \, .
}{EMaction_WL}
All of the colored terms will either be dropped or cancel amongst each other. The terms in red arise when a background Coulomb field is emitted and then reabsorbed by the heavy line.  Since these are self-energy diagrams, we drop them.  The blue, green, and purple terms vanish since the background EM field, heavy particle, and light particle satisfy their respective equations of motion. 

Last but not least, we can also drop the terms in brown which encode the dynamical propagation of the light particle.  Crucially, since the light particle action  is suppressed by an overall factor of $\mL$, which is manifestly ${\cal O}(\lambda^1)$, these terms that are quadratic in the light particle perturbations scale as ${\cal O}(\lambda^3)$ in the action, and are thus subleading to 1SF.  Thus, we enjoy an enormous simplification, which is that the deviation of the light particle from its 0SF trajectory can actually be ignored for any computation of the leading 1SF effects.

Putting together the terms in \Eq{EMaction_FF} and \Eq{EMaction_WL}  that are relevant to the leading 1SF dynamics, the action for the perturbations is
\eq{
S^{(1)} &= - \mH \int d\tau \left[\tfrac{1}{2} \delta\dotxH^2 + \zH \delta \xH^\mu \dotbarxH^\nu \dF_{\mu\nu}(\barxH) + \lambda \zL\dotbarxL^\mu \dA_\mu(\barxL) \right]  - \int d^4x\left[\tfrac14  \dF_{\mu\nu}\dF^{\mu\nu}  \right]\,.
}{}
We note again that the coupling of the light particle to the EM field has an explicit factor of $\lambda$, so we only needed to expand that term to linear order in perturbations. Typically, when expanding an action about a background solution, the terms which are linear in perturbations vanish, simply because we are expanding about a solution. In our situation however we must recall that $\bar{A}_{\mu}$ is a solution in the $\lambda\rightarrow 0$ limit, so it does not account for the light particle source.  This is why $S^{(1)}$ still has a term linear in $\delta A_{\mu}$.


The action $S^{(1)} $ describes  how the EM field deviates from the boosted Coulomb solution due to the dynamics of the heavy and light sources. 
With this in mind, let us write the  action for the perturbations as 
\eq{
S^{(1)}&= - \mH \int d\tau \left[\tfrac{1}{2} \delta\dotxH^2 + \zH \delta \xH^\mu \dotbarxH^\nu \dF_{\mu\nu}(\barxH)\right] - \int d^4x\left[ \tfrac14  \dF_{\mu\nu}\dF^{\mu\nu}  + \dA_\mu \barJL^\mu \right]\,,
}{deltaS_EM}
where the light particle 0SF current is\eq{
\barJL^\mu(x)&= \lambda \zL \mH \int d\tau \, \delta^4(x-\barxL) \dotbarxL^\mu  \, .
}{JL_EM}
We emphasize again here that for the 1SF dynamics we can completely ignore the light particle deviations away from geodesic motion.

It would be perfectly fine to simply compute observables using $S^{(1)}$.  However, if we are not interested in observables that depend directly on the state of the heavy body, we can do better.  Such observables include the light body deflection, conservative scattering angle,  and the radiated waveforms or fluxes. 
 In these cases it is more convenient to derive an even simpler effective theory by {\it integrating out the deviations of the heavy particle away from inertial motion}. 
At 0SF, the action describes a charged probe evolving in the field of an infinitely heavy source.  However, with our 1SF corrections, there is an additional effect---encoded in a term we dub the recoil operator---which accounts for the underlying dynamical propagation of the heavy particle.

Conveniently, at 1SF order $S^{(1)}$ is quadratic in $\delta \xH^\mu$, so we can integrate out this mode exactly.   Performing the path integral over $\delta\xH^\mu$ we obtain, up to constant normalization,
\eq{
\int [d \delta\xH] \exp\left(- i\mH \int d\tau \left[\tfrac{1}{2} \delta\dotxH^2 + \zH\delta \xH^\mu\dotbarxH^\nu \dF_{\mu\nu}(\barxH)\right]\right) = \exp\left(i  S_{\rm recoil}^{(1)} \right) \, ,
}{eq:1SFheavyterms}
where  $S_{\rm recoil}^{(1)}$ is the {\it electromagnetic recoil operator}, with the superscript denoting the SF order.  To compute this operator, we use the fact that for a Gaussian integral, we can simply set the perturbation to its solution under the equations of motion.  
Variation of \Eq{deltaS_EM} with respect to the heavy particle perturbation gives the equation of motion,
\eq{
\delta \ddotxH^\mu  - \zH  \dF^{\mu\nu}(\barxH) \dotbarxH{}_\nu  =0 \, ,
}{}
which, as expected, is simply the equation for the heavy worldline expanded to 1SF. Plugging this back into \Eq{deltaS_EM}, we obtain the recoil operator,
\eq{
S_{\rm recoil}^{(1)} &= -\frac12 \zH^2 \mH \int d\tau \,   \dotbarxH^{\alpha} \dF_{\alpha\mu}(\barxH) \frac{1}{\partial_\tau^2} \dotbarxH^{\beta} \dF_{\beta}^{\;\;\mu}(\barxH) \,.
}{recoil_op_EM}
This gauge invariant operator encodes the dynamical propagation of the heavy particle as a nonlocal-in-time correction to photon propagation localized exactly at the position of the heavy particle.   Note that, since $S_{\rm recoil}^{(1)} \sim \int d\tau \,  \delta E_{\mu}(\barxH) \partial_\tau^{-2} \delta E^{\mu}(\barxH)$, we can view the recoil operator as a polarizability operator on the heavy worldline that is nonlocal in time. The nonlocality makes this operator sufficiently different from true polarizability operators that we will not push on this analogy further. However, we note that while a photon does not scatter off of a fixed $1/r$ Coulomb potential, it \textit{does} scatter off the recoil operator.

The utility of the recoil operator is that we no longer need to track the heavy degree of freedom.  The only explicit source of photons is the light particle 0SF current and the recoil of the heavy body appears only in a modification to the  photon's propagator. This allows us to essentially use the background field method, supplemented by this modified propagator. The advantage of this will be more apparent in gravity, where the gravitational perturbations couple to the background.

Note that, in \Eq{recoil_op_EM}, we have not been explicit about the boundary conditions for the Greens function $\partial_\tau^{-2}$.  For the case of conservative dynamics, time reversal symmetry implies that quantities like the scattering angle or radial action can be computed using either retarded or advanced boundary conditions. For these computations it will then not matter if we use a retarded or advanced  $i\epsilon$-prescription for the propagator. However, more generally one must properly specify the boundary prescription for $\partial_\tau^{-2}$ appropriate to the calculation at hand. 

In this work we focus exclusively on conservative dynamics. However, our basic approach---setting up an effective action to be accurate order by order in a mass ratio expansion---along with the technical simplifications coming from extracting all-orders 0SF information, can be readily applied within an in-in formalism such as ~\cite{galley, Jakobsen:2022psy, Kalin:2022hph} to compute dissipative effects. In such a framework, provided that one is not computing heavy-particle observables, one can again integrate out $\dxH$ to obtain a recoil operator whose nonlocality describes a causal propagator. If one is actually interested in computing a heavy-particle observable, one would no longer integrate out $\dxH$ but would rather include it as either an external line or a cut line in perturbative diagrams.

In conclusion,  we have obtained the effective action encoding all 1SF dynamics,
\eq{
S^{(1)}_{\rm eff}&=  S_{\rm recoil}^{(1)} - \int d^4 x \left(\tfrac14   \dF_{\mu\nu} \dF^{\mu\nu}  + \dA_\mu \barJL^\mu\right) \, .
}{1SF_action_EM}
This describes photon perturbations about the background Coulomb field, which are sourced by the probe motion of the light particle and augmented by the recoil of the heavy particle via $S_{\rm recoil}^{(1)}$.  For example, for the case of Rutherford scattering, this would account for the wobble of the nucleus upon scattering.

\subsubsection{2SF Dynamics}

Our formalism can be generalized to 2SF order by expanding the action to $\OO(\lambda^{3})$. This procedure will generate terms of the schematic form,  $\dxL^{2}$ and $\dxL \dA$, as well as $(\dxH)^2\dA$. Contributions involving the light particle were already presented earlier in \Eq{EMaction_WL},
\eq{
S^{(2)}_{L} = 
 - \mL\int d\tau & \bigg[\tfrac{1}{2} \delta\dot x_L^2 + z_L \delta x_L^\mu \dot \barx_L^\nu    \dF_{\mu\nu}(\barx_L) + \tfrac{1}{2} z_L \delta x_L^\mu \delta \dot x_L^\nu \bar F_{\mu\nu}(\barx_L) + \tfrac{1}{2} z_L \delta x_L^\rho \delta x_L^\mu \dot \barx_L^\nu \partial_\rho \bar F_{\mu \nu} (\barx_L) \bigg] \, .
}{}
The first two terms encode the propagation and deflection of the light particle away from its probe motion due exchange of the photon.
These contributions are analogous to those of the heavy particle at 1SF, which precisely generated the recoil operator.  The last two terms involve the value of the background field, $\bar F_{\mu\nu}$, at the 0SF position of the light particle. The analogous terms for the heavy particle were singular self-energy divergences that were discarded in dimensional regularization. Here, these light particle terms are non-singular and describe the fact that while the light particle can be perturbed away from its 0SF trajectory by non-Coulomb effects, it still propagates in the background, $\bar F_{\mu\nu}$. As a preview, we note that in GR these kinds of light particle contributions encode the \textit{geodesic deviation} caused by the graviton perturbations.

The contributions to the heavy world-line action which are cubic in perturbations are straightforwardly computed to be
\eq{
S^{(2)}_{H}=-m_H \zH\int d\tau \bigg[&\tfrac{1}{2}\dxH^{\mu}\dotdxH^{\nu}\delta F_{\mu \nu}(\barxH)+\tfrac{1}{2}\dxH^{\rho}\dxH^{\mu}\dotbarxH^{\nu}\partial_{\rho} F_{\mu \nu}(\barxH)\\
&{\color{red}+\tfrac{1}{3}\dxH^{\rho}\dxH^{\mu} \dotdxH^{\nu}\partial_{\rho}\bar{F}_{\mu \nu}(\barxH) +\tfrac{1}{6}\dxH^{\sigma} \dxH^{\rho}\dxH^{\mu}\dotbarxH^{\nu}\partial_{\sigma}\partial_{\rho}\bar{F}_{\mu \nu}(\barxH)} \bigg]\,.
}{}
The terms in red are self-energy divergences and can again be dropped. The terms in black give rise to a double-recoil operator, i.e., an effective cubic photon operator which contains two iterated matter propagators. 

To see this, we combine all terms in the effective action through 2SF order involving the heavy particle fluctuation,
\eq{
S^{(1)}_{H}+S^{(2)}_{H} = - \mH\int d\tau \big[& \tfrac{1}{2} \delta\dotxH^2 + \zH\delta \xH^\mu\dotbarxH^\nu \dF_{\mu\nu}(\barxH) + \tfrac{1}{2} \zH \dxH^{\mu}\dotdxH^{\nu}\delta F_{\mu \nu}(\barxH)\\
&+ \tfrac{1}{2} \zH \dxH^{\rho}\dxH^{\mu}\dotbarxH^{\nu}\partial_{\rho}\delta F_{\mu \nu}(\barxH)  \big] \, ,
}{eq:2SFheavyterms}
and integrate out the heavy particle as before.  This generates the recoil operators,
\eq{
\int [d \delta\xH] \exp\left(iS^{(1)}_{H}+iS^{(2)}_{H}\right) = \exp\left(i  S^{(1)}_{\rm recoil}+iS^{(2)}_{\rm recoil}\right) \, .
}{}
Since \Eq{eq:2SFheavyterms} contains terms which are cubic in perturbations, i.e., scaling as $\lambda^{3}$, integrating out the heavy particle perturbations will generate terms at arbitrarily high SF order. Given that we started with a 2SF accurate action we are only, however, permitted to trust the obtained recoil operators up to 2SF order.

At 2SF order we have the operator,
\eq{
S^{(2)}_{\rm recoil}= - \frac{\zH^{3}\mH}{2}\int d\tau \Bigg[&\delta E^{\alpha}(\barxH)\frac{1}{\overleftarrow{\partial_{\tau}^{2}}}\delta F_{\alpha\mu}(\barxH)\frac{1}{\overrightarrow{\partial_{\tau}}}\delta E^{\mu}(\barxH)\\
& +\delta E^{\alpha}(\barxH)\frac{1}{\overleftarrow{\partial_{\tau}^{2}}}\partial_{\mu}\delta E_{\alpha}(\barxH)\frac{1}{\overrightarrow{\partial_{\tau}^{2}}}\delta E^{\mu}(\barxH)\Bigg]\,,
}{}
where $\delta E^{\mu}(\barxH)=\dotbarxH{}_{\nu}\delta F^{\mu\nu}(\barxH)$ is the electric field in the frame of the heavy particle.

In summary, the effective action describing the dynamics through 2SF order is
\eq{
S^{(1)}_{\rm eff}+S^{(2)}_{\rm eff}\ = \ \, &S^{(1)}_{\rm recoil}+S^{(2)}_{\rm recoil} - \int d^4 x \left( \tfrac14   \dF_{\mu\nu} \dF^{\mu\nu}  + \dA_\mu \barJL^\mu\right) \\
&- \mL\int d\tau\Big[\tfrac{1}{2} \delta\dotxL^2  + \zL \delta \xL^\mu \dotbarxL^\nu  \dF_{\mu\nu}(\barxL) \\
&+ \tfrac{1}{2} \zL \delta \xL^\mu \delta \dot x_L^\nu \bar F_{\mu\nu}(\barxL) + \tfrac{1}{2} \zL \delta \xL^\rho \delta \xL^\mu \dot \barx_L^\nu \partial_\rho \bar F_{\mu \nu} (\barxL) \Big]\, .
}{2SF_action_EM}

A quick way to determine the PL order of certain contributions to conservative scattering is to simple count powers of $z_{L}$. For example, tree-level exchange has one insertion of a photon on the light body and is therefore 1PL. We can readily  power count, in $\zL$, the operators appearing in \Eq{2SF_action_EM}. As we know from the 1SF section, the perturbation $\dF_{\mu\nu}$ is sourced by the light particle's probe current $\bar{J}_{L}^{\mu}$, such that $\dF_{\mu\nu}\sim \zL$. Additionally, the equation of motion for $\dxL$ is of the form
\eq{
\delta\ddot{x}_{L}\sim z_{L}\dF+\zL\dxL\bar{F}\,.
}{}
From this, and the scaling of $\dF$, we immediately determine the scaling $\dxL\sim \zL^{2}$. Looking at terms in \Eq{2SF_action_EM}, it follows that the light particle perturbations do not contribute until 4PL order, while the two terms involving $\zL\dxL^2\bar{F}$ do not contribute until 5PL. So, for the 3PL computations performed later in this work, we will only need the recoil operators $S^{(1)}_{\rm recoil}$ and $S^{(2)}_{\rm recoil}$.

\subsection{Feynman Rules} 

Physical observables are computed from the 1SF effective action in \Eq{1SF_action_EM} by performing the path integral over the one remaining degree of freedom, which is the  photon perturbation. From \Eq{1SF_action_EM}, it is straightforward to derive the associated Feynman rules for this calculation.  We first describe the Feynman rules for the propagator and vertices for the photon, and then move on to describe its sources.  
For ease of use, we have presented a table summarizing all of the 1SF Feynman rules for EM in \Fig{fig:Feynman_rules_EM}.

\subsubsection{Photon Propagators and Vertices}
To start, we choose Feynman gauge for the photon fluctuation\footnote{Since we are focused on the conservative dynamics, propagators should appear with a causal $i\epsilon$ prescription, thought we will only retain the time-symmetric component of the result.}, so
\eq{
\flatproparrow &= -\frac{i\eta_{\mu\nu}}{p^2} \, .
}{}
Meanwhile, it is easy to derive the two-point vertex for the photon fluctuation induced by the recoil operator in \Eq{recoil_op_EM}, which is
\eq{
\vcenter{\hbox{\flatrecoil}} &=   i\zH^2 \mH   \frac{\hat  \delta(\vH p_1+\vH p_2)}{(\vH p_1)(\vH p_2)} {\cal O}^{\alpha \mu_1}(\vH, p_1)  {\cal O}_\alpha^{\;\;\,\mu_2}(\vH, p_2) \, ,
}{recoil_vertex_EM}
where ${\cal O}^{\alpha \mu}(\v ,p) = \eta^{\alpha \mu} (\v p) -  \v^\mu p^\alpha$.
The denominator factor is  the nonlocal-in-time worldline propagator, $\partial_\tau^{-2}$, while the delta function encodes that the invariance of  heavy particle trajectory under translations in  $\vH$ direction. 

Note that the 1SF action in \Eq{1SF_action_EM} does not contain any explicit dependence on the background gauge field.  This is not an accident---since electromagnetism is a linear theory, the  perturbations decouple from the background about which we expand.  The same will not be true, however, for gravity.  Nevertheless, for future reference let us recall that 
\eq{
\bar A_\mu(q) = - \zH \mH \vH{}_\mu \frac{\hat{\delta}(\vH q)}{q^2}  \qquad \textrm{and} \qquad
 \bar F_{\mu\nu}(q) = -i q_{[\mu} \bar A_{\nu]}(q) \, ,
}{bkgd_field_EM}
which are the background electromagnetic gauge field and field strength in momentum space.

\subsubsection{Photon Sources}

At 1SF order, the photon can only terminate on the light particle source in \Eq{JL_EM} which, transformed into momentum space and to all PL orders, is
\eq{
&\barJL^\mu(p)  = \int d^4 x\, e^{ipx} \barJL^\mu(x) =\lambda \zL \mH  \int d\tau \, e^{ip \barxL} \dotbarxL^\mu \, .
}{}
In practice, we will be interested in a perturbative PL expansion. To compute to any given PL order, we expand the expression for the light particle trajectory,\eq{
\barxL^\mu = \sum_{k=0}^\infty  \barx_k^\mu\, ,
}{xL_PL}
where $\barx_k^\mu$ is the $k$-PL order contribution. Inserting this into the light particle source, we obtain
\eq{
\barJL^\mu(p) &=\lambda \zL \mH  \int d\tau \, e^{ip \barx_0 } e^{ip(\bar x_1+\bar x_2 +\cdots)} (\dot {\bar x}_0^\mu +\dot {\bar x}_1^\mu +\dot {\bar x}_2^\mu+\cdots) \\
&=\lambda \zL \mH  \int d\tau \, e^{ip \barx_0 }  (\dot {\bar x}_0^\mu - i(p \dot\barx_0 \delta^{\mu}_{\nu}- p_\nu  \dot\barx_0^\mu)(\barx_1^\nu+\barx_2^\nu)+ip\barx_1 \dot {\bar x}_1^\mu -\frac{1}{2} (p\barx_1)^2\dot {\bar x}_0^\mu+\cdots)  \, ,
}{}
where in the last line we have expanded up to 2PL order in terms of \Eq{xL_PL}.  Here, we have integrated by parts to make some terms look more uniform and the ellipses denote contributions that enter beyond 2PL order.
For concreteness, let us consider the light particle current at 1PL order, which can be written more compactly as
\eq{
\barJL^\mu(p)  &=\lambda \zL \mH e^{ipb} \left( \hat{\delta}(\vL p) \vL^\mu - i {\cal O}_\alpha^{\;\;\, \mu}(\vL, p) \bar x_{1}^\alpha(\vL p) +\cdots  \right)  \, ,
}{}
where we have defined the trajectories in frequency space, $ \bar x^\mu_i(\omega)=\int d\tau \, e^{i\omega\tau} \barx_i^\mu(\tau) $. 

 At this point we can use any method we like to compute the light particle trajectory in frequency space, $\bar x_1^\mu(\omega)$.   As described in \Sec{pert_resumEM}, we can either compute these probe trajectories perturbatively, or extract them from the known analytic solutions.  Taking the former approach, we expand \Eq{xL_EOM_EM} to leading PL order and transform to frequency space, 
\eq{
\bar x_1^\mu(\omega) = -\frac{\zL \vL{}_\nu}{\omega^2} \int \frac{d^4 q}{(2 \pi)^4}\, e^{-iqb}  \hat{\delta}(\omega- \vL q) \bar F^{\mu\nu}(q)\, ,
}{x1_omega_EM_A}
where the background electromagnetic gauge field and field strength in momentum space are defined in \Eq{bkgd_field_EM}.

Lastly, the Feynman rule associated with the light particle source for the photon perturbation is
\eq{
\vcenter{\hbox{\source}} \quad &= \quad \vcenter{\hbox{\sourcezero}} \qquad + \quad \vcenter{\hbox{\sourceone}} \qquad + \quad \cdots \\
&= -i\lambda \zL \mH e^{ipb} \left( \hat{\delta}(\vL p) \vL^\mu - i {\cal O}_\alpha^{\;\;\, \mu}(\vL, p) \bar x_{1}^\alpha(\vL p) +\cdots  \right)  \, .
}{}
Note that for $\bar x_1^\mu(\omega)$, we can choose either the expression in \Eq{x1_omega_EM_A} or any other representation of the trajectory.

\begin{figure*}
    \centering
    \begin{tabular}{|c|c|c|}
    \hline
    $\begin{array}{cc}
            \flatproparrow\\
             \textrm{Photon propagator}
        \end{array}$ & \(\displaystyle
            -\frac{i\eta_{\mu\nu}}{p^2}
        \) \\
    \hline
        $\begin{array}{cc}
            \flatrecoil\\
             \textrm{Recoil vertex}
        \end{array}$
         & \(\displaystyle
            i\zH^2 \mH   \frac{\hat{\delta}(\vH p_1+\vH p_2)}{(\vH p_1)(\vH p_2)} {\cal O}^{\alpha \mu_1}(\vH, p_1)  {\cal O}_\alpha^{\;\;\,\mu_2}(\vH, p_2)
        \) \\
    \hline
       $\begin{array}{cc}
            \\
            \source\\
             \textrm{Photon source}
        \end{array}$
         & \(\displaystyle
            -i\lambda \zL \mH e^{ipb} \left( \vL^\mu  \hat{\delta}(\vL p) - i {\cal O}_\alpha^{\;\;\, \mu}(\vL, p) \bar x_{1}^\alpha(\vL p) +\cdots  \right)
        \) \\
        \hline
    \end{tabular}
    \caption{Feynman rules for computing the radial action for EM at 1SF.}
    \label{fig:Feynman_rules_EM}
\end{figure*}

\subsection{Classical Resummation}\label{pert_resumEM}

A convenient byproduct of our effective field theory is that it repackages certain perturbative contributions into  the probe trajectories.  Here we discuss various methods to extract this information directly from the probe motion.

\subsubsection{From Second-Order Equations of Motion}

The most direct path to computing the probe trajectories is to solve \Eq{xL_EOM_EM} perturbatively in the coupling, which in the case of EM is the PL expansion.  This is basically the approach of \cite{Kalin:2020mvi}, albeit here restricted to the probe limit.  With this method, one plugs \Eq{xL_PL} into the light particle equations of motion in \Eq{xL_EOM_EM} to obtain the perturbative equations of motion at 0PL, 1PL, and 2PL order,
\eq{
\ddot \barx_0^\mu &=0 \, ,\\
\ddot \barx_1^\mu  &= \zL
\bar F^{\mu\nu}(\barx_0) \dot \barx_{0\nu} \, ,\\
\ddot \barx_2^\mu  &=  \zL  \left( \bar  F^{\mu\nu}(\barx_0) \dot \barx_{1\nu} + \barx_1^\rho \partial_\rho \bar  F^{\mu\nu}(\barx_0) \dot \barx_{0\nu}  \right)\, ,
}{pert_EOM_EM}
and so on and so forth.  The solution to the 0PL equation is just the straight line trajectory,
\eq{
\barx_0^\mu = b^\mu + \vL^\mu \tau \, ,
}{x0_EM}
where $\vL^\mu$ is the light particle velocity and $b^\mu$ is a space-like vector defining the impact parameter.  Plugging this back into the 1PL equation, we obtain 
\eq{
\barx_1^\mu = \frac{1}{\partial_\tau^2} \zL  
\bar  F^{\mu\nu}(b+\vL \tau) \vL{}_\nu \, ,
}{traj_EM}
expressed formally in terms of the light particle propagator, $1/\partial_\tau^2$.   The appearance of this propagator indicates the second-order differential nature of the equations of motion.

Similarly, we consider the PL expansion of photon equation of motion,
\eq{
       \partial_\mu \bar F^{\mu\nu}= \square \bar A^{\nu} = \bar J^\nu_H = z_H m_H \int d\tau u_H^\nu \delta^4(x^\mu - u_H^\mu \tau)\,,
}{}
which truncates at leading order because EM is a linear theory.
In nonlinear field theories like GR, the fields' low PM order solutions are themselves sources for higher order perturbations, and the perturbative series for the field configuration does not truncate---leading, in principle, to more and more complicated perturbative computations.

\subsubsection{From First-Order Conservation Laws}

The procedure outlined above calculates the perturbative trajectories from the second-order equations of motion.  However, since this system has several underlying symmetries, like time translation and rotational invariance, it is natural to instead study the first-order equations dictated by the associated conserved quantities.   To do so, let us consider the orbital equations in spherical coordinates, $(t,r,\phi)$, in the fixed background of the heavy particle.  For the outward branch of the scattering trajectory, corresponding to $\dot{r}\geq0$, we then have 
\eq{
\dot{t}&=\sigma+\frac{\Rc}{r}\, ,\\
\dot{\phi}&=(\sigma^{2}-1)^{1/2}\frac{b}{r^2}\, ,\\
\dot{r}&=\sqrt{\left(\sigma+\frac{\Rc}{r}\right)^2-\frac{b^2 (\sigma^{2}-1)}{r^2}-1}\, ,
}{EMpolarEOM}
so the dynamics are controlled by three first-order differential equations which are straightforwardly solved in perturbation theory.  Here one may rightly wonder whether there is any operational advantage over the above second-order approach of the previous section.   As we will see, the solutions extracted from \Eq{EMpolarEOM} have a much simplified structure.   This gain is best articulated in the language of Feynman loop integrals: the trajectories extracted from the first-order equations effectively exhibit manifest propagator pinches and a considerable reduction of tensor structures, leading to simpler numerators. In \App{app:probe_traj}, we provide details on this procedure in general $D$ dimensions, which illustrate that this method effectively performs integral-by-parts reduction on the Feynman integrals which comprise the trajectories.

It is convenient to combine the equations of motion for $(r,\phi)$ into equations of motion for the Cartesian components $(x,y)$,
\eq{
\dot{x}&=\frac{x}{r}\left[\left(\sigma+\frac{\Rc}{r}\right)^2-\frac{b^2 (\sigma^{2}-1)}{r^2}-1\right]^{1/2}-\frac{y}{r^2}(\sigma^{2}-1)^{1/2}b\,, \\
\dot{y}&=\frac{y}{r}\left[\left(\sigma+\frac{\Rc}{r}\right)^2-\frac{b^2 (\sigma^{2}-1)}{r^2}-1\right]^{1/2}+\frac{x}{r^2}(\sigma^{2}-1)^{1/2}b \, , 
}{}
where the variable, $r$, is understood to be an implicit function of $(x,y)$. These components of the trajectory can be reconstituted into a Lorentz covariant form,
\eq{
\bar{x}^{\mu}(\tau)= t(\tau) \vH^{\mu}+ x(\tau)\frac{b^{\mu}}{b}+y(\tau)\frac{\vL^{\mu}-\sigma \vH^{\mu}}{(\sigma^{2}-1)^{1/2}}\,.
}{}
As described in detail in \App{app:probe_traj}, it is straightforward to integrate the above first-order equations in the PL expansion to obtain time-domain trajectories.  Crucially, we can algorithmically rewrite these time-domain solutions in terms of iterated time integrals of powers of $R(\tau)=\sqrt{b^{2}+(\sigma^{2}-1)^{1/2}\tau^{2}}$. These solutions at 1PL order are
\eq{
t_{1}&=\Rc\frac{1}{\partial_{\tau}}\left(\frac{1}{R}\right) \, , \\
x_{1}&=-\Rc\sigma\frac{1}{\partial_{\tau}^{2}} \left(\frac{b}{R^{3}}\right)\, , \\
y_{1}&=\Rc\frac{\sigma}{(\sigma^2-1)^{1/2}}\frac{1}{\partial_{\tau}}\left(\frac{1}{R}\right)\,,
}{}
while at 2PL order they are
\eq{
t_{2}=&\Rc^{2}\frac{\sigma  \frac{1}{\partial_{\tau}}\left(\frac{1}{R}\right)}{R \left(\sigma ^2-1\right)}\, , \\
x_{2}=&\Rc^{2}\left(-\frac{ \sigma ^2 \frac{1}{\partial_{\tau}}\left(\frac{\frac{1}{\partial_{\tau}}\left(\frac{b}{R^3}\right)}{R}\right)}{\sigma ^2-1}-\frac{ \sigma ^2 \frac{1}{\partial_{\tau}}\left(\frac{b\frac{1}{\partial_{\tau}}\left(\frac{1}{R}\right)}{R^3}\right)}{\sigma ^2-1}- \frac{1}{\partial^{2}_{\tau}}\left(\frac{b}{R^4}\right)\right)\, , \\
y_{2}=&\Rc^{2}\left(-\frac{\frac{1}{\partial_{\tau}}(1) \sigma ^2}{2 b^2 \left(\sigma ^2-1\right)^{3/2}}+\frac{\frac{1}{\partial_{\tau}}\left(\frac{1}{R^2}\right)}{2 \sqrt{\sigma ^2-1}}+\frac{\sigma ^2 \frac{1}{\partial_{\tau}}\left(\frac{1}{R}\right)}{R\left(\sigma ^2-1\right)^{3/2}}\right) \, ,
}{eq:simplifiedsolutions}
and so on and so forth.  

By construction, the above expressions for the trajectories resemble the elements of a Feynman diagram. In particular, a Fourier transform to momentum space maps inverse powers of $R$ to inverse powers of the spatial momentum transfer,  $\ell^{-2}$.  At the same time, it maps $\partial_{\tau}^{-1}$ to the linearized matter propagators of the form, $(\vL\cdot\ell)^{-1}$. For example, consider
\eq{
\frac{1}{\partial_{\tau}}\left(\frac{1}{R^2}\right)=16i \pi ^2\int_{\ell_{1},\ell_{2}}\frac{e^{-ir(\ell_{1}+\ell_{2})} \hat{\delta} \left(\vH\ell_{1}\right) \hat{\delta} \left(\vH\ell_{2}\right)}{\ell_{1}^{2} \ell_{2}^{2} \left(\vL\ell_{1}+   \vL\ell_{2}+i\epsilon\right)}=16i\pi^{2}\times  \vcenter{\hbox{\traj}}\,.
}{eq:EIHinsertion}
The general $D$-dimensional Fourier transforms are given in  \Eqs{eq:oneloopfan}{eq:fourierR} of \App{app:probe_traj}. Each insertion of $R^{3-D}$ and $b R^{1-D}$ corresponds to the insertion of a background photon, with products of these building blocks such as $b R^{4-2D}$ and $R^{6-2D}$ corresponding to one-loop insertions\footnote{The factor of $\frac{1}{\partial_{\tau}}(1)$ in \Eq{eq:simplifiedsolutions} actually corresponds to the $D=4$ limit of $\frac{1}{\partial_{\tau}}(R^{8-2D})$, as illustrated in \App{app:probe_traj}. Here $R^{8-2D}$ comes from the product of $R^{3-D}$ and $R^{5-D}$, with the latter coming from the insertion of a ``doubled'' photon propagator, $(\ell^{2})^{-2}$. See \App{app:probe_traj}, and particularly \Eq{traj_eg} for details.}, and higher products corresponding to multi-loop insertions.

We thus observe that the trajectories are simply Feynman integrals, where inverse powers of $R$ correspond to background photon insertions, and powers of $\partial_{\tau}^{-1}$ correspond to matter propagators. With this identification, we can ascertain which Feynman integral topologies are encoded in the trajectories, for example as shown in \Eq{eq:EIHinsertion}. The marginal advantage of this approach in EM is relatively minor. However, we will see later in GR that there is a upside to this approach because the resulting integral topologies will be the same as in EM.

\subsection{Results and Checks}

To compute the on-shell radial action, we simply perform the path integral over the fluctuation degrees of freedom.  In particular, the radial action, $I_{\rm EM}$, is defined up to a constant normalization by
\eq{
\exp(i I_{\rm EM}) = \int [d \delta\xH] [d \delta \xL] [d \dA] \exp(i S) \, .
}{}
Since we are interested only in classical physics, the path integral serves only as an organizational tool for our perturbation theory. Working up to 0SF and 1SF, we can ignore the perturbations of the light particle, $\delta \xL^\mu$.  Meanwhile, performing the path integral over the heavy particle perturbation, $\delta \xH^\mu$, simply yields the recoil operator which appears in the effective action, $S_{\rm eff}^{(1)}$, as defined in \Eq{1SF_action_EM}.  Thus, we obtain
\eq{
\exp(i I_{\rm EM}) =  \int  [d \dA] \exp(i \bar S + i S_{\rm eff}^{(1)} +\cdots) \, ,
}{}
where the ellipses denote contributions beyond 1SF.   Decomposing the radial action according to the SF expansion, we find that
\eq{
I_{\rm EM} &= I^{(0)}_{\rm EM} + I^{(1)}_{\rm EM}+ \cdots \, ,
}{}
where the superscripts denote the SF order of a given contribution and
\eq{
I^{(0)}_{\rm EM} &= \bar S \, ,\\
I^{(1)}_{\rm EM} &= -i\log  \int  [d \dA] \exp( i S_{\rm eff}^{(1)} )  \, .
}{}
The above manipulations imply, as is well-known, that the 0SF radial action, $I^{(0)}_{\rm EM}$, is simply the action evaluated on the probe solution, $\bar S$, as given in 
\Eq{eq:EMproberadialactiontext}.  Furthermore, the 1SF radial action, $I^{(1)}_{\rm EM}$, is computed by summing all tree-level connected 1SF Feynman diagrams that arise from integrating out the photon perturbation about the Coulomb background. 

\begin{figure*}
    \centering
    $\vcenter{\hbox{\brecoilEM}} \quad \rightarrow \quad \vcenter{\hbox{\recoilone}} \qquad \vcenter{\hbox{\recoiltwoone}}$
    \caption{Diagram contributing to the 1SF electromagnetic on-shell radial action. The left-hand side is composed of the source probe trajectory, photon propagators, and the 1SF electromagnetic recoil operator and the right-hand side shows diagrams contributing at 2PL and 3PL.}
    \label{fig:Feynman_diagrams_EM}
\end{figure*}

In what follows, we compute the 0SF and 1SF actions in $D=4$ spacetime dimensions, expanded in the PL expansion. Concretely, we express
\eq{
I_{\rm EM}^{(i)}=\sum_{j=i+1}^\infty I_{\rm EM}^{(i,j)} \, ,
}{}
where $i$ and $j$ denote the SF and PL orders of a given contribution. 
We will focus on scattering dynamics, in which case the kinematic data is the asymptotic impact parameter, $b$, and the relative boost factor, $\sigma=\vH\vL $. It follows from dimensional analysis that the SF+PL expansion of the radial action has the form,
\eq{
I_{\rm EM}^{(i,j)}=\lambda^{i}\,\mL\Rc\left(\frac{\Rc}{b}\right)^{j-1}\mathcal{I}_{{\rm EM}}^{(i,j)}(\sigma) \, ,
}{}
which is the electromagnetic analog of the good mass polynomiality in gravity \cite{Damour:2019lcq}.

\subsubsection{Scattering Electric Charges}

At last, we are equipped to compute the 1SF radial action for the scattering of electrically charged particles in EM. This is computed by summing Feynman diagrams in our 1SF order effective theory.

We are interested in only the conservative contributions to the scattering dynamics, leaving a treatment of radiative losses for future work. Through 3PL order, these conservative contributions to the radial action are entirely accounted for by the diagram in \Fig{fig:Feynman_diagrams_EM}. Beyond 3PL order, but still at 1SF, one must also consider multi-insertions of the recoil operator, however such diagrams vanish in the potential region.

The leading, 2PL, contribution evaluates to
\eq{
I_{\rm EM}^{(1,2)}&=\frac{\lambda^2}{2} \zL^2 \zH^2 \mH^3 \int_{\ell_1,\ell_2} \frac{e^{-ib(\ell_1+\ell_2)b}\hat{\delta}(\vH (\ell_1+\ell_2))\hat{\delta}(\vL \ell_1)\hat{\delta}(\vL \ell_2)}{\ell_1^2 \ell_2^2} \left( 1+ \frac{\sigma^2 (\ell_1 \ell_2)}{(\vH \ell_1)(\vH \ell_2)} \right) \\
&=\lambda \mL\Rc\frac{\Rc}{b}\frac{\pi}{2\sqrt{\sigma^2-1}}   \, ,
}{integrand_2PM_EM}
where the matter poles carry a principal value prescription unless otherwise specified.
The two integrations correspond to a one-loop integral over the internal photon momentum, followed by a Fourier transform to impact parameter space.  As expected, $I_{\rm EM}^{(1,2)}$ is the same as the probe result, $I_{\rm EM}^{(0,2)}$, upon swapping the light and heavy particles,  where $I_{\rm EM}^{(0,2)}$ is given by \Eq{eq:EMproberadialactiontext}.

As a check, we have also performed the 2PL calculation in general spacetime dimension, $D$, yielding 
\eq{
I_{\rm EM}^{(1,2)}&= \frac{\lambda\mL\Rc^{2}}{b^{2D-7}} \frac{\Gamma(D-7/2)\Gamma(D/2-3/2)^2}{\pi^{D-7/2} \Gamma(D-3)}   \frac{(2D-7)\sigma^2-1}{2(\sigma^2-1)^{3/2}} \, .
 }{} 
We have verified that it is consistent using known results in the probe limit. 
See \App{app:probe_action} for technical details.

Moving on to the next order, we compute the 3PL Feynman diagrams in \Fig{fig:Feynman_diagrams_EM} and integrate them via the methods described thoroughly in \cite{Parra-Martinez:2020dzs, Dlapa:2023hsl}. Here we have employed both {\tt FIRE6} and {\tt LiteRed} to reduce the loop integrals onto a small basis of master integrals via integration-by-parts relations~\cite{Chetyrkin:1981qh, Laporta:2000dsw, Lee:2013mka, Smirnov:2019qkx}. The $b$ dependence of the integrals is fixed by dimensional analysis, leaving only the nontrivial velocity dependence to be determined  using the method of differential equations~\cite{Henn:2013pwa}. The boundary conditions for the differential equations are simply ``static'' integrals, which describe the particles at zero relative velocity. As we are focused on the conservative dynamics up to 3PM order, we expand these integrals in the potential region.

The integrand and final integrated answer is
\eq{
I_{\rm EM}^{(1,3)}&=  -(\mH\mL)^2(\zH\zL)^3\int_{\ell_{1}, \ell_{2}, \ell_{3}} e^{ib(\ell_{1}+\ell_{2}+\ell_{3})}\hat{\delta} (\vH(\ell_{1}+\ell_{2}+\ell_{3}))\frac{\hat{\delta} (\vL \ell_{1})\hat{\delta} (\vH \ell_{2})\hat{\delta} (\vL \ell_{3}) }
   {\ell_{1}^2 \ell_{2}^2 \ell_{3}^2 (\vH \ell_{1})^2 (\vL \ell_{2})^2}\\
&\times 
   \left(-(\ell_{1} \ell_{3})(\ell_{2} \ell_{3}) \sigma^3-\tfrac{1}{2}q^2(\vH \ell_{1})(\vL \ell_{2}) \sigma^2 +(\ell_{2} \ell_{3}) (\vH \ell_{1})^2\sigma+ (\ell_{1} \ell_{3}) (\vL \ell_{2})^2\sigma\right. \\
   &\;\;\;\;\;+\left. (\vL \ell_{2})(\vH \ell_{1})^3 +(\vH \ell_{1}) (\vH \ell_{2})^3\right) \\
&=  -\lambda\mL\Rc \left(\frac{\Rc}{b}\right)^2 \frac{2   \left(\sigma ^4-3 \sigma ^2+3\right)}{3 \left(\sigma
   ^2-1\right)^{5/2}} ,
}{}
which agrees exactly with the calculation from \cite{Bern:2021xze}. 

We also computed the 2SF radial action to 3PM order. The integrand and integral is again quite simple,
\eq{
I_{\rm EM}^{(2,3)}=&-32 \pi ^3 \lambda ^2 \mL \Rc^3 \sigma\int_{\ell_{1},\ell_{2},\ell_{3}} e^{ib(\ell_{1}+\ell_{2}+\ell_{3})}\hat{\delta}(\vH(\ell_{1}+\ell_{2}+\ell_{3}))\frac{\hat{\delta}(\vL\ell_{1}) \hat{\delta}(\vL\ell_{2}) \hat{\delta}(\vL \ell_{3})  }{{\ell_{1}^2 \ell_{2}^2 \ell_{3}^2 (\vH\ell_{1})^2 (\vH\ell_{2})^2}}\\
&\times\left((\ell_{1}\ell_{3}) (\ell_{2}\ell_{3}) \sigma
   ^2-(\ell_{1}\ell_{3}) (\vH\ell_{2})^2-(\ell_{2}\ell_{3}) (\vH\ell_{1})^2\right)\\
   =&\frac{\lambda^{2}\mL \Rc^3 \sigma  \left(2
   \sigma ^2-3\right)}{3 b^2 \left(\sigma
   ^2-1\right)^{5/2}}\,.
}{}
This agrees with the 0SF result, \Eq{eq:EMproberadialactiontext}, providing a crucial check on the consistency of our effective field theory description.

\medskip

\subsubsection{Scattering Dyonic Charges} 

It is straightforward to generalize the above results to the scattering of particles with both electric and magnetic charges. The action for the point-particle effective theory is then
\eq{
S_{\rm dyon} &= \sum_{i=H,L} \int d\tau \left[- \tfrac{1}{2}m_i  \dot x_i^2- q_i\dot x_i^\mu A_\mu(x_i)- g_i\dot x_i^\mu B_\mu(x_i) \right]+   \int d^4x\left[  -\tfrac14 F_{\mu\nu}(x)F^{\mu\nu}(x)\right]\, ,
}{S_dyon}
where $q_{i}$ and $g_{i}$ are the electric and magnetic charges, respectively.  Here the fields, $A_\mu$ and  $B_{\mu}$, are the photon and dual photon field \cite{zwanziger}. Of course, these degrees of freedom are not actually independent, since they are related to the field strength tensor by 
\eq{
\partial_{[\mu}A_{\nu]}&=F_{\mu\nu}\,, \nonumber \\
\partial_{[\mu}B_{\nu]}&=\ast F_{\mu\nu}.
}{AB_field_strength}
Here $\ast$ denotes the Hodge star operation, so $\ast^2 = -1$ in Lorentzian signature.  

The derivation of the 1SF effective action for dyons is essentially the same as for the pure electric case, so we will not repeat those details here.  However, the upshot of this exercise is that the 1SF dyon action can be obtained by performing a set of EM duality rotations on the 1SF action in the pure electric case. Conveniently, this procedure is relatively straightforward because the electric framework is written in terms of the field strength, which rotates simply under EM duality.

In particular, under EM duality the light particle equation of motion transforms to
\eq{
\mL \ddotbarxL^\mu  - \qL \dotbarxL{}_\nu  \bar F^{\mu\nu}(\xL)=0 \quad \rightarrow\quad 
\mL \ddotbarxL^\mu  - \dotbarxL{}_\nu (\qL+g_{L}\ast)\bar F^{\mu\nu}(\xL)=0\,,
}
{}
while the background field strength transforms as
\eq{
\bar{F}_{\mu\nu}(k) =i \qH \frac{\hat{\delta}(\vH k)}{k^{2}}k_{[\mu} \vH{}_{\nu]}  \quad \rightarrow \quad i\frac{\hat{\delta}(\vH k)}{k^{2}}(\qH-g_{H}\ast )k_{[\mu} \vH{}_{\nu]} \, .
}{}
Last but not least, in the recoil operator, we send 
\eq{
\qH\delta F_{\mu\nu} \quad \rightarrow \quad  (\qH + g_{H}\ast)\delta F_{\mu\nu}\, ,
}{fluctuation_dual_photons}
which is simply an EM duality transformation of the photon perturbation.

In the previous sections, we expressed all quantities in terms of the charge to mass ratios, $z_{i}=q_{i}/m_{i}$. To draw a structure which is parallel to the pure electric case, here it will be convenient to define the electric and magnetic charges in terms of angles,
\eq{
q_{i}= m_{i}z_{i}\,\cos\theta_i\qquad {\rm and} \qquad
g_{i}= m_{i}z_{i}\,\sin\theta_i\,,
}{}
so that the total magnitude of the EM charge of each particle is still given by $m_{i}z_{i}$. An EM duality rotation rotates these angles simultaneously and a check on our final results is that they depend only on the duality invariant, $\Delta\theta=\theta_L-\theta_H$. 

The above EM duality transformations mechanically generate the 1SF dynamics for interacting dyons, which we now summarize.  The equation of motion for the light dyon in the probe limit is
\eq{
\ddotbarxL^\mu  - \zL\dotbarxL{}_\nu (\cos\Delta\theta + \sin\Delta\theta \,\ast)\bar F^{\mu\nu}(\xL)=0 \, ,
}
{dyon_EoM}
where $\bar F_{\mu\nu}$ is given by \Eq{bkgd_field_EM}.
Meanwhile, the 1SF effective action for the scattering of dyons is given by
\eq{
\delta S_{\rm dyons}^{\rm eff}&=  {S_{\textrm{dyons}}^{\textrm{recoil}}}+ \int d^4 x \left( - \tfrac14   \dF_{\mu\nu} \dF^{\mu\nu}  - (\cos\theta_L\dA_\mu+\sin\theta_L\delta B_\mu )\barJL^\mu\right) \, ,
}{1SF_action_dyons}
where the dyonic recoil operator is
\eq{
{S_{\textrm{dyons}}^{\textrm{recoil}}} &= -\frac12 \zH^2 \mH \int d\tau \,   \dotbarxH^{\alpha} \delta G_{\alpha\mu}(\barxH) \frac{1}{\partial_\tau^2} \dotbarxH^{\beta} \delta G_{\beta}^{\;\;\mu}(\barxH) \, .
}{recoil_op_dyon}
Here $\bar{J}^{\mu}_{L}$ is the point-particle vector current given in \Eq{JL_EM} with the particle following a solution to \Eq{dyon_EoM}, and $\delta G_{\mu\nu}=(\cos\theta_H+\sin\theta_H \ast)\delta F_{\mu\nu}$ is the appropriately EM duality rotated field strength.

We have not yet addressed how to compute in perturbation theory in a formalism containing both the photon and dual photon, $\delta A_\mu$ and $\delta B_\mu$. In general, this is not trivial because these are not independent degrees of freedom.  However, our formalism evades this complication rather nicely.
In particular, we can ignore Wick contractions between $\delta A_\mu$ and $\delta B_\mu$ because they correspond to quantum self-energy contributions that are dropped in the classical limit.   Instead, $\delta A_\mu$ and $\delta B_\mu$ only contract with $\delta F_{\mu\nu}$ in the recoil operator.  These Wick contractions can be performed using the definition in \Eq{AB_field_strength}.   Said another way, when computing Feynman diagrams we only ever encounter contractions between gauge potentials and field strengths and never between two gauge potentials. Consequently, the manifest gauge invariance of the recoil operator allows us to straightforwardly compute the radial action for the scattering of dyonic charged particles.

We find that the 1SF radial action for the scattering of a pair of dyons is
\eq{
I_{\textrm{EM}}^{(1,2)} \quad \rightarrow  \quad I_{\textrm{EM}, \Delta\theta}^{(1,2)}&=  \lambda\mL\Rc\frac{\Rc}{b}\frac{\pi  \left(\cos^2\theta_H+\sin^2\theta_H\right)
   \left(\cos^2\theta_L+\sin^2\theta_L\right)}{2 \sqrt{\sigma ^2-1}}\,, \\ 
I_{\textrm{EM}}^{(1,3)} \quad \rightarrow \quad I_{\textrm{EM}, \Delta\theta}^{(1,3)}&=  -\lambda\mL\Rc \left(\frac{\Rc}{b}\right)^2 \frac{ \cos\Delta\theta \left(\sigma ^2 \cos2\Delta\theta +2 \sigma ^4-7 \sigma ^2+6\right)}{3  \left(\sigma ^2-1\right)^{5/2}} \, ,
}{}
which is a new 3PL result. The above expression is consistent with known results, where they overlap.
For example, $\theta_L$ and $\theta_H$ drop out of the 2PL radial action, so this expression for dyonic scattering is exactly the same as for pure electric scattering. This agrees with the probe limits computed in \cite{Kol:2021jjc}, wherein pure electric and dyonic scattering are identical at 2PL.  Furthermore,  the 3PL radial action vanishes for a relative angle $\Delta\theta = \pi/2$, corresponding to a pure electric charge scattering against a pure magnetic charge. This is consistent with general expectations from computations of probe motion \cite{Kol:2021jjc}. 




\medskip

\section{General Relativity} 
\label{sec:GR}

In this section we derive an effective field theory for GR in the extreme mass ratio expansion.  Our manipulations will  parallel all of the steps taken in our analysis of EM.  
In particular, we begin with the worldline action for a pair of massive scalar particles interacting gravitationally.   At 0SF\footnote{For the purposes of this paper, 0SF refers to a test-particle traversing through a background Schwarzschild metric while higher SF orders simply count contributions at increasing orders in the mass ratio, $\lambda = m_L/m_H$, with 1SF being of the order, $\lambda^2$. This is distinct from counting using the symmetric mass ratio, $\nu=m_1 m_2/(m_1 + m_2)^2$. See \Sec{greft} for details.}, one particle produces a background gravitational field described by the Schwarzschild metric while the other particle, whose mass is comparatively negligible, evolves as a test body moving in this background.

As before, at 1SF the heavy particle fluctuates dynamically while the light particle fluctuations can be conveniently ignored.  Integrating out the former, we derive the gravitational recoil operator for the effective field theory.  This operator encodes the recoil of the Schwarzschild background against the orbiting mass.  An important difference between GR and EM is that gravitons are self-interacting.  As we will see, these effects are encoded entirely in the usual framework describing gravitons propagating in a Schwarzschild background.  While an analytic expression for the propagator of such gravitons is not known, we can leverage the known background metric to straightforwardly compute this in the PM expansion. 

As a crucial test of our framework, we compute the radial action for massive particle scattering at 0SF, 1SF, and 2SF, through 3PM order in general relativity. The 0SF radial action is already known to all PM orders, as recapitulated in \App{app:probe_action}, and our results at 1SF and 2SF find exact agreement with known results \cite{Cachazo:2017jef,BBDamFestPlaVan,PM1,PM2}.

To demonstrate the versatility of our framework, we also compute the 2PM and 3PM radial actions for gravitational theories in which the light particle sources a scalar or vector field which itself interacts gravitationally.  The former agrees with existing calculations, while the latter is a new result.

\subsection{Effective Theory}\label{greft}

As our starting point, we consider the action for a pair of massive, gravitationally interacting scalar particles in GR. Once again, as shown in \App{app:probe_traj}, we can fix the worldline einbein so that the action takes the simple form,
\eq{
S &= - \sum_{i=H,L} m_i  \int d\tau  \left[\tfrac{1}{2}+\tfrac{1}{2}  \dot x_i^\mu \dot x_i^\nu g_{\mu\nu}(x_i)\right] - \int d^4x  \sqrt{-g}\left[\tfrac{1}{16\pi G}  R \right] \, ,
}{S_GR}
where $x_i^\mu(\tau)$ are the particle trajectories and $g_{\mu\nu}(x)$ is the metric field.  Our gauge fixing implies the on-shell condition, $x_i^2(\tau)=1$.

The geodesic equation and Einstein field equations are
\eq{
&\ddot x^\mu_i  +  \Gamma^{\mu}_{\;\; \rho\sigma}(x_i) \dot x_{i}^\rho \dot x_{i}^\sigma =0 \qquad \textrm{and} \qquad R_{\mu\nu} -\tfrac12 g_{\mu\nu} R=8\pi G T_{\mu\nu}  \, ,
}{EOM_GR}
where the energy-momentum tensor is
\eq{
T^{\mu\nu}(x) &= \sum_{i=H,L} T_i^{\mu\nu}(x) = \sum_{i=H,L}   m_i \int d\tau \frac{1}{\sqrt{-g(x_i)}}\delta^4(x-x_i)\dot x_i^\mu \dot x_i^\nu \, .
}{}
Our effective field theory will provide an efficient way to solve these equations of motion as an expansion in the mass ratio. As has been noted previously \cite{mininggeodesic}, the Schwarzschild metric itself encodes infinite PM orders, simply by virtue of the fact that it is a formula that holds at finite gravitational coupling.  We will describe how the Schwarzschild metric, together with the known geodesics of particles within it, can be used to incorporate all orders in PM information into a systematic calculational framework.


\subsubsection{Self-Force Expansion in Curved Space}

At 0SF, the light particle propagates as a nongravitating probe.  Consequently,  the heavy particle moves in a straight line, 
\eq{
\barxH^\mu(\tau) = \vH^\mu \tau \, ,
}{}
providing a stress-energy density,
\eq{
\sqrt{-g(\bar{x})}\bar{T}^{\mu\nu}_{H}(x)=\mH\int d\tau \delta^{4}(x-\vH\tau)\vH^{\mu}\vH^{\nu}\,,
}{}
which sources the 0SF accurate geodesic equation and Einstein field equations,
\eq{
&\ddotbarxL^\mu  +  \bar{\Gamma}^{\mu}_{\;\; \rho\sigma}(\xL) \dotbarxL^\rho \dotbarxL^\sigma =0 \qquad \textrm{and} \qquad \bar{R}_{\mu\nu} -\tfrac12 \bar{g}_{\mu\nu} \bar{R}=8\pi G \bar{T}_{H\,\mu\nu}  \, .
}{eq:1SFGREOM}
Technically, the heavy particle also follows a geodesic according to the equation,
\eq{
\ddotbarxH^\mu  +  \bar\Gamma^{\mu}_{\;\; \rho\sigma}(\barxH) \dotbarxH^\rho \dotbarxH^\sigma =0 \, .
}{}
Since $\bar\Gamma^{\mu}_{\;\; \rho\sigma}(\barxH)$ is the Christoffel symbol for the Schwarzschild metric evaluated at the point, $r=0$, it is formally divergent. However, just as in the case of EM, we interpret this as a potential graviton mode that has been emitted and then reabsorbed by the heavy particle.  This divergent self-energy contribution can be absorbed by a counterterm, and can effectively be dropped. Consequently, the effective equation of motion for the heavy particle is $\ddotbarxH^\mu=0$, which defines a straight line trajectory.

As before, one could try to solve the equations of motion perturbatively by PM expanding the trajectory and metric and solving the equations iteratively at each order. Alternatively, thanks to the symmetries of the problem, we can solve these equations exactly---the gravitational background is just the boosted Schwarzschild metric and $\barxL$ is just a geodesic within it. The solution can be written in isotropic coordinates,
\eq{
\bar g_{\mu\nu}(x) =  \left(1+\tfrac{r_S}{4r}\right)^4 \eta_{\mu\nu} + \left[ \left(\frac{1-\frac{r_S}{4r}}{1+\frac{r_S}{4r}}\right)^2 - \left(1+\tfrac{r_S}{4r}\right)^4 \right]\vH{}_\mu \vH{}_\nu \, ,
}{}
where $r=\sqrt{(\vH x)^2 - x^2}$ is the boosted radius and $r_S = 2G\mH$ is the Schwarzschild radius. Since we will recast the Schwarzschild background as a resummation of an infinite class of flat space diagrams, it will be useful to define the deviation from flat space,
\eq{
 \bar \gamma_{\mu\nu}(x) &= \bar g_{\mu\nu}(x)-\eta_{\mu\nu}  = \frac{r_S}{r}  (\eta_{\mu\nu} -2 \vH{}_\mu  \vH{}_\nu) + \frac{1}{8} \left(\frac{r_S}{r}\right)^2  (3 \eta_{\mu\nu} + \vH{}_\mu  \vH{}_\nu) +\mathcal{O}(\RS^{3}) \, ,
}{bkgd_GR}
allowing us to trivially read off PM data order by order. Details on the light particle geodesic will be provided in upcoming sections, where we will again see that there is no need to explicitly study the graviton-graviton field interactions and we can instead read off PM data from simple position space expressions.

Beyond their dynamics at 0SF, the Schwarzschild metric and particle trajectories are perturbed by corrections,
\eq{
x^\mu_i  = \bar x^\mu_i + \dx_i^\mu \qquad \textrm{and} \qquad
g_{\mu\nu} = \bar g_{\mu\nu} + \dg_{\mu\nu} \, ,
}{pert_GR}
where $\dg_{\mu\nu}$ is the graviton propagating on a Schwarzschild background.  Since the solutions, $\bar x_i^\mu$ and $\bar g_{\mu\nu}$, are valid at 0SF, we know that
\eq{
\delta x_i^\mu \sim \dg_{\mu\nu} \sim {\cal O}(\lambda^1) \, ,
}{}
for on-shell configurations of the fluctuation degrees of freedom.

\subsubsection{0SF Dynamics}

As before, we compute the on-shell action by inserting SF expanded trajectories and fields in \Eq{pert_GR} into the action in \Eq{S_GR}, yielding
\eq{
S = \bar S +\sum_{n=1}^{\infty} S^{(n)} \, ,
}{}
where $n$ denotes the SF order.  Here $\bar S$ is the 0SF action evaluated on solutions in the probe limit,
\eq{
\bar S &= - \mH \int d\tau \left[\tfrac{1}{2}+\tfrac{1}{2} \eta_{\mu\nu}\dotbarxH^{\mu}\dotbarxH^{\nu} + \lambda\left(\tfrac{1}{2}+ \tfrac{1}{2}\bar{g}_{\mu\nu}(\barxL)\dotbarxL^{\mu}\dotbarxL^{\nu}\right) \right]\, ,
}{}
after renormalizing away self-energy divergences in dimensional regularization.  

Again focusing on the 0SF dynamics in $D=4$, we can ignore the heavy particle while the light particle trajectory is governed by the action,
\eq{
S=-\mL\int d\tau\left[\frac{1}{2}+\frac{1}{2}\bar{g}_{\mu\nu}(\xL)\dotxL^{\mu}\dotxL^{\nu}\right]\,.
}{}
The background metric sourced by the heavy  particle is the Schwarzschild metric, expressed here in isotropic coordinates,
\eq{
\bar{g}_{00}=\left(\frac{f_{-}(r)}{f_{+}(r)}\right)^{2}\quad \textrm{and} \quad \bar{g}_{ij} = -\delta_{ij}f_{+}(r)^{4}\,,
}{}
where we have defined
\eq{
f_{\pm}(r)=1\pm\frac{r_S}{4r}\,.
}{}
Solutions to the equations of motion satisfy the curved space on-shell condition, $g_{\mu\nu}\dot{x}^{\mu}\dot{x}^{\nu}=1$. 

Since the metric is static and isotropic, we can restrict to motion in the equatorial plane, subject to conserved energy and angular momentum. We will again label the trajectories with ($\sigma,b$), in terms of which the equations of motion are
\eq{
    \dot{t}=\sigma \frac{f_{+}(r)^{2}}{f_{-}(r)^{2}}\qquad\textrm{and}\qquad \dot{\phi}=\frac{b(\sigma^{2}-1)^{1/2}}{r^{2}}f_{+}(r)^{-4}\,.
}{}
The radial equation of motion arises from the on-shell condition,
\eq{
\dot{r}=\left[f_{+}(r)^{-4}\left(\frac{f_{+}(r)^{2}}{f_{-}(r)^{2}}\sigma^{2}-1\right)-\frac{b^{2}(\sigma^{2}-1)}{r^{2}}f_{+}(r)^{-8}\right]^{1/2}\,.
}{}
As discussed in \App{app:probe_action}, the 0SF radial action is 
\eq{
\bar{S}=2\int_{r_{\textrm{min}}}^{\infty}\,dr\,|p_{r}(r,E,J)|\,.
}{}
Using the equations of motion and conservation laws, we obtain the probe radial momentum,
\eq{
|p_{r}|=m\left[f_{+}(r)^{4}\left(\frac{f_{+}^{2}(r)}{f_{-}(r)^{2}}\sigma^{2}-1\right)-\frac{b^{2}(\sigma^{2}-1)}{r^{2}}\right]^{1/2}
\,.
}{}
As shown in \App{app:probe_action}, we then integrate the radial action to obtain
\eq{
\bar{S}=&-\pi  b \mL \sqrt{\sigma ^2-1}+G\mL \mH \frac{\left(\left(2-4 \sigma ^2\right) \log \left(\tilde{\mu}be^{\frac{-1}{D-4}}\right)+1\right)}{\sqrt{\sigma ^2-1}}+  G^2 \mL \mH^2 \frac{3 \pi\left(5 \sigma ^2-1\right)}{4 b \sqrt{\sigma ^2-1}}\\
&+G^3 \mL \mH^3\frac{ \left(64 \sigma ^6-120 \sigma ^4+60 \sigma ^2-5\right)}{3 b^2 \left(\sigma ^2-1\right)^{5/2}}\, + \mathcal{O}(G^4),
}{}
expanded here up to 3PM order.

\subsubsection{1SF Dynamics}

Now, let us move on to the 1SF gravitational dynamics.  As before, we will derive the 1SF effective action by integrating out the deviations of the heavy particle about its probe motion. Again, the 0SF action receives contributions at $\mathcal{O}(\lambda^0)$ and $\mathcal{O}(\lambda^1)$, corresponding to factors of the heavy and light particle masses in the action.  We then expand up to ${\cal O}(\lambda^2)$ to extract the 1SF contributions.

Expanding the Einstein-Hilbert action to 1SF order about the Schwarzschild background, we obtain the usual action for a graviton perturbation in curved spacetime,
\eq{
- \tfrac{1}{16\pi G}\int d^4 x \sqrt{-g}& \left[R - \tfrac12 F_\mu F^\mu \right] = \\
- \tfrac{1}{16\pi G}\int d^4 x \sqrt{-\bar g} &\left[\bar R  {\color{blue} - (\bar R_{\mu\nu} -\tfrac12 \bar g_{\mu\nu} \bar R) \dg^{\mu\nu} } \right. \\
&\; - \left.\tfrac14 \bar\nabla_\rho \dg_{\mu\nu}\bar\nabla^\rho \dg^{\mu\nu} + \tfrac18 \bar\nabla_\rho  \dg \bar\nabla^\rho \dg  +\tfrac12 \dg_{\mu\nu} \dg_{\rho\sigma} \bar R^{\mu\rho\nu\sigma} \right. \\
&  \;+
\left.\tfrac12 (\dg_{\mu\rho} \dg_\nu^{\;\;\rho}  - \dg_{\mu\nu} \dg )\bar R^{\mu\nu} - \tfrac14 (\dg_{\mu\nu} \dg^{\mu\nu} -\tfrac12 \dg^2)\bar R \right]
\, ,
}{GRaction_R}
where $\delta g = \delta g^{\mu}_{\ \mu}$ and we have include a harmonic gauge fixing term for the graviton perturbation defined by $F_\mu = \bar\nabla^\nu \delta g_{\mu\nu} -\tfrac12 \bar\nabla_\mu \delta g$. All raising, lowering, and contractions of indices are performed with the background metric, $\bar g_{\mu \nu}$.  

At this point let us comment on an important subtlety in the above action.  Typically, on a Schwarzschild background we can set $\bar R_{\mu\nu} = \bar R=0$ because the metric satisfies the vacuum Einstein field equations.  This would naively imply that we should drop the terms in blue and in the final line of \Eq{GRaction_R}.  However, this is actually not correct in a point-particle effective field theory.  Specifically, our equations of motion in \Eq{EOM_GR} involve worldline sources rather than a vacuum, so in perturbation theory, $\bar R_{\mu\nu}$ and $\bar R$ are zero except on the support of the heavy worldline.  This crucial distinction implies that we must actually retain all such terms for any perturbative calculation.
For the same reason, the terms highlighted in blue in \Eq{GRaction_R} are not identically zero in perturbation theory, but rather cancel exactly with other terms in the worldline action on the support of the Einstein field equations since, together, they form the complete Einstein's equations with a source.

Next, let us consider the worldline actions, 
\eq{
 - m_i \int d\tau \left[\tfrac12 + \tfrac12 \dot x_i^\mu \dot x_i^\nu g_{\mu\nu}(x_i) \right] = - m_i & \int d\tau \left[\tfrac12+ \tfrac12 \dot \barx_i^2 
  - \delta x_i \ddot\barx_i  \right. \\
& \;   \left. - \delta x_i^\rho  \dot \barx_i^\mu \dot \barx_i^\nu \bar\Gamma_{\rho\mu\nu} (\barx_i )+\tfrac12 \dot\barx_i^\mu \dot\barx_i^\nu \dg_{\mu\nu}(\barx_i)  \right. \\
& \;   \left. + \tfrac12 \dot\barx_i^\mu \dot\barx_i^\nu \bar\nabla_\mu \delta x_i^\rho \bar\nabla_\nu \delta x_{i\rho} + \tfrac12 \delta x_i^\rho \delta x_i^\sigma \dot\barx_i^\mu \dot\barx_i^\nu \bar R_{\nu\rho\sigma\mu} (\barx_i)  \right. \\
& \;   \left. -    \delta x_{i\rho}  \dot \barx_i^\mu \dot \barx_i^\nu \delta \Gamma^{\rho}_{\mu\nu} (\barx_i ) +  \cdots \right]  \, ,
}{}
where the ellipses denote contributions that are higher than quadratic order in the fluctuations and we have defined the difference of connections, $\delta \Gamma^\rho_{\mu\nu}  = \Gamma^\rho_{\mu\nu}-\bar\Gamma^\rho_{\mu\nu} = \tfrac{1}{2} \bar g ^{\rho \sigma}(\bar\nabla_\mu \delta g_{\sigma \nu}+\bar\nabla_\nu \delta g_{\sigma \mu}-\bar\nabla_\sigma \delta g_{\mu \nu} )$, which is a tensor with respect to the background metric. As before, the worldline actions for the heavy and light particles are treated differently, so
\eq{
 - m_H \int d\tau \left[\tfrac12+\tfrac12 \dotxH^\mu \dotxH^\nu g_{\mu\nu}(\xH) \right] = - m_H  &\int d\tau \left[\tfrac12+\tfrac12 \dotbarxH^2 
{\color{green} - \delta \xH \ddotbarxH }   \right. \\
& \;   \left.{\color{red} -    \delta \xH^\rho  \dotbarxH^\mu \dotbarxH^\nu \bar\Gamma_{\rho\mu\nu} (\barxH ) } {\color{blue} +\tfrac12 \dotbarxH^\mu \dotbarxH^\nu \dg_{\mu\nu}(\barxH) }\right. \\
& \;   \left. + \tfrac12 \dot\barx_H^\mu \dot\barx_H^\nu \bar\nabla_\mu \delta x_H^\rho \bar\nabla_\nu \delta x_{H\rho} {\color{red}  + \tfrac12 \delta x_H^\rho \delta x_H^\sigma \dot\barx_H^\mu \dot\barx_H^\nu \bar R_{\nu\rho\sigma\mu} (\barx_H)}  \right. \\
& \;   \left. -    \delta x_{H\rho}  \dotbarxH^\mu \dotbarxH^\nu \delta \Gamma^\rho_{\mu\nu} (\barxH) +  \cdots \right]  \, ,\\
- m_L \int d\tau \left[\tfrac12+\tfrac12 \dotxL^\mu \dotxL^\nu g_{\mu\nu}(\xL) \right] =   - m_L & \int d\tau \left[\tfrac12+\tfrac12 \dotbarxL^2 
{\color{purple}   -  \delta \xL \ddotbarxL} \right. \\
& \;   \left.{\color{purple} -   \delta \xL^\rho  \dotbarxL^\mu \dotbarxL^\nu \bar\Gamma_{\rho\mu\nu} (\barxL ) } + \tfrac12 \dotbarxL^\mu \dotbarxL^\nu \dg_{\mu\nu}(\barxL) \right. \\
& \;   \left. {\color{brown}+ \tfrac12 \dot\barx_L^\mu \dot\barx_L^\nu \bar\nabla_\mu \delta x_L^\rho \bar\nabla_\nu \delta x_{L\rho}  + \tfrac12 \delta x_L^\rho \delta x_L^\sigma \dot\barx_L^\mu \dot\barx_L^\nu \bar R_{\nu\rho\sigma\mu} (\barx_L)}  \right. \\
& \;   \left. {\color{brown}-    \delta x_{L\rho}  \dotbarxL^\mu \dotbarxH^\nu \delta \Gamma^\rho_{\mu\nu} (\barxL)} +  \cdots \right]
\, ,
}{GRaction_WL}
where all colored terms can be dropped.   The terms in red contain the background metric evaluated at the position of the heavy particle, and they include divergent self-energy contributions from the emission and reabsorption of potential gravitons from the heavy particle.  
More generally, all terms involving $\bar \Gamma_{\rho\mu\nu}(\barxH)$ or $\bar R_{\rho \sigma\mu\nu}(\barxH)$ are self-energy contributions. Furthermore, whenever there is an implicit insertion of $\bar g_{\mu\nu}(\barxH)$, the components which deviate from flat space again correspond to self-energy contributions. 
As described in detail in \Sec{reg}, all of these self-energy contributions can be renormalized away, and in fact, in dimensional regularization they are identically zero. Hence, in perturbation theory, the metric evaluated on the heavy trajectory, $\bar g_{\mu\nu}(\barxH) $, can be effectively set to the flat metric, $\eta_{\mu\nu}$.

Furthermore, the blue, green, and purple terms all vanish on the support of the equations of motion for the background metric, heavy trajectory, and light trajectory.  Finally, the terms in brown can be dropped because the light particle action is suppressed by $\mL$, so these contributions enter as ${\cal O}(\lambda^3)$ and are subleading at 1SF.

Combining the terms relevant terms from \Eq{GRaction_R} and \Eq{GRaction_WL}, we obtain the 1SF action for fluctuations,
\eq{
S^{(1)} =& \; - \mH \int d\tau \left[ \tfrac12 \delta \dotxH^2  -    \delta \xH^\rho  \dotbarxH^\mu \dotbarxH^\nu \delta \Gamma_{\rho\mu\nu} (\barxH)  +\tfrac12 \lambda \dotbarxL^\mu \dotbarxL^\nu \dg_{\mu\nu}(\barxL)\right] \\
&+ \int d^4x \sqrt{-\bar g}\left[  \tfrac{1}{32\pi G} (\tfrac12 \bar\nabla_\rho \dg_{\mu\nu}\bar\nabla^\rho \dg^{\mu\nu} - \tfrac14 \bar\nabla_\rho  \dg \bar\nabla^\rho \dg +\cdots )   \right]\,.
}{}
The 1SF action describes a dynamical graviton which is sourced by the light particle geodesic motion, and whose propagator is corrected by the motion of the heavy particle. We can rewrite the action in the form
\eq{
S^{(1)} =& \; - \mH \int d\tau \left[\tfrac12 \delta \dotxH^2  -    \delta \xH^\rho  \dotbarxH^\mu \dotbarxH^\nu \delta \Gamma_{\rho\mu\nu} (\barxH) \right] \\
&+ \int d^4x \sqrt{-\bar g}\left[  \tfrac{1}{32\pi G} (\tfrac12 \bar\nabla_\rho \dg_{\mu\nu}\bar\nabla^\rho \dg^{\mu\nu} - \tfrac14 \bar\nabla_\rho  \dg \bar\nabla^\rho \dg +\cdots )   - \tfrac12 \dg_{\mu\nu}\bar T^{\mu\nu} \right] \,,
}{deltaS_GR}
where the light particle 0SF energy-momentum tensor is\eq{
\bar T^{\mu\nu} (x)&= \lambda  \mH \int d\tau \, \frac{\delta^4(x-\barxL)}{\sqrt{-\bar{g}}} \dotbarxL^\mu  \dotbarxL^\nu  \, .
}{TL_GR}
\Eq{deltaS_GR} describes a graviton propagating in a Schwarzschild background, sourced by a light particle geodesic, together with the fluctuations of the heavy particle.
Just like in the case of electromagnetism, we see that for general relativity, we only need the geodesic for the light particle in order to compute at 1SF order.

Next, let us integrate out the heavy particle fluctuation, $\delta \xH^\mu$, to obtain a 1SF effective action that depends solely on the graviton.  The path integral over $\delta \xH^\mu$ yields
\eq{
\int [d \delta \xH] \exp\left(- i  \mH \int d\tau \left[\tfrac12 \delta \dotxH^2  -   \delta \xH^\rho  \dotbarxH^\mu \dotbarxH^\nu \delta \Gamma_{\rho\mu\nu} (\barxH) \right]\right) = \exp\left(i  S_{\rm recoil}^{(1)} \right) \, ,
}{} 
where  $S_{\rm recoil}^{(1)}$ is the {\it gravitational recoil operator}.  To do the path integral, we plug the solution for the heavy particle back into the action.   As a consistency check we see that variation of the action with respect to $\delta \xH^\mu$ gives
\eq{
\delta \ddotxH^\mu  +  \delta\Gamma^{\mu}_{\;\; \alpha\beta}(\barxH) \dotbarxH^\alpha \dotbarxH^\beta =0 \, ,
}{}
which is the geodesic equation expanded to 1SF.  Inserting this solution back into \Eq{deltaS_GR} yields the recoil operator,
\eq{
S_{\rm recoil}^{(1)} &=   -\frac12  \mH \int d\tau \,   \dotbarxH^{\alpha} \dotbarxH^{\beta} \dG^\mu_{\;\;\alpha\beta}(\barxH) \frac{1}{\partial_\tau^2} \dotbarxH^{\gamma} \dotbarxH^{\delta} \dG_{\mu\gamma\delta}(\barxH) \, .
}{recoil_op_GR}
We emphasize again that this operator is gauge invariant because $\delta \Gamma^\rho_{\mu\nu}$ is a difference of connections, and thus a tensor with respect to the background metric at each of the two spacetime points where it is evaluated in \Eq{recoil_op_GR}.  Furthermore, in our renormalization scheme the background metric evaluated at the location of the heavy body is effectively flat and the tensor indices are thus unambiguously parallel transported between the two spacetime points without the need for a gravitational Wilson line.  

Here we emphasize the subtle difference between the Schwarzschild metric as the field generated by a massive point particle source in a perturbative description of GR, as opposed to a vacuum solution in full nonperturbative GR.  This caveat
was also emphasized by Pfenning and Poisson in the context of PN dynamics ~\cite{Pfenning:2000zf}, and later by Gralla and Lobo for 2PM scattering~\cite{Gralla:2021qaf}. Those authors referred to the contribution from the stress-energy of the heavy body at 1SF order as a ``matter-mediated-force''. The present article can be interpreted as an alternative approach to systematically deriving these effects within the framework of effective field theory. Crucially, our work prescribes precisely which ``matter'' operators must be appended to the background field action at a given SF order to allow for consistent computations at arbitrary PM order.

To summarize, we have defined a 1SF effective action for general relativity,
\eq{
S^{(1)}&=  S_{\rm recoil}^{(1)}+  \int d^4x \sqrt{-\bar g}\left[  \tfrac{1}{32\pi G} (\tfrac12 \bar\nabla_\rho \dg_{\mu\nu}\bar\nabla^\rho \dg^{\mu\nu} - \tfrac14 \bar\nabla_\rho  \dg \bar\nabla^\rho \dg +\cdots )   -\tfrac12 \dg_{\mu\nu}\bar T^{\mu\nu}_L \right] \, ,
}{1SF_action_GR}
which describes a graviton in a Schwarzschild background which is sourced by a light particle geodesic through $\bar T^{\mu\nu}_L$ in \Eq{TL_GR} and perturbed by the recoil operator, $S_{\rm recoil}^{(1)}$, defined in \Eq{recoil_op_GR}.  This operator  encodes the fluctuations of the Schwarzschild metric entering at 1SF due to the motion of the orbiting light particle.

\subsubsection{2SF Dynamics}

\label{sec:2SF_dyn}

Just like in EM, the 2SF dynamics of GR are determined by expanding the action to $\OO(\lambda^{3})$. Contributions to the light particle worldline were shown in brown in \Eq{GRaction_WL},
\eq{
S^{(2)}_{L} =
- \int d\tau &\bigg[\tfrac12 \dot\barx_L^\mu \dot\barx_L^\nu \bar\nabla_\mu \delta x_L^\rho \bar\nabla_\nu \delta x_{L\rho}  + \tfrac12 \delta x_L^\rho \delta x_L^\sigma \dot\barx_L^\mu \dot\barx_L^\nu \bar R_{\nu\rho\sigma\mu} (\barx_L) \\
& \;  -    \delta x_{L\rho}  \dotbarxL^\mu \dotbarxL^\nu \delta \Gamma^\rho_{\mu\nu} (\barxL) \bigg]
\, .
}{}
The first two terms describe \textit{geodesic deviation} of the light particle trajectory caused by fluctuation gravitons which kick the body off its 0SF motion due to the interaction shown in the last term. This interaction of fluctuation gravitons with the light body is analogous to the recoil experienced by the heavy body at 1SF.

Contributions to the heavy particle worldline at 2SF consist of terms of the form, $\delta \xH \delta \xH \delta g$, which are cubic in perturbations,
\eq{
S^{(2)}_{H} =  - \int d\tau &\bigg[ \tfrac{1}{2} \delta g_{\mu \nu} \delta \dotxH^{\mu} \delta \dotxH^{\nu} + \delta \bar \Gamma_{\mu \rho \nu} \delta x_H^\rho \delta \dotxH^\mu \dot \barx_H^\nu +\tfrac{1}{2} \delta \bar R_{\mu \rho \sigma \nu} \delta x_H^{\rho} \delta x_H^{\sigma} \dot \barx_H^\mu \dot \barx_H^\nu \bigg] \, ,
}{}
such that the total heavy particle effective action to 2SF order is given by
\eq{
S^{(1)}_{H} + S^{(2)}_{H} = - \mH\int d\tau \big[ &\tfrac12 \delta \dotxH^2  -    \delta \xH^\rho  \dotbarxH^\mu \dotbarxH^\nu \delta \Gamma_{\rho\mu\nu} (\barxH) +\tfrac{1}{2} \delta g_{\mu \nu} \delta \dotxH^{\mu} \delta \dotxH^{\nu} + \delta \bar \Gamma_{\mu \rho \nu} \delta x_H^\rho \delta \dotxH^\mu \dot \barx_H^\nu \\
&+\tfrac{1}{2} \delta \bar R_{\mu \rho \sigma \nu} \delta x_H^{\rho} \delta x_H^{\sigma} \dot \barx_H^\mu \dot \barx_H^\nu \bigg] \, .
 }{}
The additional terms give rise to a new recoil operator at 2SF which is determined, as before, by integrating the heavy particle perturbations,
\eq{
\int [d \delta\xH] \exp\left(iS^{(1)}_{H}+iS^{(2)}_{H}\right) = \exp\left(i  S^{(1)}_{\rm recoil}+iS^{(2)}_{\rm recoil} + \cdots \right) \, ,
}{}
where the 2SF recoil operator is
\eq{
S^{(2)}_{\rm recoil}= - \mH \int d\tau \Big[& \frac{1}{2}\dot \barx_H^{\alpha} \dot \barx_H^{\beta} \delta \Gamma^{\mu}_{\alpha \beta}\frac{1}{\overleftarrow{\partial_\tau}}\delta g_{\mu \nu}\frac{1}{\overrightarrow{\partial_{\tau}}}\dot \barx_H^{\gamma} \dot \barx_H^{\delta} \delta \Gamma^{\mu}_{\gamma \delta}\\
& + \dot \barx_H^{\alpha} \dot \barx_H^{\beta} \delta \Gamma^{\mu}_{\alpha \beta}\frac{1}{\overleftarrow{\partial_\tau}}\dot \barx_H^{\rho} \delta \Gamma_{\mu \nu \rho}\frac{1}{\overrightarrow{\partial_{\tau}^{2}}}\dot \barx_H^{\gamma} \dot \barx_H^{\delta} \delta \Gamma^{\mu}_{\gamma \delta}\\
&+\frac{1}{2} \dot \barx_H^{\alpha} \dot \barx_H^{\beta} \delta \Gamma^{\mu}_{\alpha \beta}\frac{1}{\overleftarrow{\partial_{\tau}^{2}}}\dot \barx_H^{\rho} \dot \barx_H^{\sigma} \delta R_{\rho \mu \nu \sigma}\frac{1}{\overrightarrow{\partial_{\tau}^{2}}}\dot \barx_H^{\gamma} \dot \barx_H^{\delta} \delta \Gamma^{\mu}_{\gamma \delta}\Big]
\, .
}{}
With this new recoil operator, the full effective action at 2SF is
\eq{
S^{(1)} + S^{(2)} = & S^{(1)}_{\rm recoil}+S^{(2)}_{\rm recoil} + \int d^4 x \left.\frac{1}{3!} \frac{\delta^3 S}{\delta g_{\mu \nu} \delta g_{\rho \sigma} \delta g_{\alpha \beta}} \right\rvert_{\bar g} \delta g_{\mu \nu} \delta g_{\rho \sigma} \delta g_{\alpha \beta}\\&+\mL\int d\tau\Big[\tfrac12 \dot\barx_L^\mu \dot\barx_L^\nu \bar\nabla_\mu \delta x_L^\rho \bar\nabla_\nu \delta x_{L\rho}  + \tfrac12 \delta x_L^\rho \delta x_L^\sigma \dot\barx_L^\mu \dot\barx_L^\nu \bar R_{\nu\rho\sigma\mu} (\barx_L) \\
& \; \qquad  \qquad \quad -    \delta x_{L\rho}  \dotbarxL^\mu \dotbarxL^\nu \delta \Gamma^\rho_{\mu\nu} (\barxL)\Big] \, .
}{2SF_action_GR}
where the curved space graviton cubic self-interaction terms are calculated in a PM expansion where necessary.
\subsection{Feynman Rules}

Armed with the 1SF effective action in \Eq{1SF_action_GR}, we can now calculate physical observables by computing the path integral over the graviton perturbations about the Schwarzschild background. Ideally, we would choose to work directly with the Schwarzschild graviton propagator, however there is unfortunately no {\it simple} closed form expression for this quantity.  The absence of any notion of conserved momenta for particles also makes technical calculations in curved spacetime quite difficult.  

Instead, we opt to further decompose the dynamics in the PM expansion, which describes perturbative corrections away from flat space.  In this approach, the Schwarzschild graviton propagator is equal to the flat space propagator corrected by terms involving interactions with the background, as shown in \Fig{fig:gravprop}.  We also PM expand the background fields, treating them as corrections to flat space. The advantage of extracting PM corrections from the known Schwarzschild background will be discussed in \Sec{GR_class_resum}.

\begin{figure*}
    \centering
\begin{align*}
    &\vcenter{\hbox{\bgprop}}
    = \vcenter{\hbox{\flatprop}} +
    \vcenter{\hbox{\propone}} \nonumber + \vcenter{\hbox{\proptwo}} + \ \ \cdots\\[10pt]
    &\raisebox{0.3pt}{\hbox{\propone}}
    \raisebox{2.8pt}{$=$} \vcenter{\hbox{\gbackgroundoneone}} \raisebox{2.8pt}{$+$}\vcenter{\hbox{\gbackgroundonetwo}} \raisebox{2.8pt}{$+$} \vcenter{\hbox{\gbackgroundonethree}}\raisebox{2.8pt}{$+ \ \ \cdots$}
\end{align*}
    \caption{The curved space graviton propagator can be thought of as a sum of the flat space propagator and corrections involving interactions with the background. These interactions, depicted as insertions on the flat space propagator, organize into a PM expansion of the background gravitational field.}
    \label{fig:gravprop}
\end{figure*}

\subsubsection{Graviton Propagator and Vertices}

Given our choice of harmonic gauge for the background field action for the graviton in \Eq{GRaction_R}, the 0PM limit yields a flat space graviton propagator in deDonder gauge, so
\eq{
\flatproparrow &= \frac{32 \pi i G }{p^2}  \left( \frac{\eta_{\mu\rho}\eta_{\nu\sigma}+\eta_{\mu\sigma}\eta_{\nu\rho}}{2} - \frac{\eta_{\mu\nu} \eta_{\rho\sigma}}{2}\right) \, .
}{deDonder}
Meanwhile, it is trivial to compute the two-point vertex for the graviton from the recoil operator $S^{(1)}_{\rm recoil}$  in \Eq{recoil_op_GR}, giving
\eq{
\vcenter{\hbox{\flatrecoil}} &= {  \frac{i\mH}{2}   \frac{\hat{\delta}(\vH p_1+ \vH p_2)}{(\vH p_1)(\vH p_2)} {\cal O}^{\alpha \mu_1 \nu_1}(\vH, p_1)  {\cal O}_\alpha^{\;\; \mu_2 \nu_2}(\vH, p_2) } \, ,
}{recoil_vertex_GR}
where we have defined ${\cal O}^{\alpha \mu\nu}(\v ,p) =\frac12 ( (\v^\mu \eta^{\nu\alpha} + \v^\nu \eta^{\mu \alpha}) (\v p) -  \v^\mu \v^\nu p^\alpha )$.  Notice the similarity of the above equation to the corresponding Feynman vertex in EM in \Eq{recoil_op_GR}.

Since we will compute in a PM expansion about flat space, we must treat the background metric insertions as perturbations.  At leading nontrivial PM order, the background metric and Christoffel connection are
\eq{
\bar \gamma_{\mu\nu}(p) &= - \frac{8\pi G \mH ( \eta_{\mu\nu}-2\vH{}_\mu \vH{}_\nu)}{p^2} \hat{\delta}(\vH p)+\cdots  \, ,\\
\bar \Gamma_{\mu\alpha\beta}(p) &= - \frac{i}{2} \left( p_{\alpha}  \bar \gamma_{\beta\mu}(p)+p_{\beta}  \bar \gamma_{\alpha\mu}(p) -p_{\mu}  \bar \gamma_{\alpha\beta}(p)  \right) +\cdots\, .
}{bkgd_field_GR}
These should be inserted into flat space Feynman diagrams as sources.   To obtain the corresponding flat space interaction vertices, we insert \Eq{bkgd_field_GR} into the 1SF action in \Eq{1SF_action_GR} and then expand in PM.  For example, after PM expanding, we find a  three-point vertex linking two fluctuation gravitons and one linearized insertion of the background field, and so on and so forth.  

Notably, at 3PM and lower orders, the background metric insertions beyond 1PM order are not needed\footnote{This follows from the happy accident that the 2PM background metric, i.e.~the Einstein-Infeld-Hoffman correction to the Newtonian potential, enters at 3PM via a two-loop diagram that does not contribute classically.  In particular, it arises through diagram \#8 in Fig.~14 of \cite{PM3}.}.  However, starting at 4PM order, these contributions will become relevant.  The value added by
resumming perturbative data directly from the background metric will grow exponentially with PM order.

\subsubsection{Graviton Sources}

From the 1SF effective action, we see that gravitons are sourced solely by the light particle geodesic in \Eq{TL_GR}. Taking the Fourier transform of the stress-energy density, we define
\eq{
&\bar{T}_L^{\mu\nu}(p)  = \int d^4 x\, e^{ipx} \sqrt{- \bar g(x)}\barTL^{\mu\nu}(x) =\lambda  \mH  \int d\tau \, e^{ip \barxL}  \dotbarxL^\mu\dotbarxL^\nu \\
&=\lambda  \mH  \int d\tau \, e^{ip \barx_0 } \times e^{ip(\bar x_1 + \cdots)} (\dot {\bar x}_0^\mu +\dot {\bar x}_1^\mu + \cdots) (\dot {\bar x}_0^\nu +\dot {\bar x}_1^\nu +\cdots) \\
&=\lambda  \mH  \int d\tau \, e^{ip \barx_0 } \times  (\dot {\bar x}_0^\mu  \dot {\bar x}_0^\nu  - i(p \dot\barx_0 (\dot\barx_0^\mu \delta^{\nu}_{\rho}+\dot\barx_0^\nu \delta^{\mu}_{\rho})- p_\rho  \dot\barx_0^\mu  \dot\barx_0^\nu)\barx_1^\rho+\cdots)  \, ,
}{}
where we have expanded up to 1PM order via \Eq{xL_PM} and reorganized the expression using integration by parts.  In terms of the 1PM light particle trajectory, this expression is 
\eq{
\bar{T}_L^{\mu\nu}(p)  &=\lambda  \mH e^{ipb} \left( \vL^\mu  \vL^\nu \hat{\delta}(\vL p) - 2i {\cal O}_\alpha^{\;\;\, \mu \nu}(\vL, p) \bar x_{1}^\alpha(\vL p) +\cdots  \right)  \, ,
}{}
where we have again defined the frequency domain trajectory, $ \bar x^\mu_i(\omega)=\int d\tau \, e^{i\omega\tau} \barx_i^\mu(\tau) $.   The 1PM trajectory can be obtained perturbatively from \Sec{sec:traj_GR},
\eq{
\bar x_1^\mu(\omega) = \frac{1}{\omega^2} \int \frac{d^4 q}{(2 \pi)^4} \, e^{-iqb} \bar \Gamma^{\mu}_{\;\;\alpha\beta}(q) \vL^\alpha \vL^\beta \hat{\delta}(\omega- \vL q) \, ,
}{x1_omega_GR_A}
where the background metric and Christoffel symbol in momentum space are defined in \Eq{bkgd_field_GR}.  Putting everything together, the Feynman vertex for the graviton source is

\eq{
\vcenter{\hbox{\source}} \quad & = \quad \vcenter{\hbox{\sourcezero}} \qquad + \quad \vcenter{\hbox{\sourceone}} \qquad + \quad \cdots \\
&= { -i   \lambda  \mH e^{ipb} \left( \tfrac12 \vL^\mu \vL^\nu \hat{\delta}(\vL p) - i {\cal O}^{\alpha \mu\nu}(\vL, p) \barx_{1\alpha}(\vL p) +\cdots  \right)} \, .
}{}
See \Fig{fig:Feynman_rules_GR} for a convenient summary of the 1SF Feynman rules for the flat space graviton propagator, recoil vertex, and light particle source for the graviton.

\begin{figure*}
    \centering
    \begin{tabular}{|c|c|}
    \hline
        $\begin{array}{cc}
            \flatproparrow\\
             \textrm{Graviton propagator}
        \end{array}$
         &
           \(\displaystyle
           \frac{32 \pi i G }{p^2}  \left( \frac{\eta_{\mu\rho}\eta_{\nu\sigma}+\eta_{\mu\sigma}\eta_{\nu\rho}}{2} - \frac{\eta_{\mu\nu} \eta_{\rho\sigma}}{2}\right)
           \)\\
    \hline
        $\begin{array}{cc}
            \flatrecoil\\
             \textrm{Recoil vertex}
        \end{array}$
        & \(\displaystyle
            \frac{i\mH}{2}   \frac{\hat{\delta}(\vH p_1+ \vH p_2)}{(\vH p_1)(\vH p_2)} {\cal O}^{\alpha \mu_1 \nu_1}(\vH, p_1)  {\cal O}_\alpha^{\;\; \mu_2 \nu_2}(\vH, p_2)
        \) \\
    \hline
        $\begin{array}{cc}
            \\
            \source\\
             \textrm{Graviton source}
        \end{array}$
         & \(\displaystyle
            -i   \lambda  \mH e^{ipb} \left( \tfrac12 \vL^\mu \vL^\nu \hat{\delta}(\vL p) - i {\cal O}^{\alpha \mu\nu}(\vL, p) \barx_{1\alpha}(\vL p) +\cdots  \right)
        \) \\
        \hline
    \end{tabular}
    \caption{Feynman rules for the flat space graviton propagator, recoil vertex, and graviton vertex that enter into the radial action for GR at 1SF. }
    \label{fig:Feynman_rules_GR}
\end{figure*}

\medskip

\subsection{Classical Resummation}
\label{GR_class_resum}

Paralleling our analysis of EM, we now explore how the classical probe dynamics of GR also encodes perturbative data to all orders.  We first discuss resummation in the context of the Schwarzschild background metric, and then move on to the case of geodesic trajectories.

\subsubsection{Background Field Vertices}

Let us focus on the Feynman rules for a gravitationally coupled scalar propagating in a nontrivial background field,
\begin{equation}
  S =   \frac12 \int d^Dx \sqrt{-\bar g} \bar g^{\mu\nu}(x) \partial_\mu\varphi \partial_\nu\varphi \, .
\end{equation}
The flat space propagator receives corrections,
\begin{align}
    \vcenter{\hbox{\sbgprop}}
    & = \vcenter{\hbox{\sflatprop}} +
    \vcenter{\hbox{\spropone}} \nonumber + \vcenter{\hbox{\sproptwo}} + \ \ \cdots
\end{align}
and the Feynman rules are then very simple, 
\begin{align}
\scalarprop  & = \frac{i}{k^2}\,, \\[0.5cm]
 \vcenter{\hbox{\backgroundonearrow}} &=   -i (\sqrt{-\bar g} \bar g^{\mu\nu}(q) - \eta^{\mu\nu}) k_{1\mu} k_{2\nu} \, ,
\end{align}
with momentum conservation requiring that  $k_1+k_2 +q = 0$ on the vertex.  

The background-field Feynman rule described above simply encodes the following sum over flat-space Feynman diagrams, 
\begin{align}
\begin{split}
    \vcenter{\hbox{\backgroundone}}
    & \quad = 
    \vcenter{\hbox{\tree}}
    +
    \vcenter{\hbox{\contacttriangle}}
    +
    \vcenter{\hbox{\triangl}} \\[0.5cm]
    &+
    \vcenter{\hbox{\wdiagram}}
    +
    \vcenter{\hbox{\triY}}
    +
    \vcenter{\hbox{\othertriY}}
    + \ \ \cdots \, ,
\end{split}
\end{align}
which is the perturbative solution of the Einstein field equations with a point-like source,
\begin{equation}
    T^{\mu\nu}(x) = \mH \int d\tau \vH^\mu \vH^\nu \delta^{(D)}(x^\mu- \vH^\mu \tau)\,,
\end{equation}
or, equivalently, in momentum space,
\begin{equation}
    T^{\mu\nu}(q) = \mH\vH^\mu \vH^\nu \hat{\delta}(\vH q) \, .
\end{equation}
The crucial observation is that we do not need to compute the sum over diagrams, as we already know that they just compute the metric. One has to choose coordinates in which to write the background metric, but we need not restrict ourselves to traditional choices such as harmonic gauge.   Here a convenient choice of background field coordinates is isotropic coordinates in $D$ dimensions, where
\begin{align}
    \bar g_{\mu\nu}(x) &=    \left(1+\frac{\mu}{4|\bm r|^{D-3}}\right)^{4/(D-3)}  (\eta_{\mu\nu}  - \vH{}_{\mu}\vH{}_{\nu})
    +
    \left(\frac{1-\frac{\mu}{4|\bm r|^{D-3}}}{1+\frac{\mu}{4|\bm r|^{D-3}}}\right)^2 \vH{}_{\mu}\vH{}_{\nu} \, ,
\end{align} 
where the ``mass parameter'' is 
\begin{equation}
    \mu = \frac{4\pi \Gamma(\frac{D-1}{2}) }{(D-2)\pi^{(D-1)/2}} \times 2G \mH =  \tilde R \frac{\Gamma \left(\frac{D-3}{2}\right)}{4 \pi^{\frac{D-1}{2}}}\,, \quad \text{and} \quad  |\bm r| = \sqrt{(\vH\cdot x)^2-x^2}\,,
\end{equation}
with $ \tilde R = 16\pi G M (D-3)/(D-2) $ defined  for later convenience.
Expanded perturbatively in the PM expansion, the metric becomes
\begin{align}
    \bar g_{\mu\nu}(x) &=  \eta_{\mu\nu} - \frac{D-2}{D-3} \left( \vH{}_{\mu}\vH{}_{\nu} - \frac{1}{D-2}\eta_{\mu\nu} \right) \frac{\mu}{|\bm r|^{D-3}} + \cdots \, .
\end{align} 
More generally, the position-space background field vertex has an expansion of the form,
\begin{equation}
    1 + a_0 \frac{\mu}{|\bm r|^{D-3}} + a_1 \frac{\mu^2}{|\bm r|^{2(D-3)}} + a_2  \frac{\mu^3}{|\bm r|^{3(D-3)}} + \cdots \, .
\end{equation}
The Fourier transform of a given power is
\begin{equation}
    F^L(\bm q) = \int d^{D-1} \bm r  e^{-i \bm q\cdot \bm r} \frac{\mu^{L+1}}{|\bm r|^{(L+1)(D-3)}} =  \frac{\tilde R^{L+1}}{(\bm q^2)^{1-\frac{D-3}{2}L}} \frac{\Gamma \left(\frac{D-3}{2}\right)^{L+1} \Gamma \left(1-\frac{D-3}{2}L\right)}{(4\pi)^{\frac{D-1}{2}L}\Gamma \left(\frac{D-3}{2}(L+1)\right)}\,.
\end{equation}
Let us compare the Fourier transform of a power to the ``fan'' integral,
\begin{align}
    I_{\rm fan}^L(\bm q) = \vcenter{\hbox{\scalarfan}}
&=\int \prod_i^{L}\frac{d^D\ell_i} {(2\pi)^{D}} \frac{\hat{\delta}(\vH\cdot \ell_1)\hat{\delta}(\vH\cdot \ell_2)\cdots \hat{\delta}(\vH\cdot \ell_L)}{\ell_1^2\ell_2^2\cdots \ell_{L}^2 (q-\sum\ell_i)^2}\,,
\end{align}
where $\vH\cdot q=0$, such that $q^{\mu}=(0,\bm q)$ in the rest frame of the heavy object. The integral is then evaluated by going to this rest frame,
\begin{align}
    I_{\rm fan}^L(\bm q) &=(-1)^{L+1} 
    \int \prod_i^{L}\frac{d^{D-1}\bm{\ell}_i} {(2\pi)^{D-1}} \frac{1}{\bm\ell_1^2\bm\ell_2^2\cdots \bm\ell_{L}^2 (\bm q-\sum\bm\ell_i)^2}  = \frac{(-1)^{L+1}}{(\bm q^2)^{1-\frac{D-3}{2}L}} \frac{\Gamma \left(\frac{D-3}{2}\right)^{L+1} \Gamma \left(1-\frac{D-3}{2}L\right)}{(4\pi)^{\frac{D-1}{2}L} \Gamma \left(\frac{D-3}{2}(L+1)\right)} \nonumber \,,
\end{align}
where the last equality is obtained using iteratively the identity,
\begin{equation}
   \int\frac{d^{D-1}\bm{\ell}} {(2\pi)^{D-1}} \frac{1}{\bm\ell^2 [(\bm p-\bm \ell)^2]^a} = \frac{1}{(\bm p^2)^{a-(D-3)/2}} \frac{\Gamma \left(\frac{D-3}{2}\right) \Gamma \left(a-\frac{D-3}{2}\right)
   \Gamma \left(\frac{D-1}{2}- a\right)}{(4\pi)^{\frac{D-1}{2}} \Gamma (a) \Gamma (D-2-a)}\,.
\end{equation}
It is then easy to see that 
\begin{equation}
    F^L(\bm q) = (-\tilde R)^{L+1} I_{\rm fan}^L(\bm q) \, .
\end{equation}
This means that we can rewrite any background field insertion in terms of simple loop integrals without bulk graviton vertices.  The above procedure is, in some sense, the reverse algorithm of the constructions of \cite{Duff,Damgaard:2024fqj}, which calculated relatively complicated multi-loop integrals in GR to obtain the metric.  Here we instead extract simplified multi-loop integrands from the known expression for the metric.

In particular, taking the PM expansion of
\begin{align}
\begin{split}
    \sqrt{-\bar g} \bar g^{\mu\nu}(x) - \eta^{\mu\nu} &= \mu |\bm r|^{3-D}  \frac{(D-2) \vH^{\mu}\vH^{\nu}}{D-3}+ \mu ^2 |\bm r|^{6-2 D} \left(\frac{(D (4 D-17)+19) \vH^{\mu}\vH^{\nu} }{8 (D-3)^2}-\frac{1}{16} \eta^{\mu\nu}  \right)\\
    &+ \mu ^3 |\bm r|^{9-3 D}\frac{(D-2) (D (3 D-13)+16) \vH^{\mu}\vH^{\nu}}{24 (D-3)^3} +\cdots \, ,
   \end{split}
\end{align}
we find that the PM expansion of the isotropic gauge background insertions on the scalar propagator
\begin{align}
    \raisebox{0.3pt}{\hbox{\backgroundone}}
    &\raisebox{2.8pt}{$=$} \vcenter{\hbox{\backgroundoneone}} \raisebox{2.8pt}{$+$}\vcenter{\hbox{\backgroundonetwo}} \raisebox{2.8pt}{$+$} \vcenter{\hbox{\backgroundonethree}}\raisebox{2.8pt}{$+ \ \ \cdots$}\\
    &=i\tilde R \frac{D-2}{D-3}(\vH k_{1})(\vH k_{2})\hat{\delta}(\vH q)\times\vcenter{\hbox{\trees}}\nonumber \\[0.5cm]
    &-
    i\tilde R^2\left(\frac{D(4D-17)+19}{8(D-3)^2}(\vH k_{1})(\vH k_{2})-\frac{1}{16}(k_{1}k_{2}) \right)\hat{\delta}(\vH q) \times \vcenter{\hbox{\contacttriangles}} 
    \\[0.5cm]
    &+
    i\tilde R^3\frac{(D-2) (D (3 D-13)+16)}{24 (D-3)^3}(\vH k_{1})(\vH k_{2})\hat{\delta}(\vH q) \times\vcenter{\hbox{\wdiagrams}}
   \\[0.5cm]
   & + \ \cdots \nonumber \, .
\end{align}
As mentioned in the previous sections, we do not have an analytic expression for the propagator of a field in a Schwarzschild background. Nonetheless, we have demonstrated in this section how one can build such a propagator to a desired perturbative order rather efficiently. Without resorting to summing complicated trees of self-interacting gravitons, we simply extract the perturbative insertions by expanding the exactly known metric.  This effectively sidesteps numerous complicated tree diagrams in favor of simple scalar fan integrals, which are known to arbitrary loop order. The expansion of the Schwarzschild graviton propagator using the known all order metric inherits simplifications in the same manner.

\subsubsection{From Second-Order Equations of Motion}\label{sec:traj_GR}

It is straightforward to compute the particle trajectories by solving the second-order equations of motion in the PM expansion \cite{Kalin:2020mvi}.  To do so, one expands the worldline and background fields in a PM series,
\eq{
\barxL^\mu = \sum_{k=0}^\infty \barx_k^\mu \, ,
}{xL_PM}
\eq{
\bar\Gamma^\mu_{\alpha\beta} = \sum_{k=1}^\infty \bar\Gamma^\mu_{k\,\,\alpha\beta} \, ,
}{Gamma_PM}
in which case the equations of motion become
\eq{
\ddot \barx_0^\mu &=0 \, ,\\
\ddot \barx_1^\mu  &=  -  \bar \Gamma^\mu_{\alpha\beta}  \dot \barx_{0}^\alpha   \dot \barx_{0}^\beta \, , \\
\ddot \barx_2^\mu  &= - \left(\dot \barx_{0}^\alpha   \dot \barx_{1}^\beta+\dot \barx_{1}^\alpha   \dot \barx_{0}^\beta +\dot \barx_{0}^\alpha   \dot \barx_{0}^\beta\barx_{1}^{\nu}\partial_\nu\right)\, \bar \Gamma^\mu_{1\,\,\alpha\beta}(\barx_{0}) - \dot \barx_{0}^\alpha   \dot \barx_{0}^\beta \,\, \bar \Gamma^\mu_{2\,\,\alpha\beta}(\barx_{0})\,, 
}{}
and so on and so forth.  Rather quickly, there becomes a proliferation of complexity arising from the number of derivatives acting on background metric, the various independent index contractions, and the powers of $\partial_{\tau}^{-2}$.  When inserting these solutions into Feynman diagrams to compute a observables such as the radial action, we are led to increasingly complicated loop integrands. 

The complete loop integrand is physically equivalent to infinitely many other integrands, thanks to various linear relations among integrals in dimensional regularization. One could hope to find, without solving complicated linear systems of equations, a simpler form of the integrand via a more direct route. In the previous section, we showed precisely how this can be done for background field insertions on the propagator of a  field. In the following section we will show how this can be done for the trajectory of the light particle. Further details are given in \App{app:probe_traj}.


\subsubsection{From First-Order Conservation Laws}

We have seen above how to derive the particle trajectories from the second-order equations of motion.  However, it will pay dividends to instead consider a first-order formulation.   Again, using conservation of energy and angular momentum, we go to $(t,r,\phi)$ coordinates and obtain the equations,
\eq{
    \dot{t}=&\sigma \frac{f_{+}(r)^{2}}{f_{-}(r)^{2}}\\
    \dot{\phi}=&\frac{b(\sigma^{2}-1)^{1/2}}{r^{2}}f_{+}(r)^{-4} \\
\dot{r}=&\left[f_{+}(r)^{-4}\left(\frac{f_{+}(r)^{2}}{f_{-}(r)^{2}}\sigma^{2}-1\right)-\frac{b^{2}(\sigma^{2}-1)}{r^{2}}f_{+}(r)^{-8}\right]^{1/2}\,.
}{}
As before, we go to Cartesian coordinates to obtain
\eq{
\dot{x}&=\frac{x}{r}\left[f_{+}(r)^{-4}\left(\frac{f_{+}(r)^{2}}{f_{-}(r)^{2}}\sigma^{2}-1\right)-\frac{b^{2}(\sigma^{2}-1)}{r^{2}}f_{+}(r)^{-8}\right]^{1/2}-\frac{y}{r^2}f_{+}(r)^{-4}(\sigma^{2}-1)^{1/2}b\,,\\
\dot{y}&=\frac{y}{r}\left[f_{+}(r)^{-4}\left(\frac{f_{+}(r)^{2}}{f_{-}(r)^{2}}\sigma^{2}-1\right)-\frac{b^{2}(\sigma^{2}-1)}{r^{2}}f_{+}(r)^{-8}\right]^{1/2}+\frac{x}{r^2}f_{+}(r)^{-4}(\sigma^{2}-1)^{1/2}b\, 
}{}
which are three first-order differential equations.  Again, writing the trajectory in a Lorentz covariant form,
\eq{
\bar{x}^{\mu}(\tau)= t(\tau) \vH^{\mu}+ x(\tau)\frac{b^{\mu}}{b}+y(\tau)\frac{\vL^{\mu}-\sigma \vH^{\mu}}{(\sigma^{2}-1)^{1/2}}\, ,
}{}
we solve for the trajectories perturbatively in the time domain.   Concretely, at 1PM order we obtain
\eq{
t_{1}&=2 (G\mH) \sigma  \frac{1}{\partial_{\tau}}\left(\frac{1}{R}\right) \, , \\
x_{1}&=(G\mH) \left(1-2 \sigma ^2\right) \frac{1}{\partial^{2}_{\tau}}\left(\frac{b}{R^3}\right)\, , \\
y_{1}&=\frac{(G\mH) \frac{1}{\partial_{\tau}}\left(\frac{1}{R}\right)}{\sqrt{\sigma ^2-1}}\,,
}{}
while at 2PM we have
\eq{
t_{2}&=6 (G\mH)^2 \sigma  \frac{1}{\partial_{\tau}}\left(\frac{1}{R^2}\right)+\frac{2 (G\mH)^2 \sigma  \frac{1}{\partial_{\tau}}\left(\frac{1}{R}\right)}{R \left(\sigma ^2-1\right)}\, , \\
x_{2}&=\frac{(G\mH)^2 \left(1-2 \sigma ^2\right) \frac{1}{\partial_{\tau}}\left(\frac{\frac{1}{\partial_{\tau}}\left(\frac{b}{R^3}\right)}{R}\right)}{\sigma ^2-1}+\frac{ (G\mH)^2 \left(1-2 \sigma ^2\right) \frac{1}{\partial_{\tau}}\left(b\frac{\frac{1}{\partial_{\tau}}\left(\frac{1}{R}\right)}{R^3}\right)}{\sigma ^2-1}\\
   &\;\;\;\;+\frac{3}{2}  (G\mH)^2 \left(1-5 \sigma ^2\right) \frac{1}{\partial^{2}_{\tau}}\left(\frac{b}{R^4}\right)\, , \\
y_{2}&=-\frac{\frac{1}{\partial_{\tau}}(1) (G\mH)^2 \left(1-2 \sigma ^2\right)^2}{2 b^2 \left(\sigma ^2-1\right)^{3/2}}+\frac{3 (G\mH)^2 \left(3 \sigma ^2+1\right) \frac{1}{\partial_{\tau}}\left(\frac{1}{R^2}\right)}{4 \sqrt{\sigma
   ^2-1}}+\frac{(G\mH)^2 \frac{1}{\partial_{\tau}}\left(\frac{1}{R}\right)}{R \left(\sigma ^2-1\right)^{3/2}} \, ,
}{}
and so on and so forth.  Here we have defined $R(\tau)=\sqrt{b^{2}+(\sigma^{2}-1)^{1/2}\tau^{2}}$. The above formulas apply to $D=4$ but the analogous expressions for general $D$ are given in \App{app:probe_traj}.  Much like in the case of EM, we have recast the PM trajectories in the form of Feynman integrands.  In particular, a Fourier transform to momentum space maps inverse powers of $R$ to inverse powers of the spatial momentum transform, $\ell^{-2}$, and sends $\partial_\tau^{-1}$ to linearized matter propagators, $(\vL\cdot\ell)^{-1}$.

Note that the Feynman integrand topologies encoded in the above trajectories for GR are exactly the same in EM.  This is a remarkable simplification, considering the fact that the graviton has far more complicated interaction vertices than the photon.

\begin{figure*}
    \centering
\begin{align*}
    &\vcenter{\hbox{\bprop}} \rightarrow \vcenter{\hbox{\proponethree}} \quad \vcenter{\hbox{\proptwotwo}}\quad \vcenter{\hbox{\proptwofour}} \quad 
    \vcenter{\hbox{\proptwofive}} \\[5pt]
    &\vcenter{\hbox{\brecoil}}
    \rightarrow
    \vcenter{\hbox{\recoilone}} \quad
    \vcenter{\hbox{\recoiltwoone}} \quad  \vcenter{\hbox{\recoiltwothree}}
\end{align*}
    \caption{The 1SF radial action in GR is computed from the Feynman diagrams on the left.  The corresponding Feynman propagators and vertices were derived in our effective field theory and summarized in \Fig{fig:gravprop} and \Fig{fig:Feynman_rules_GR}.  Meanwhile, by expanding these expressions to 3PM order, we obtain the flat space Feynman diagrams on the right.  
    }
    \label{fig:1SF_GR}
\end{figure*}

\subsection{Results and Checks} 

In order to calculate the radial on-shell action for GR, we compute the path integral over graviton perturbations about the Schwarzschild background.  Following our procedure in EM, we compute
\eq{
\exp(i I_{\rm GR}) = \int [d\delta \xH] [d\delta \xL] [d\dA] \exp(iS_{\rm GR}) = \int  [d \dA] \exp(i \bar S + i \delta S^{(1)}_{\rm eff} +\cdots) \, .
}{}
Order by order in the SF expansion, the radial action is
\eq{
I_{\rm GR} &= I^{(0)}_{\rm GR} + I^{(1)}_{\rm GR}+ \cdots \, ,
}{}
where the leading few contributions are
\eq{
I^{(0)}_{\rm GR} &= \bar S \, ,\\
I^{(1)}_{\rm GR} &= -i\log  \int  [d \dA] \exp( i \delta S^{(1)}_{\rm eff} )  \, .
}{}
Each SF contribution to the radial action is then further expanded in PM, yielding
\eq{
I_{\rm GR}^{(i)}=\sum_{j=i+1}^\infty I_{\rm GR}^{(i,j)} \, .
}{}
We will compute this SF and PM expanded radial action for an array of gravitational theories. Note that by dimensional analysis, we find that
\eq{
I_{\rm GR}^{(i,j)}=\lambda^{i}\,\mL\RS\left(\frac{\RS}{b}\right)^{j-1}\mathcal{I}_{{\rm GR}}^{(i,j)}(\sigma) \, ,
}{}
ensuring mass polynomiality of the expansion.

\begin{figure*}
    \centering
    $\vcenter{\hbox{\fpropone}} \qquad \vcenter{\hbox{\fproptwo}}  \qquad \vcenter{\hbox{\fpropsix}}  \qquad \vcenter{\hbox{\fpropseven}}$ \\[20pt]
    $\vcenter{\hbox{\fpropthree}} \qquad \vcenter{\hbox{\fpropfour}}   \qquad \vcenter{\hbox{\fpropfive}}$ \\[20pt]
    $\vcenter{\hbox{\frecone}}  \qquad \vcenter{\hbox{\frectwo}}   \qquad \vcenter{\hbox{\frecthree}}
    $
    \caption{Flat space Feynman diagrams that contribute to the 1SF action to 3PM order. The dotted lines depict static massive sources and the solid straight lines represent matter propagators.}
    \label{fig:1SF_flat}
\end{figure*}

\subsubsection{Scattering Masses}

Let us now compute the 0SF, 1SF, and 2SF radial actions for GR up to 3PM order.  Using  the methodology reviewed in \App{app:probe_action}, we obtain the 0SF radial action for GR,
\eq{
I^{(0)}_{\rm GR} = & \mL\RS\bigg[\left(\frac{1}{D-4}-\log\left(b\mL(\sigma^{2}-1)^{1/2}\right)\right)\frac{(2 \sigma^2 -1)}{\sqrt{\sigma^2 -1}} + \frac{\RS}{b} \frac{3\pi (5\sigma^2-1)}{16\sqrt{\sigma^2-1}}  \\
&+ \frac{\RS^{2}}{b^2} \frac{64 \sigma^6 -120 \sigma^4 + 60 \sigma^2 -5}{24 (\sigma^2-1)^{\frac{5}{2}}} + \ \cdots\bigg] \, ,
}{}
shown here to the first few orders in the PM expansion.

Moving on to 1SF order, we use the Feynman rules described in \Fig{fig:gravprop} and  \Fig{fig:Feynman_rules_GR} to compute the radial action, $I_{\rm GR}^{(1)}$ from the Feynman diagrams depicted in \Fig{fig:1SF_GR}.  A mechanical calculation yields the 2PM contribution,
\eq{
I_{\rm GR}^{(1,2)}=\lambda\mL\RS\frac{\RS}{b }  \frac{3\pi(5\sigma^2-1)}{4\sqrt{\sigma^2-1}}\, .
}{}
The flat space diagram topologies that need to be evaluated for 1SF 3PM computations are shown in \Fig{fig:1SF_flat} for comparison. 
As an additional consistency check, we have performed the 2PM calculation in a more general gauge fixing defined by 
\Eq{GRaction_R} but with the choice, $F_\mu = \zeta_1 \bar\nabla^\nu \delta g_{\mu\nu} -\tfrac12 \zeta_2 \bar\nabla_\mu \delta g$.  For general $\zeta_1$ and $\zeta_2$, the flat space graviton propagator deviates from the deDonder form in \Eq{deDonder}, an in fact has spurious $1/p^4$ and $1/p^6$ poles.  Working in this gauge, we find that the contributions from pure background field method diagrams are not gauge invariant, nor is the contribution from the recoil operator.  However, their sum is gauge invariant, and yields the correct 2PM expression.
As yet another check, we have also done this calculation in general spacetime dimension $D$, yielding 
\eq{
I_{\rm GR}^{(1,2)}&= \lambda\mL \RS^2 b^{7-2 D} \frac{\pi ^{\frac{7}{2}-D} \left((2 D-5) \sigma ^2 \left((2 D-3) \sigma ^2-6\right)+3\right) \Gamma \left(D-\frac{7}{2}\right) \Gamma \left(\frac{D-1}{2}\right)^2}{(D-2)^2 \left(\sigma ^2-1\right)^{3/2} \Gamma (D-2)}
 \, ,
}{} 
which agrees with known results \cite{Cristofoli:2020uzm}.

Next, we compute the 1SF Feynman diagrams at 3PM order in \Fig{fig:1SF_GR} to obtain the 3PM radial action,
\eq{
I_{\rm GR}^{(1,3)}=\lambda \mL\RS \left(\frac{\RS}{b}\right)^2 \left(\frac{ \sigma  \left(36 \sigma ^6-114 \sigma ^4+132 \sigma ^2-55\right)}{12\left(\sigma
   ^2-1\right)^{5/2}}-\frac{\left(4 \sigma ^4-12 \sigma ^2-3\right) \textrm{arccosh}\,\sigma }{2\left(\sigma ^2-1\right)}\right) \, ,
}{I_GR}
using the integration methods described in \cite{Parra-Martinez:2020dzs,Dlapa:2023hsl}. The above results agree exactly with the known 2PM and 3PM expressions \cite{Cachazo:2017jef,BBDamFestPlaVan,PM1,PM2}.

\begin{figure*}
    \centering
    $\vcenter{\hbox{\bint}}
    \qquad   \vcenter{\hbox{\brecoiltwoone}}
    \qquad
    \vcenter{\hbox{\brecoiltwotwo}}$
    \caption{The 2SF 3PM radial action in GR is computed from these Feynman diagrams.  The first diagram involves cubic graviton vertices in a Schwarzschild background, the second involves the 2SF recoil operator, and the third involves 1SF recoil operator in combination with the cubic graviton vertex. Diagrams with multiple recoil operator insertions vanish in the potential region at 3PM order. When considering the radiation region or computing at higher PM orders, however, such diagrams can contribute and must be added to this list.}
    \label{fig:2SF_GR}
\end{figure*}

Last but not least, as a highly nontrivial check of our 2SF formalism, we also compute the 2SF radial action expanded to 3PM order, $I_{\rm GR}^{(2,3)}$. This calculation computes the Feynman diagrams in  \Fig{fig:2SF_GR} using the Feynman rules derived from the 2SF action in \Sec{sec:2SF_dyn}.  
Importantly, $I_{\rm GR}^{(2,3)}$ is exactly equal to $I_{\rm GR}^{(0,3)}$ under the exchange of the heavy and light particle masses---and indeed we find that our final answer precisely exhibits this feature.

\subsubsection{Scattering Scalar Charged Masses}

It is trivial to incorporate additional fields in our framework.  In particular, let us consider an additional scalar field that couples directly to the light particle but only gravitationally to the heavy particle.  Such theories have been explored in SF studies \cite{poissontoy,scalartoy} as a toy model for full gravity.   The action for this theory is
\eq{
S_{\rm scalar} &=  \int d^4 x  \sqrt{-\bar g}  \left[ \tfrac12 \bar\nabla_\mu \Phi \bar\nabla^\mu \Phi + \tfrac{1}{2} \xi \bar R \Phi^2 - \Phi J \right]\, ,
}{}
where, for maximum generality, we have included a nonminimal coupling, $\xi$, and the scalar couples to the current,
\eq{
J(x)&= \yL \mL \int d\tau \frac{\delta^4(x-\barxL)}{\sqrt{-\bar g}}    \, ,
}{}
which only involves the light particle.

Crucially, in this theory the heavy particle does not accrue any additional interactions.  Consequently, the gravitational recoil operator in \Eq{recoil_op_GR} is completely unchanged.   Thus, to compute the radial action we need only include the additional background field diagram depicted in \Fig{fig:toy_models}.

\begin{figure*}
    \centering
    $\hbox{\bpropscalar}$ \raisebox{15pt}{$\rightarrow$}
   $\hbox{\bspropone} \quad
   \hbox{\bsproptwothree} \quad \hbox{\bsproptwoone} \quad \hbox{\bsproptwotwo}$
    \caption{The 1SF radial action receives contributions from a field which couples to the light body but not the heavy body, so there is no recoil operator insertion. The double line denotes the propagator of this additional field, here taken to be a scalar or vector, in the Schwarzschild background.  The diagrams on the right correspond to the flat space expansion of these contributions to 3PM order. }
    \label{fig:toy_models}
\end{figure*}

Including this Feynman diagram, we find that the 1SF radial action for the scattering of scalar charged masses is
\eq{
I_{\rm scalar}^{(1,2)}&= -\lambda\mL\RS \left(\frac{\Rsc}{b}\right)\left(\frac{\pi}{8}\frac{\sigma ^2-1+4 \xi}{ \sqrt{\sigma ^2-1}}\right) \, ,\\
I_{\rm scalar}^{(1,3)} &= -\lambda \mL\RS\left(\frac{\RS\Rsc}{b^{2}}\right)\frac{ \sigma  \left(2 \sigma ^4-\sigma ^2-1+\xi  \left(6 \sigma ^2-3\right)\right)}{6 \left(\sigma
   ^2-1\right)^{3/2}} \, ,
}{}
where we've defined the scalar charge radius,
\eq{
\Rsc=\frac{\yL^2 \mH}{4\pi} \, .
}{}
The above expression agrees with the results of \cite{scalartoy}.  Note that the contribution from the nonminimal coupling, $\xi$, constitutes a new calculation.

An interesting check of this result can be performed by computing the probe action in a background given by a solution to the Einstein field equations in the presence of a gravitationally coupled massless scalar field \cite{einsteinscalar1}. Our expressions precisely match this result, under the swap of the heavy and light bodies. We have also verified that for the case of a conformally coupled scalar, $\xi=1/6$, our results are in agreement with the probe action for the conformal scalar solution in \cite{einsteinconf1, einsteinconf2}.

\subsubsection{Scattering Vector Charged Masses} The procedure described above can also be applied to derive results for a theory in which an additional vector field couples to the light particle but only gravitationally to the heavy particle.   This theory is described by the action,
\eq{
S_{\rm vector} & - =  \int d^4 x  \sqrt{-\bar g}  \left[\tfrac14 F_{\mu\nu} F^{\mu\nu} + A_\mu J^\mu \right] \, ,
}{}
where the vector current couples only to the light particle,
\eq{
J^\mu(x)&= \zL \mL \int d\tau \frac{\delta^4(x-\barxL)}{\sqrt{-\bar g}}   \dotbarxL^\mu \, .
}{}
As before, the recoil operator is unchanged, and now the only additional diagram is a background field loop of the vector field shown in \Fig{fig:toy_models}. Computing this Feynman diagram and integrating, we obtain the 1SF radial action for the scattering of vector charged masses,
\eq{
I_{\rm vector}^{(1,2)}&= -\lambda \mL\RS\left(\frac{\Rvc}{ b}\right)\left(\frac{\pi}{8}\frac{ 3 \sigma ^2-1}{\sqrt{\sigma ^2-1}}\right) \, ,\\
I_{\rm vector}^{(1,3)} &= -\lambda \mL\RS\left(\frac{\RS\Rvc}{b^{2}}\right) \left(\frac{\sigma  \left(8 \sigma ^4-28 \sigma ^2+23\right)}{12  \left(\sigma
   ^2-1\right)^{3/2}}+\frac{ \left(2 \sigma ^2+1\right) \textrm{arccosh}\,\sigma }{ \left(\sigma ^2-1\right)}\right) \, ,
}{}
where we have defined the vector charge radius,
\eq{
\Rvc=\frac{\zL^2 \mH}{4\pi} \, .
}{}
The above expression is a new result. We find that the expression for the probe radial action in the Reissner-Nordstr\"om metric when it is linearized in the vector charge and under the swap of the heavy and light bodies, matches with the expression above.

\medskip

\section{Conclusions}

In this paper we have presented a systematic effective field theory describing the dynamics of two interacting bodies expanded in powers of their mass ratio, $\lambda = \mL/\mH$.   To derive this formulation, we expanded the light and heavy particle trajectories about their geodesics in a Schwarzschild background.  By integrating out the geodesic deviation of the heavy particle, we systematically constructed an effective field theory whose sole degrees of freedom are the light particle and the graviton fluctuation.   A key ingredient in our setup is the fact that classical solutions---like the Schwarzschild metric together with the span of all of probe geodesics---carry information that is effectively all orders in perturbation theory from the point of the field theory constructed in a trivial background.  

The main technical result of our paper is a precise characterization of those perturbative contributions which are {\it not encoded} in the background fields and geodesics that constitute the 0SF theory.  In particular, these ``leading corrections to the background field method’’ enter at 1SF and are  accounted for by a recoil operator describing the wobble of the heavy particle sourcing the background field.   The sole effect of this operator is a nonlocal-in-time correction to the two-point function of the force carrier.  Importantly, higher-order corrections to the effective action can also be systematically derived, and we present those contributions at 2SF.

Applying these ideas to EM and GR, we use our effective field theory framework to compute the conservative radial action for scattering particles in various systems.  Here the time-domain probe trajectories in EM and GR can be used to derive explicit loop integrand contributions for the corresponding scattering processes.   We have verified that our framework correctly reproduces the conservative dynamics in a number of familiar scenarios.  We also present a few new calculations.

The present work leaves many directions for future study.   First and foremost is the question of whether our results can be made at all useful for existing approaches to the SF problem \cite{PoissonReview,PoundReview,BarackReview},  which are inherently nonperturbative and often numerical.  While we have shown how the Schwarzschild metric and geodesics avail information at all orders in the PM expansion, this data takes the form of perturbative loop {\it integrands}.  In all honesty, it will be daunting, if not outright impossible, to integrate these contributions to all loop orders.   

A related obstacle is that our construction is fundamentally built from an effective field theory of {\it point particles} interacting through long range forces.  While resumming diagrams reproduces the classical backgrounds such as the Schwarzschild metric, we should nevertheless interpret the resulting background as a fundamentally perturbative field sourced by sources, rather than a vacuum solution to the Einstein field equations.  
 In the approach of existing SF methods, however, there is no heavy point source.  Hence, dimensional regularization is not an option, and self-energy contributions must instead be dealt with in a substantially different way.  For this reason, it would be useful to try to port our results to this different approach.  For example, it would worthwhile to try to adapt our recoil operator, which is defined naturally in dimensional regularization, to other regulators.

A second avenue for exploration is the generalization of our results to other types of binary systems relevant to gravitational wave astronomy. Our formalism has already been used to compute the two-loop matching and renormalization-group running of scalar analogs to dynamical Love numbers, which describe the response of a black hole to scalar disturbances~\cite{Ivanov:2024sds}. It also proved to be efficient in studying the dynamics of a binary system of charged Reissner-Nordstr\"{o}m black holes~\cite{Wilson-Gerow:2023syq}, though this example is not relevant to astrophysics. A more phenomenologically interesting target would be spinning black hole binaries, which at 0SF are described by a spinning probe in a Kerr background.  Here the 1SF sector should also be corrected by a recoil operator corresponding to the back-reaction on the heavy spinning source.  

A third promising direction relates to the application of our effective field theory to quantum mechanical processes.  While we have applied our formalism to classical GR, there is nothing intrinsically classical about the derivation of our construction.  In particular, one can consider a heavy particle worldline that sources a Schwarzschild background and furthermore couples quantum mechanically to fluctuating gravitons and matter fields.  In such a setup, it should be possible to repeat the seminal calculation of Hawking \cite{Hawking:1975vcx}, albeit including the effect of black hole recoil from the emitted radiation.

Last but not least, it would be interesting to apply our methodology to systems other than EM and GR.  Here a natural candidate for exploration is fluid dynamics, which exhibits gapless ``force carriers'' in the form of perturbations of the fluid velocity field.  The analog of perturbation theory is the so-called Wyld formalism \cite{Wyld:1961gqg}, which is equivalent to solving the Navier-Stokes equations as a $1/\nu$ expansion about the diffusion equation, where $\nu$ is the viscosity.  Furthermore, there exist classical vortex solutions to the Navier-Stokes equations that are regular at small $1/\nu$, and can, in principle, be reconstructed in perturbation theory in the spirit of the Duff approach for the Schwarzschild metric.   As a result, it may be possible to construct an effective field theory of light and heavy particles interacting via a fluid medium.


\medskip

\appendix

\section{Time-Domain Trajectories}\label{app:probe_traj}

In this appendix we describe how to extract loop integrands from the time-domain solutions of the probe particle equations of motion. We cover the cases of both electromagnetism and gravity. The general strategy is as follows: {\it i}) use the conserved charges of the probe system to write simple first-order ordinary differential equations for the motion, {\it ii}) integrate these equations perturbatively in the time domain, {\it iii}) express these time-domain solutions in terms of the zeroth order radial trajectory,
\eq{
R(\tau)=\sqrt{b^{2}+(\sigma^2-1)\tau^{2}}\,,
}{}
and then {\it iv}) recast them into momentum space integrals using the identity,
\eq{
R^{2\alpha+1-D}=\frac{(-4)^{\alpha}\pi^{\tfrac{D-1}{2}}\Gamma(\alpha)}{\Gamma(\tfrac{D-1}{2}-\alpha)}\int\frac{d^{D}\ell}{(2\pi)^{D}}e^{-i\ell(b+\vL\tau)}\frac{\hat{\delta}(\vH\ell)}{(\ell^{2})^{\alpha}}\,,
}{eq:oneloopfan}
together with
\eq{
b^{\mu}R^{2\alpha-1-D}=i\frac{(-4)^{\alpha}\pi^{\tfrac{D-1}{2}}\Gamma(\alpha)}{2\Gamma(\tfrac{D-3}{2}-\alpha)}\int\frac{d^{D}\ell}{(2\pi)^{D}}e^{-i\ell(b+\vL\tau)}\frac{\hat{\delta}(\vH\ell)\Pi^{\mu\nu}\ell_{\nu}}{(\ell^{2})^{\alpha}}\,,
}{eq:fourierR}
where $\alpha$ is a positive integer we have defined
\eq{
\Pi^{\mu\nu}=\eta^{\mu\nu}-(\sigma^{2}-1)^{-1}(\sigma\vH^{\mu}-\vL^{\mu})\vL^{\nu}-(\sigma^{2}-1)^{-1}(\sigma\vL^{\mu}-\vH^{\mu})\vH^{\nu},
}{}
which is the projector orthogonal to both four-velocities.\footnote{Since $\vH\ell=0$ in the integrand, $\Pi^{\mu\nu}$ as written has redundant terms.}

\subsection{Electromagnetism}
To begin, we review the solution for a relativistic charged probe trajectory in a Coulomb potential. In $D=4$ dimensions this can be done exactly, as in \cite{Landau:1975pou, sommerfeld}. However, since our formalism naturally uses dimensional regularization for divergences, we will want results in general $D$.  Unfortunately, there is no closed-form solution for $D$-dimensional charged probe trajectories.  However, we will see how to mechanically compute such a solution to a desired order in the PL expansion in terms of iterated integrals involving ${}_2F_{1}$ hypergeometric functions. While the latter are naively quite cumbersome, we demonstrate how corresponding $D$-dimensional loop integrands are readily extracted from the expressions. The resulting expressions, especially in the gravitational case, are considerably more compact than expressions that appear using standard Feynman diagrams.   Moreover, from our final expressions one will see that this detour through position space effectively performs integration by parts reduction automatically.

Consider the einbein action for the charged probed particle in a Coulomb potential,
\eq{
S_{\rm EM} &= - \int d\tau  \left[ \frac{1}{2} e^{-1} \dot x^2+ \frac{1}{2} e m^2 +q \dot x^\mu  A_\mu(x) \right] \, .
}{}
To reduce notational clutter, we will drop subscripts and bars denoting this to be the light particle evolving in a background EM field. The einbein equation of motion gives $e=\sqrt{\dot x^2}/m$. Let us define the components $x^\mu = (t,r,\theta,\phi)$, where $\theta=\pi/2$ for scattering in the equatorial plane. Next, we gauge fix $e=1/m$, which imposes the on-shell condition, 
\eq{
1=\dot x^2= \dot t^2 -\dot r^2 - r^2 \dot \phi^2 \, ,
}{}
on the space of solutions. 
As usual, we take the background EM field $A_\mu$ to be the $D$-dimensional Coulomb potential.   Working in Lorenz gauge and in the rest frame of the source, we find that the gauge potential is
\eq{
\frac{q}{m}A_{0}(r)=-\frac{k_{D}}{r^{D-3}}=-\frac{\Rc}{r^{D-3}}\left(\frac{\Gamma(\frac{D-3}{2})}{\pi^{\frac{D-3}{2}}}\right)\,,
}{}
where $\Rc$ is the charge radius
defined in \Eq{def_rc} and $k_{D}$ notation introduced to condense upcoming expressions. The quantity in parentheses is unity in $D=4$. 

Since $A_\mu$ is time-independent and spherically symmetric, we can impose conservation of energy and angular momentum,
\eq{
E= m\dot t + q A_0 \qquad \textrm{and} \qquad J= m r^2 \dot \phi \, .
}{def_E_J}
More convenient variables for scattering processes are
\eq{
\sigma=\frac{E}{m} \qquad \textrm{and} \qquad b = \frac{J}{m(\sigma^{2}-1)^{1/2}}\,,
}{def_sig_b}
which are readily defined in a Lorentz invariant manner from the asymptotic inertial trajectories of the two bodies.  Eliminating $\dot t$ and $\dot \phi$ via \Eq{def_sig_b}, the on-shell condition becomes
\eq{
1=\left(\sigma-\frac{q}{m}A_{0}(r)\right)^2-\dot{r}^2-\frac{b^2(\sigma^{2}-1)}{r^{2}}\,.
}{EM_probe_on-shell_condition}
On the outward branch of the scattering trajectory, corresponding to $\dot{r}\geq0$, we obtain
\eq{
\dot{t}&=\sigma+\frac{k_D}{r^{D-3}}\, ,\\
\dot{\phi}&=(\sigma^{2}-1)^{1/2}\frac{b}{r^2}\, ,\\
\dot{r}&=\sqrt{\left(\sigma+\frac{k_{D}}{r^{D-3}}\right)^2-\frac{b^2 (\sigma^{2}-1)}{r^2}-1}\,.
}{EOM_gen_D}
So by the integrability of the probe motion, the equations of motion have reduced to three first-order differential equations. 

As mentioned above, we do not have a closed-form $D$-dimensional solution to \Eq{EOM_gen_D}.  However, these equations are very simple to integrate perturbatively. To this end, it will be more convenient to combine the equations of motion for $(r,\phi)$ into equations of motion for the Cartesian components $(x,y)$,
\eq{
\dot{x}&=\frac{x}{r}\left[\left(\sigma+\frac{k_{D}}{r^{D-3}}\right)^2-\frac{b^2 (\sigma^{2}-1)}{r^2}-1\right]^{1/2}-\frac{y}{r^2}(\sigma^{2}-1)^{1/2}b \, ,\\
\dot{y}&=\frac{y}{r}\left[\left(\sigma+\frac{k_{D}}{r^{D-3}}\right)^2-\frac{b^2 (\sigma^{2}-1)}{r^2}-1\right]^{1/2}+\frac{x}{r^2}(\sigma^{2}-1)^{1/2}b\, ,
}{EMcartesianEOM}
where hereafter we take $r$ to be an implicit function of $(x,y)$.
Next, we perform a PL expansion of the trajectory,
\eq{
(t(\tau),x(\tau),y(\tau))=\sum_{n=0}^{\infty}(t_{n}(\tau),x_{n}(\tau),y_{n}(\tau)),
}{}
where the $n$ labels terms that are $\mathcal{O}(k_{D}^{n})$.  We then solve \Eq{EMcartesianEOM} order by order in $k_{D}$. This yields the general solution, which expressed in Lorentz covariant form is
\eq{
\bar{x}^{\mu}(\tau)= t(\tau) \vH^{\mu}+ x(\tau)\frac{b^{\mu}}{b}+y(\tau)\frac{\vL^{\mu}-\sigma \vH^{\mu}}{(\sigma^{2}-1)^{1/2}}\,.
}{eq:covtrajectory}
The leading order solution at $n=0$ is the straight line trajectory,
\eq{
    t_{0}=\sigma\tau\,,\qquad x_{0}=b\,,\qquad y_{0}=(\sigma^{2}-1)^{1/2}\tau\,.
}{leadingtrajectory}
Meanwhile, for $n\geq1$ the resulting equations of motion take the form
\eq{
\frac{d}{d\tau}t_{n}(\tau)&=T_{n}(\tau)\, ,\\
\frac{d}{d\tau}\left(\frac{x_{n}(\tau)}{\tau}\right)&=\frac{X_{n}(\tau)}{\tau}\, ,\\
\frac{d}{d\tau}y_{n}(\tau)&=Y_{n}(\tau)\,,
}{EMODEs}
where $(T_{n},X_{n},Y_{n})$ depend on $\tau$ only through the lower order solutions $(t_{m},x_{m},y_{m})$ for $m<n$. As written these equations are now exact differentials, so their solutions are given by  integrals of the right-hand side. The functions $(T_{n},X_{n},Y_{n})$ are computed by expanding \Eq{EMcartesianEOM} to the desired order, and the solutions $(t_{m},x_{m},y_{m})$ for $m<n$ on which they depend are written as integrals of the lower order $(T_{m},X_{m},Y_{m})$. One then starts from the inertial trajectory, and iteratively solves these equations of motion order-by-order. The trajectory solutions at a given order then take the form of  iterated integrals of various functions of proper time.

Since our goal is to extract loop integrands from these trajectories we do not want explicit $\tau$ dependence in the solutions. To match the structure of Feynman loop integrals, all $\tau$ dependence should come entirely from the lowest order solution $\bar{x}^{\mu}_{0}(\tau)$. Indeed, as we will outline momentarily, the equations of motion can be manipulated in such a way that all $\tau$ dependence is implicit, through powers of $R$ and $\partial_{\tau}R$. As a result the structure of the solutions will be iterated integrals of powers of $R$, which can be readily mapped to Feynman integrals using \Eqs{eq:oneloopfan}{eq:fourierR}.

As previously mentioned, the functions $(T_{n},X_{n},Y_{n})$ already depend on $\tau$ only through the lower order trajectories, so the $(t_n, y_n)$ solutions follow immediately,
\eq{
t_{n}(\tau)&=\frac{1}{\partial_{\tau}}\bigg(T_{n}(\tau)\bigg)\, ,\\
y_{n}(\tau)&=\frac{1}{\partial_{\tau}}\bigg(Y_{n}(\tau)\bigg)\,.
}{eq:TYsolns}
Here inverse powers of $\partial_\tau$ are shorthand for $d\tau$ integrals.  As we will see later on, these are precisely matter propagators, so the solutions $(t_n, y_n)$ automatically take the form of Feynman loop integrals.

Unfortunately, a bit more work is needed in order to massage the $x_{n}$ solution into loop integral form.  Here it will be useful to differentiate the $x_{n}$ equation in \Eq{EMODEs} once more and then use \Eq{EMODEs} to eliminate $\dot{x}_{n}$. The resulting expression is a second order ordinary differential equation with no factors of $\tau$ remaining on the left-hand side,
\eq{
\frac{d^{2}}{d\tau^{2}}{x}_{n}(\tau)=\frac{1}{\tau}\frac{d}{d\tau}\bigg(\tau X_{n}(\tau)\bigg)\,.
}{eq:ddxeqn}
The above expression will be useful because it will turn out that the combination of factors $\tau X_{n}$ is more readily expressed in terms of $R$ than other choices like $X_{n}$ or $\tau^{-1}X_{n}$.

The solution to \Eq{eq:ddxeqn} can be expressed in two convenient forms.  The first is
\eq{
x_{n}(\tau)=\frac{1}{\partial_{\tau}^{2}}\left[\frac{1}{\tau}\frac{d}{d\tau}\big(\tau X_{n}(\tau)\big)\right]\,,
}{eq:xsolutionouter}
which is related by integration by parts to the second,
\eq{
x_{n}(\tau)=-\tau X_{n}(\tau)+\tau \frac{1}{\partial_{\tau}}\left[\frac{1}{\tau}\frac{d}{d\tau}\big(\tau X_{n}(\tau)\big)\right]\,.
}{eq:xsolutioninner}
Just as the right-hand side of \Eq{eq:TYsolns} is a function of $\tau$ only through its dependence on $R$, so too is the right-hand side of \Eq{eq:xsolutionouter}, in spite of appearances. 
While \Eq{eq:xsolutionouter} and \Eq{eq:xsolutioninner} are mathematically equal, it will be advantageous to use one or the other at various steps in the calculation. 

To construct a solution $x_{n}$ which takes the form of a loop integrand, one should start with \Eq{eq:xsolutionouter}.
We will refer to this as the ``outer layer'' form of the solution, since it should be used in the very first step taken to construct the solution for $x_{n}$. Next, we need to evaluate the contribution from $X_n$ entering in \Eq{eq:xsolutionouter}.  Here $X_n$ depends on the lower-order solutions $x_{m}$ for $m<n$.  To evaluate these lower-order expressions, it turns out that we should plug in  \Eq{eq:xsolutioninner}, which we hence dub the ``inner layer'' form of the solution.  This choice has two advantages: it simplifies the resulting expressions without needing integration by parts identities, and it exhibits fewer proper time integrals, leading to fewer matter propagators in the final result.

To be concrete, let us compute the explicit trajectories to a few nontrivial orders. Starting at $n=1$, the right-hand sides of \Eq{EMODEs} are
\eq{
T_{1}&=k_{D}R^{3-D}\, ,\\
X_{1}&=k_{D}\frac{b \sigma}{\sigma^2-1}\frac{R^{3-D}}{\tau}\, ,\\
Y_{1}&=k_{D}\frac{\sigma}{(\sigma^2-1)^{1/2}}R^{3-D}\,,
}{}
which are integrated using \Eqs{eq:TYsolns}{eq:xsolutionouter} to give the expressions,
\eq{
t_{1}&=k_{D}\frac{1}{\partial_{\tau}}\left(R^{3-D}\right)\, ,\\
x_{1}&=-k_{D}b(D-3)\sigma\frac{1}{\partial_{\tau}^{2}}\left(R^{1-D}\right)\, ,\\
y_{1}&=k_{D}\frac{\sigma}{(\sigma^2-1)^{1/2}}\frac{1}{\partial_{\tau}}\left(R^{3-D}\right)\,.
}{}
The inverse powers of $\partial_\tau$ are shorthand for $d\tau$ integrals whose explicit expressions are
\eq{
\int d\tau\, R^{3-D}&=b^{3-D} \tau\,\,{}_{2}F_{1}(\tfrac{1}{2}, \tfrac{D-3}{2}, \tfrac{3}{2}, -\tfrac{(\sigma^{2}-1) \tau^2}{b^2}) \, ,\\
\int d\tau\int^{\tau} d\tau'\,R^{1-D} &= \frac{b^{3-D}-R^{3-D}}{(3-D)(\sigma^{2}-1)}+ b^{1-D}\tau^{2}\,\,{}_{2}F_{1}(\tfrac{1}{2}, \tfrac{D-1}{2}, \tfrac{3}{2}, -\tfrac{(\sigma^{2}-1) \tau^2}{b^2})\,.
}{}
In $D=4$, these reduce to trigonometric and algebraic functions of $\tau$, and the iterated integrals required to build the higher order solutions are expressible in closed form. In general $D$, however, one is left with iterated integrals involving hypergeometric functions.  Consequently, it is more notationally economical to leave the expressions above in terms of formal $d\tau$ integrals of powers of $R$. 


For $n>1$ and above, one can algorithmically solve the equations of motion to directly yield a loop integrand.  The procedure for this is as follows. 
First, let us assume that we already have solutions for $(t_{m},x_{m},y_{m})$ for $m<n$ in terms of integrals of powers of $R$ and have used these to obtain the functions $(T_{n},X_{n},Y_{n})$ on the right-hand sides of \Eq{EMODEs} at order $n$. Crucially, as we noted above, the $x_{m}$ solution one should be using here is best written in the form of the inner layer solution \Eq{eq:xsolutioninner}. Take these functions and use them to write the first order equations of motion \Eq{EMODEs} for $(t_{n},y_{n})$ and the second order equation of motion \Eq{eq:ddxeqn} for $x_{n}$. The right-hand sides of these equations are now in a form which can be simplified mechanically to write as a loop integrand.

 These expressions will have various powers of $\tau$ in both the numerator and denominator. We want to eliminate explicit appearance of $\tau$ in favor of $R$ and its derivatives so that we can trivially go to momentum space. If $\tau$ appears in the combination $b^{2}+(\sigma^{2}-1)\tau^{2}$, then we simply replace it with $R^2$. If $\tau$ appears alone in the denominator, so there are terms with a common factor $\tau^{-l}$ for $l>0$, then we take the coefficient of $\tau^{-l}$ and express all of factors of $R$ within it explicitly in terms of $\tau$ and then simplify the expression.  This will introduce overall factors of $\tau$ in the numerator that will cancel  the factor of $\tau^{-l}$, yielding only positive powers of $\tau$ to be dealt with. For even-positive powers we write $\tau^{2l}=(\sigma^2-1)^{-l}(R^2-b^2)^{l}$ while for odd-positive powers we write $\tau^{2l+1}=(\sigma^2-1)^{-l}(R^2-b^2)^{l}\tau$. At this point, the only explicit $\tau$ dependence will be linear,  moreover, it will necessarily be in the form of a factor $\tau R^{q}$ for some $q$. Here we simply rewrite such terms as 
\eq{
\tau R^{q}=\frac{\partial_{\tau}\left(R^{q+2}\right)}{(q+2)(\sigma^{2}-1)}\,.
}{}
The expression is now expressed as a string of iterated integrals and derivatives, with respect to proper time, of various power of $R$. From here the solutions are readily obtained as $d\tau$ integrals of the newly simplified equations.

While the above prescription succeeds in expressing the solutions solely in term of $R(\tau)$, there may still be another manipulation to perform in order to extract a loop integral. For the solution at order $n$ in the PL expansion, neglecting the various integrals and derivatives, the powers of $R$ that appear in each term will be in the form,
\eq{
R^{c+n(1-D)}\,,
}{}
for some number $c$. To apply the general formulae \Eqs{eq:oneloopfan}{eq:fourierR}, we write this as 
\eq{
R^{2(\alpha_{1}+\cdots +\alpha_{n})+n(1-D)}\,,
}{}
for some choice of $\alpha_{j}$ such that $c=\sum_{j=1}^{n} \alpha_{j}$. The choice is not unique, so one should aim to chose a minimal set with all $\alpha_{j}$ the smallest possible positive integers.

It may be the case, however, that some terms can only be written with some $\alpha_{j}$  being non-positive integers, precluding the use of \Eq{eq:oneloopfan}, because of the singular $\Gamma(\alpha)$ factor. There is a simple fix for all such terms, simply replace the offending factors using the formula,
\begin{equation}
    R^{j-D}=\frac{j+3-D}{b^2 (j+2-D)}R^{j+2-D}-\frac{1}{b^2 v^2 (j+2-D) (j+4-D)}\partial_{\tau}^{2}R^{j+4-D}\,,
\end{equation}
which will raise the power of the offending $\alpha$ until it is a positive integer. The result will be an expression for the trajectories which is immediately expressible in terms of loop integrals.

There are further simplifications one can apply to the result to make it even more compact. We utilize the following simple identities,
\eq{
\partial_{\tau}\left(R^{a}\right)\partial_{\tau}\left(R^{b}\right)&=ab(\sigma^{2}-1)\left(R^{a+b-2}-b^{2}R^{a+b-4}\right)\,,\\
\partial_{\tau}\left(R^{a}\right)\frac{1}{\partial_{\tau}}\left(R^{b}\right)&=\partial_{\tau}\left(R^{a}\frac{1}{\partial_{\tau}}R^{b}\right)-R^{a+b}\,,\\
\frac{1}{\partial_{\tau}}\left(R^{a}\right)\frac{1}{\partial_{\tau}}\left(R^{b}\right)&=\frac{1}{\partial_{\tau}}\left(R^{a}\frac{1}{\partial_{\tau}}R^{b}+R^{b}\frac{1}{\partial_{\tau}}R^{a}\right)\,.
}{}
The first formula is nothing other than an explicit evaluation of the derivatives. In terms of Feynman integrals it is rather simplifying though, because it is eliminating two powers of momenta from the numerator. The second formula is just integration by parts. The third equation is a way to planarize diagrams, which is also a simple integration by parts in the time domain.  In terms of momentum space propagators, it is just the partial fraction decomposition discussed by~\cite{Driesse:2024xad}.

With this strategy one can systematically compute solutions to a desired PL order. For example, to compute the radial action to 4PL order, one needs only 0SF and 1SF contributions. To compute the 1SF contribution one needs only the 1PL and 2PL trajectories, with the former given above and the latter given by,
\eq{
t_{2}=&k_{D}^{2}\bigg(\frac{2 (D-4) \sigma  \frac{1}{\partial_{\tau}}\left(R^{6-2 D}\right)}{\sigma ^2-1}-\frac{(D-5) \sigma  R^{3-D} \frac{1}{\partial_{\tau}}\left(R^{3-D}\right)}{\sigma ^2-1}\bigg) \, ,\\
x_{2}=&k_{D}^{2}\bigg(-\frac{b (D-3) \sigma ^2 \frac{1}{\partial_{\tau}}\left(R^{3-D} \frac{1}{\partial_{\tau}}\left(R^{1-D}\right)\right)}{\sigma ^2-1}+\frac{b (D-5) (D-3) \sigma ^2 \frac{1}{\partial_{\tau}}\left(R^{1-D} \frac{1}{\partial_\tau}\left(R^{3-D}\right)\right)}{\sigma ^2-1}\\
   &-\frac{b (D-3) \left((2 D-7) \sigma ^2-1\right) \frac{1}{\partial_{\tau}^{2}}\left(R^{4-2 D}\right)}{\sigma ^2-1}\bigg) \, ,\\
y_{2}=&k_{D}^{2}\bigg(-\frac{(D-5) \sigma ^2 R^{3-D}\frac{1}{\partial_{\tau}}\left(R^{3-D}\right) }{\left(\sigma^2-1\right)^{3/2}}+\frac{(13-3 D) \sigma ^2 \frac{1}{\partial_{\tau}}\left(R^{8-2 D}\right)}{2 b^2 (D-5) \left(\sigma^2-1\right)^{3/2}}+ \frac{(D-4) \sigma ^2 R^{5-D}\frac{1}{\partial_{\tau}}\left(R^{3-D}\right)}{b^2 (D-5) \left(\sigma ^2-1\right)^{3/2}}\\
   &-\frac{(D-4)^2 \sigma ^2 \frac{1}{\partial_{\tau}^{2}}\left(R^{3-D} \frac{1}{\partial_{\tau}}\left(R^{3-D}\right)\right)}{b^2 \sqrt{\sigma ^2-1}}+\frac{\left((4 D-15) \sigma^2-1\right) \frac{1}{\partial_{\tau}}\left(R^{6-2 D}\right)}{2 \left(\sigma ^2-1\right)^{3/2}}\bigg)\,.
}{eq:2PLEMtrajectory}
This expression is readily covariantized using \Eq{eq:covtrajectory} and expressed as loop integrals using \Eq{eq:fourierR}. The corresponding Feynman diagram topologies can be read off the time-domain expressions by noting that the $\partial_{\tau}^{-1}$ and $R$ have straightforward interpretations as background field insertions and matter propagators. For example,
\eq{
k_{D}^{2}\frac{1}{\partial_{\tau}}\left(R^{8-2 D}\right)=-i32(5-D)\pi^{2}\Rc^{2}\int_{\ell_{1}\,\ell_{2}}\frac{e^{-ir(\ell_{1}+\ell_{2})}\hat{\delta}(\vH\ell_{1})\hat{\delta}(\vH\ell_{2})}{\ell_{1}^{2}(\ell_{2}^{2})^{2}(\vL\ell_{1}+\vL\ell_{2})}\,,
}{traj_eg}
corresponds to the diagram topology shown in \Fig{fig:traj}.
The explicit factors of $D$ in the expressions \Eq{eq:2PLEMtrajectory} together with doubled propagators such as the factor of $(\ell_{2}^{2})^{-2}$ are both structures which do not arise in a standard Feynman diagrammatic computation and suggest that integral reduction has, at least partially, been automatically performed by passing through the time domain.

\begin{figure*}
    \centering
    \traj
    \caption{Diagram topology corresponding to \Eq{traj_eg}.}
    \label{fig:traj}
\end{figure*}

Covariantizing these expressions, and Fourier transforming, we find for the 1PL trajectory
\eq{
x^{\mu}_{1}=\frac{-i \pi\Rc}{\left(\sigma ^2-1\right)}\int_{\ell}e^{-ir \ell} \hat{\delta} \left(\vH\ell\right)\frac{ \left(4 (\vL\ell) (\vH^{\mu}-\sigma  \vL^{\mu})-(D-5) (D-3) \sigma  \left(\sigma ^2-1\right) \Pi^{\mu\nu}\ell_{\nu}\right)}{\ell^{2}(\vL\ell)^{2}  }\,.
}{}
To keep the expressions concise we present only the leading in $D=4$ contribution to the 2PL trajectory integrand,
\eq{
x^{\mu}_{2}=&-\frac{i4\pi^2\Rc^{2}}{b^{2}(\sigma^{2}-1)^{2}}\int_{\ell_{1},\ell_{2}}\frac{e^{-ir(\ell_{1}+\ell_{2})}\hat{\delta}(\vH\ell_{1})\hat{\delta}(\vH\ell_{2})}{\ell_{1}^{2}(\ell_{2}^{2})^{2}(\vL\ell_{2})(\vL\ell_{1}+\vL\ell_{2})^{2}}\bigg(4 b^2 (\vL\ell_{1}+\vL\ell_{2})^2 \ell_{2}^{2} \sigma  (\vH^{\mu}-\sigma  \vL^{\mu})\\
&+(\vL\ell_{1}+\vL\ell_{2}) \left(b^2 \ell_{2}^{2} \left(\sigma ^2-1\right) \sigma ^2 \Pi^{\mu\nu}(\ell_{1}{}_{\nu}+\ell_{2}{}_{\nu})+2
   (\vL\ell_{2}) \left(b^2 \ell_{2}^{2} \left(\sigma ^2-1\right)-2 \sigma ^2\right) (\sigma  \vH^{\mu}-\vL^{\mu})\right)\\
   &+b^2 \ell_{2}^{2} \left(\sigma ^2-1\right)^2 (\vL\ell_{2}) \Pi^{\mu\nu}\ell_{2}{}_{\nu}+\mathcal{O}(D-4)\bigg)\,.
}{}
The procedure continues straightforwardly to higher PL orders.  Note that the trajectory is only needed up to 2PL order in order to compute scattering at 4PL order.

\subsection{General Relativity}\label{app:trajGR}

The strategy of solving probe equations of motion in order to extract loop integrands is nearly identical between GR and EM. Indeed, this highlights an advantage of our approach, which is that the trajectories in  GR results are  similar in form and complexity to the those of EM, despite gravity being a nonlinear field theory.

We start from the worldline action for the probe,
\eq{
S=-\int d\tau\left[\frac{1}{2}e^{-1}g_{\mu\nu}(x)\dot{x}^{\mu}\dot{x}^{\nu}+\frac{1}{2}em^{2}\right]\,,
}{}
again dropping the subscripts and bars that would indicate we are describing the light particle in a background metric.  Here  the background metric sourced by a heavy particle without spin is just the Schwarzschild-Tangherlini solution,
\eq{
g_{00}=\frac{f_{-}(r)}{f_{+}(r)}^{2}\quad \textrm{and} \quad g_{ij} = -\delta_{ij}f_{+}(r)^{\frac{4}{D-3}}\,,
}{}
presented here in isotropic coordinates. We have defined the function,
\eq{
f_{\pm}(r)=1\pm\frac{\mu}{4r^{D-3}}\,,
}{}
in terms of a generalization of the mass parameter,
\eq{
\mu &= \frac{4\pi \Gamma(\frac{D-1}{2}) }{(D-2)\pi^{(D-1)/2}} \times 2G \mH \, ,
}{}
which is not to be confused a renormalization scheme subtraction scale.
Like before, gauge fixing the einbein to $e=1/m$ imposes the curved space on-shell condition, $g_{\mu\nu}\dot{x}^{\mu}\dot{x}^{\nu}=1$, on the space of solutions. 

The background is time-independent and spherically symmetric, so we can restrict to motion in the equatorial plane with dynamics constraints by the conserved energy and angular momentum. We will again prefer to label trajectories by ($\sigma,b$) using \Eq{def_sig_b}, in terms of which we have equations of motion,
\eq{
    \dot{t}=\sigma \frac{f_{+}(r)^{2}}{f_{-}(r)^{2}}\qquad\textrm{and}\qquad \dot{\phi}=\frac{b(\sigma^{2}-1)^{1/2}}{r^{2}}f_{+}(r)^{-\frac{4}{D-3}}\,,
}{}
with the radial equation of motion coming from the on-shell condition,
\eq{
\dot{r}=\left[f_{+}(r)^{-\frac{4}{D-3}}\left(\frac{f_{+}(r)^{2}}{f_{-}(r)^{2}}\sigma^{2}-1\right)-\frac{b^{2}(\sigma^{2}-1)}{r^{2}}f_{+}(r)^{-\frac{8}{D-3}}\right]^{1/2}\,.
}{}
As before, we will prefer to write equation of motion for the Cartesian components, 
\eq{
\dot{x}&=\frac{x}{r}\left[f_{+}(r)^{-\frac{4}{D-3}}\left(\frac{f_{+}(r)^{2}}{f_{-}(r)^{2}}\sigma^{2}-1\right)-\frac{b^{2}(\sigma^{2}-1)}{r^{2}}f_{+}(r)^{-\frac{8}{D-3}}\right]^{1/2}-\frac{y}{r^2}f_{+}(r)^{-\frac{4}{D-3}}(\sigma^{2}-1)^{1/2}b \, ,\\
\dot{y}&=\frac{y}{r}\left[f_{+}(r)^{-\frac{4}{D-3}}\left(\frac{f_{+}(r)^{2}}{f_{-}(r)^{2}}\sigma^{2}-1\right)-\frac{b^{2}(\sigma^{2}-1)}{r^{2}}f_{+}(r)^{-\frac{8}{D-3}}\right]^{1/2}+\frac{x}{r^2}f_{+}(r)^{-\frac{4}{D-3}}(\sigma^{2}-1)^{1/2}b\, .
}{GRcartesianEOM}
The solutions at order $n$ in the PM expansion scale as $\mathcal{O}(\mu^n)$.  As before the leading order solution is a straight line,
\eq{
    t_{0}=\sigma\tau\,,\qquad x_{0}=b\,,\qquad y_{0}=(\sigma^{2}-1)^{1/2}\tau\,,
}{}
while the higher order equations of motion are
\eq{
\frac{d}{d\tau}t_{n}(\tau)&=T_{n}(\tau) \, ,\\
\frac{d}{d\tau}\left(\frac{x_{n}(\tau)}{\tau}\right)&=\frac{X_{n}(\tau)}{\tau} \, ,\\
\frac{d}{d\tau}y_{n}(\tau)&=Y_{n}(\tau)\,.
}{}
The exact same strategy outlined in the previous section for EM can be used in GR. For example, the trajectory at 1PM order is
\eq{
t_{1}=&\mu \sigma  \frac{1}{\partial_{\tau}}\left(R^{3-D}\right)\, ,\\
x_{1}=&-\mu \frac{1}{2} b \left((D-2) \sigma ^2-1\right) \frac{1}{\partial^{2}_{\tau}}\left(R^{1-D}\right)\, ,\\
y_{1}=&\frac{\mu \left((D-4) \sigma ^2+1\right) }{2 (D-3) \sqrt{\sigma ^2-1}}\frac{1}{\partial_{\tau}}\left(R^{3-D}\right)\,,
}{}
while at 2PM we obtain
\eq{
t_{2}&=-\mu^{2}\frac{\sigma\left((D-4) \sigma ^2-1\right) }{2 \left(\sigma ^2-1\right)}R^{3-D}\frac{1}{\partial_{\tau}}\left(R^{3-D}\right)+\mu^{2}\frac{\sigma  \left((2 D-5) \sigma ^2-3\right)}{2\left(\sigma ^2-1\right)} \frac{1}{\partial_{\tau}}\left(R^{6-2 D}\right) \, ,\\
x_{2}&=-\mu^{2}\frac{b \left((D-4) \sigma ^2+1\right) \left((D-2) \sigma ^2-1\right) }{4 (D-3) \left(\sigma ^2-1\right)}\frac{1}{\partial_{\tau}}\left(R^{3-D} \frac{1}{\partial_{\tau}}\left(R^{1-D}\right)\right)\\
&+\mu^{2}\frac{b \left((D-4) \sigma
   ^2-1\right) \left((D-2) \sigma ^2-1\right) }{4 \left(\sigma ^2-1\right)}\frac{1}{\partial_{\tau}}\left(R^{1-D} \frac{1}{\partial_{\tau}}\left(R^{3-D}\right)\right)\\
   &-\mu^{2}\frac{b \left((2 D-5) \sigma ^2 \left((2 D-3) \sigma
   ^2-6\right)+3\right) }{8 \left(\sigma ^2-1\right)}\frac{1}{\partial^{2}_{\tau}}\left(R^{4-2 D}\right)\, ,\\
y_{2}&=-\mu^{2}\frac{(D-4)^2 \left((D-2) \sigma ^2-1\right)^2 }{4 b^2 (D-3)^2 \sqrt{\sigma ^2-1}}\frac{1}{\partial^{2}_{\tau}}\left(R^{3-D} \frac{1}{\partial_{\tau}}\left(R^{3-D}\right)\right)\\
&-\mu^{2}\frac{(3 D-13) \left((D-2) \sigma ^2-1\right)^2
   }{8 b^2 (D-5) (D-3)^2 \left(\sigma ^2-1\right)^{3/2}}\frac{1}{\partial_{\tau}}\left(R^{8-2 D}\right)+\mu^{2}\frac{ \left(1-(D-4)^2 \sigma ^4\right)}{4 (D-3) \left(\sigma ^2-1\right)^{3/2}}R^{3-D}\frac{1}{\partial_{\tau}}\left(R^{3-D}\right)\\
&+\mu^{2}\frac{(D-4)  \left((D-2) \sigma ^2-1\right)^2}{4 b^2 (D-5) (D-3)^2
   \left(\sigma ^2-1\right)^{3/2}}R^{5-D}\frac{1}{\partial_{\tau}}\left(R^{3-D}\right) \\   
&+\mu^{2}\frac{\left(\left(D \left(8 D^2-76 D+241\right)-251\right) \sigma ^4-2 (D (2 D-15)+31)
   \sigma ^2-7 D+25\right) }{16 (D-3)^2 \left(\sigma ^2-1\right)^{3/2}}\frac{1}{\partial_{\tau}}\left(R^{6-2 D}\right)\,.
}{}
Here we emphasize that we obtain the same topologies as in EM, without additional structures describing the graviton self-interactions of GR.

The transformation to momentum space was already outlined in the previous section. Here we present the $D$-dimensional trajectory at 1PM order,
\eq{
x_{1}^{\mu}=&-\frac{2 i \pi  G \mH }{(D-2)  \left(\sigma ^2-1\right) }\int_{\ell}\frac{e^{-ir \ell} \hat{\delta} \left(\vH\ell\right)}{\ell^{2}(\vL\ell)^2}\bigg(
(D-5) (D-3) \left(\sigma ^2-1\right) \left((D-2) \sigma ^2-1\right) \Pi^{\mu\nu}\ell_{\nu}\\
&+4 \sigma  (\vL\ell) \left((D-2) \sigma ^2-2 D+5\right) \vH^{\mu}+4 (\vL\ell) \left((D-4) \sigma ^2+1\right) \vL^{\mu}\bigg)\,.
}{}
Again, since the $D$-dependent coefficients in the numerator are lengthy, we will present the loop integrand for the 2PM trajectory in $D=4$,
\eq{
x_{2}^{\mu}=&-\frac{2 i \pi ^2 G^2 \mH^2 }{b^2  \left(\sigma ^2-1\right)^2}\int_{\ell_{1},\ell_{2}}\frac{e^{-ir \left(\ell_{1}+\ell_{2}\right)} \hat{\delta} \left(\vH\ell_{1}\right) \hat{\delta} \left(\vH\ell_{2}\right)}{ \ell_{1}^{2} (\ell_{2}^{2})^2(\vL\ell_{2})
   \left(\vL\ell_{1}+\vL\ell_{2}\right){}^2}\bigg(
   3 b^2 \ell_{2}^{2} \left(\sigma ^2-1\right)^2 \left(5 \sigma ^2-1\right) (\vL\ell_{2}) \Pi^{\mu\nu}\ell_{2}{}_{\nu}\\
   &-8 b^2 \ell_{2}^{2} \left(\vL\ell_{1}+\vL\ell_{2}\right){}^2 \left(\sigma  \left(2 \sigma ^2-3\right)
   \vH^{\mu}+\vL^{\mu}\right)\\
   &+2 b^2 \ell_{2}^{2} \left(2 \sigma ^4-3 \sigma ^2+1\right) \left(\vL\ell_{1}+\vL\ell_{2}\right) \Pi^{\mu\nu}(\ell_{1}{}_{\nu}+\ell_{2}{}_{\nu})\\
   &-8 \left(1-2 \sigma ^2\right)^2 (\vL\ell_{2}) \left(\vL\ell_{1}+\vL\ell_{2}\right) (\sigma  \vH^{\mu}-\vL^{\mu})\\
   &-6 b^2 \ell_{2}^{2} \left(\sigma ^2-1\right) (\vL\ell_{2}) \left(\vL\ell_{1}+\vL\ell_{2}\right) \left(\sigma  \left(\left(5 \sigma ^2-9\right) \vH^{\mu}+3 \sigma  \vL^{\mu}\right)+\vL^{\mu}\right)+\mathcal{O}(D-4)\bigg)\,.
}{}
Note that in GR, the trajectories are only needed up to 2PM order in order to compute scattering up to 4PM order. 

\section{Probe Radial Actions}\label{app:probe_action}

\subsection{On-shell Action and Radial Action}
This section will review some aspects of classical mechanics. Consider the generally covariant action for a massive charged particle,
\eq{
S=-\int d\lambda \left[\frac{e^{-1}}{2}\dot{x}^{\mu}\dot{x}^{\nu}g_{\mu\nu}(x)+\frac{e}{2}m^{2}+q \dot{x}^{\mu}A_{\mu}(x)\right]\,,
}{}
where we have included the worldline einbein $e(\tau)$ for manifest reparameterization invariance. If we define the conjugate momentum,
\eq{
p_{\mu}=-\frac{d L}{d\dot{x}^{\mu}},
}{}
then the action can be expressed in the first-order form,
\eq{
S&=\int d\lambda\left[-p_{\mu}\dotxL^{\mu}-e\mathcal{H}\right]\, ,
}{}
where the reparameterization generator for this problem is\footnote{One can also consider problems with a more general $\mathcal{H}(p,x)$, and what follows will continue to hold.}
\eq{
\mathcal{H}=\frac{1}{2}\left(m^2-(p_{\mu}-q A_{\mu}(x))(p_{\nu}-q A_{\nu}(x))g^{\mu\nu}\right)\,.
}{}
As written, the above action exhibits a sensible variational principle if prescribe Dirichlet boundary conditions for the $x^{\mu}$, but not for the $p^{\mu}$. If we instead desire an action suitable for Dirichlet conditions on the momenta we could simply add the boundary counter term,
\eq{
S_{\partial}=\int d\lambda\,\frac{d}{d\lambda}(p_{\mu}x^{\mu})\, .
}{}
This does not change the equations of motion, but it allows for a well-defined variational principle and modifies the value of the on-shell action. Since we are interested in scattering dynamics, we  will fix the asymptotic momenta. The upshot of this choice is that the correct action for the problem of scattering from a static isotropic source is
\eq{
S&=\int d\lambda\left[\dot{p}_{r}\xL^{r}+\dot{p}_{t}\xL^{t}+\dot{p}_{\phi}\xL^{\phi}-e\mathcal{H}\right]\,.
}{}
This expression simplifies considerably in the on-shell limit because \textit{i})
energy and angular momentum are conserved, and \textit{ii}) the einbein is a Lagrange multiplier enforcing the vanishing of the reparameterization generator on physical solutions. This leaves the radial contribution to the on-shell 0SF action,
\eq{
\bar{S}&=\int\,d\lambda \dot{p}_{r}(r,E,J)r.
}{}
Integrating by parts and using time reversal symmetry, we recast this as
\eq{
\bar{S}=\lim_{r_{\rm max}\rightarrow\infty}\left(2r_{\textrm{max}}p_{r}(r_{\max})-2\int_{r_{\textrm{min}}}^{r_{\textrm{max}}}\,dr\,p_{r}(r,E,J)\right)\,.
}{}
where the radial momentum is
\eq{
p_{r}=e^{-1}\dot{x}^{\mu}g_{r\mu}(x)+q A_{r} \, .
}{}
Thus we discover that $\bar S$ is nothing but the radial action, suitably subtracted to be finite as $r_{\textrm{max}}\rightarrow\infty$. This subtraction will affect only the ``free-particle'' contribution to the radial action, and leaves ``scattering'' contributions unaffected, so we will omit the former in our subsequent discussion.

The total change in azimuthal angle follows straightforwardly from Hamilton's equation. Since $J=-p_{\phi}$ is a constant of motion, we have
\eq{
-\frac{d\bar{S}}{dJ}=\left(\int d\lambda\, e\frac{d}{dp_{\phi}}\mathcal{H} \right)\bigg|_{\textrm{on-shell}}=\int d\lambda \dot{\phi}=\Delta \phi\, ,
}{}
which can be used to compute the scattering angle.

\subsection{General Perturbative Radial Action Integral}
For a general $D$-dimensional theory,  the on-shell radial action for a scattering solution is 
\eq{
\bar{S} &= 2\int^{\infty}_{r_{\textrm{min}}}dr\, |p_{r}(r)|
\, ,
}{probeRadialAction}
where 
$r_{\textrm{min}}$ its positive  real zero of $p_r$. In this work, we are only interested in scattering from static spherically symmetric backgrounds. To solve for $p_{r}$, we first write the momentum as
\eq{
p_\mu = (\sqrt{p^2+m^2}, \vec p\,(p,r)) \, ,
}{P_def}
with $p$ the asymptotic spatial momentum, and then impose the on-shell condition $\dot x^2=1$.
which implies the following form for the spatial momentum,
\eq{
{\vec p}^{\,2}(p,r) =  p^2  + \sum_{k=1}^\infty \frac{\epsilon^k N_k(p)}{r^{k(D-3)}} \,.
}{general_onshell_condition}
which applies with and without the inclusion of $A_\mu$. Then we write
\eq{
p_r= \sqrt{{\vec p}^{\,2}-p_\phi^2},
}{}
To compute \Eq{probeRadialAction} perturbatively in $\epsilon$ requires some care. Since the radial momentum is solved for from the quadratic equation \Eq{general_onshell_condition}, it is singular at the turning point $r_{\textrm{min}}$, and a na\"{i}ve expansion in $\epsilon$ yields divergent integrals. A treatment of this is given in \cite{goldstein:mechanics,damourschaefer1988}, and the upshot is that one should: \textit{i}) integrate to the unperturbed turning point $r_{\textrm{min}}(\epsilon=0)$, \textit{ii}) Taylor expand $p_{r}$ in $\epsilon$, and \textit{iii}) impose a hard cut-off scheme $r_{\rm min}=b+\delta$, and simply discard power-law divergences in $1/\delta$, retaining only the finite part of the integrals. The fact that this simple renormalization scheme is indeed correct follows from knowledge of the analytic structure of the exact radial action integrand.
 

Following this prescription yields the well defined series expansion for the radial action
\eq{
\bar{S}=\sum_{k=0}^{\infty}\sum_{q=0}^{k}\epsilon^{k}c_{k,q}(p)p^{1-2q}\,\int_{b+\delta}^{\infty}dr\,\frac{(r^{2}-b^{2})^{1/2-q}}{r^{1-2q+k(D-3)}}\, ,
}{}
where $\delta$ is to be taken to zero after discarding power divergent terms, $b=J p^{-1}$, $J$ is the conserved angular momentum, and the $c_{k,q}(\sigma)$ are simple monomials in the $N_{k}(p)$. Through $\mathcal{O}(\epsilon^{3})$ the non-zero coefficients are
\eq{
c_{0,0}&=2\hspace{39pt} c_{1,1}=N_{1}\hspace{54pt} c_{2,2}=-\tfrac{1}{4}N_{1}^2 \qquad
c_{2,1}=N_{2} \nonumber \\
c_{3,3}&=\tfrac{1}{8}N_{1}^3 \qquad c_{3,2}=-\tfrac{1}{2}N_{1}N_{2} \qquad c_{3,1}=N_{3} \, .
}{}
Evaluating the integral yields the general result
\eq{
\bar{S}=\frac{1}{2}\sum_{k=0}^{\infty}\sum_{q=0}^{k}\epsilon^{k}c_{k,q}(p)\frac{p^{k(D-3)-2q}}{J^{k(D-3)-1}} B(\tfrac{1}{2}k(D-3)-\tfrac{1}{2},\tfrac{3}{2}-q) \,,
}{generalProbeRadialAction}
where $B$ is the Euler beta function.

\subsection{Electromagnetism}

For electromagnetism, the on-shell condition is
\eq{
(p_\mu - \qL A_\mu)^2 = \mL^2 \, .
}{on-shell_EM}
The $D$-dimensional attractive Coulomb potential is 
\eq{
\frac{\qL}{\mL} A_0 = - \frac{\Gamma(\frac{D-3}{2})}{\pi^{\frac{D-3}{2}}}\frac{\Rc}{r^{D-3}} \, ,
}{}
where $\alpha$ is the fine structure constant.  Inserting \Eqs{P_def}{general_onshell_condition} with $p_\phi^2=\frac{J^2}{r^2}= \frac{p^2 b^2}{r^2}$, and solving for ${\vec p}^{\,2}(p,r)$ we can extract the $N_i(p)$.  Plugging back into the general probe radial action \Eq{generalProbeRadialAction} under the identification $\epsilon = \Rc$, and writing $p=\mL(\sigma^{2}-1)^{1/2}$ to conform with our notation throughout the main text, we obtain
\eq{
I_{\rm EM}^{(0)} &= -\pi b\mL\sqrt{\sigma ^2-1}+\mL\Rc b^{4-D}\,\frac{\pi ^{2-\frac{D}{2}} \sigma   \Gamma \left(\frac{D}{2}-2\right)}{\sqrt{\sigma ^2-1}} \nonumber \\
&+\mL\Rc^{2}b^{7-2 D}\frac{(2\pi) ^{4-D } \left(2 D  \sigma ^2-7 \sigma ^2-1\right) \Gamma \left(D -\frac{7}{2}\right) \Gamma \left(\frac{D -3}{2}\right)^{2}}{2 \left(\sigma ^2-1\right)^{3/2} \Gamma
   \left(D-3\right)} \nonumber \\
   &+\mL\Rc^{3}b^{10-3 D }\frac{\pi ^{5-\frac{3 D }{2}} \sigma   \left((3 D -10) \sigma ^2-3\right) \Gamma \left(\frac{D -3}{2}\right)^3 \Gamma \left(\frac{3 D }{2}-5\right)}{3 \left(\sigma ^2-1\right)^{5/2} \Gamma
   \left(\frac{3 D }{2}-\frac{11}{2}\right)}
    \, ,
}{}
expanding up to 3PL order.  These expressions hold for general spacetime dimension $D$.

\subsection{General Relativity}

For gravity, the on-shell condition is
\eq{
p_\mu p_\nu g^{\mu\nu} = \mL^2 \, ,
}{on-shell_GR}
where $g_{\mu\nu}$ is the background metric in isotropic coordinates. For example the $D$-dimensional Schwarzschild metric in isotropic coordinates is
\eq{
g_{00} = \left(\frac{1-\frac{\mu}{4r^{D-3}}}{1+\frac{\mu}{4 r^{D-3}}}\right)^2 \quad \textrm{and} \quad g_{ij} = -\delta_{ij} (1+\frac{\mu}{4r^{D-3}})^{\frac{4}{D-3}} \, ,
}{metric_iso_D}
where we have defined the mass parameter,
\eq{
\mu &= \frac{4\pi \Gamma(\frac{D-1}{2}) }{(D-2)\pi^{(D-1)/2}} \times 2G \mH \, .
}{}
Solving \Eq{on-shell_GR}, we obtain the corresponding $N_i(p)$.   Identifying $\epsilon= \RS$, we obtain the probe radial action,
\eq{
I_{\rm GR}^{(0)} &= -\pi b \mL \sqrt{\sigma ^2-1}+G\mH\mL b^{4-D}\frac{2 \pi ^{2-\frac{D}{2}} \left((D-2) \sigma ^2-1\right) \Gamma \left(\frac{D}{2}-2\right)}{(D-2) \sqrt{\sigma ^2-1}} \nonumber \\
&+(G\mH)^{2}\mL b^{7-2 D}\frac{(2 \pi )^{4-D}  \left((2 D-5) \sigma ^2 \left((2 D-3) \sigma ^2-6\right)+3\right) \Gamma \left(D-\frac{7}{2}\right) \Gamma \left(\frac{D-1}{2}\right)}{(D-2) \left(\sigma ^2-1\right)^{3/2} \Gamma
   \left(\frac{D}{2}\right)} \nonumber \\
   &+(G\mH)^{3}\mL b^{10-3 D} \bigg(\frac{16 \pi ^{5-\frac{3 D}{2}} \left((3 D-8) \sigma ^2 \left((D-2) \sigma ^2 \left((3 D-4) \sigma ^2-15\right)+15\right)-5\right) }{(D-2)^3 \left(\sigma ^2-1\right)^{5/2} \Gamma \left(\frac{3 D}{2}-\frac{7}{2}\right)}\nonumber \\
   &\qquad\qquad\qquad\qquad\qquad\times \Gamma \left(\frac{D-1}{2}\right)^3 \Gamma \left(\frac{3  D}{2}-5\right)\bigg)
\, 
}{}
expanding up to 3PM order. 

More generally, if a metric can be put in isotropic coordinates,
\eq{
g_{00} = A(\epsilon/r^{D-3}) \quad \textrm{and} \quad g_{ij} = -\delta_{ij} B(\epsilon/r^{D-3}) \, ,
}{}
with $A(0)=B(0)=1$, then one can express the spatial momentum in the form \Eq{general_onshell_condition} with the first few $N_{i}(p)$ given by
\eq{
N_{1}(p)&=\left(m^2+p^2\right) \left(B'(0)-A'(0)\right)-m^2 B'(0) \nonumber \\
N_{2}(p)&=\frac{1}{2} \left(m^2+p^2\right) \left(-2 A'(0) B'(0)+2 A'(0)^2-A''(0)+B''(0)\right)-\frac{1}{2} m^2 B''(0) \nonumber \\
N_{3}(p)&=\frac{1}{6} \left(m^2+p^2\right) \left(A'(0) \left(6 A''(0)-3 B''(0)\right)-3 A''(0) B'(0)+6 A'(0)^2 B'(0)-6 A'(0)^3\right. \nonumber \\
&\left.-A^{(3)}(0)+B^{(3)}(0)\right)-\frac{1}{6} m^{2}B^{(3)}(0) \,.
}{}

\vspace{\baselineskip}

\medskip
\noindent {\bf Acknowledgments:} C.C., N.S., and J.W.-G. are supported by the Department of Energy (Grant No. DE-SC0011632) and by the Walter Burke Institute for Theoretical Physics. J.W.-G is also supported by a Presidential Postdoctoral Fellowship and the Simons Foundation (Award No. 568762). I.Z.R. is supported by the Department of Energy (Grant No. DE-FG02-04ER41338 and FG02-06ER41449). I.Z.R is grateful to the Burke center for theoretical physics for its hospitality.

\vspace{\baselineskip}

\vspace{-5mm}

\bibliographystyle{utphys-modified}
\bibliography{refs}

\end{document}